\documentclass[aps,preprint, superscriptaddress,nofootinbib]{revtex4}
\usepackage{amsmath}
\usepackage{graphicx}
\usepackage{subfigure}
\usepackage{amssymb}
\usepackage{xcolor}
\usepackage{multirow}
\usepackage{cancel}
\usepackage{color}
\usepackage{ulem}
\usepackage{listings}
\usepackage{xcolor}
\usepackage{float}
\usepackage{threeparttable}

\usepackage[colorlinks,citecolor=blue]{hyperref}
\usepackage{amsmath}%\ge=\geqslant; \le=\leqslant
\usepackage{slashed}

\newcommand{\be}{\begin{equation}}
\newcommand{\ee}{\end{equation}}
\newcommand{\beq}{\begin{equation}}
\newcommand{\eeq}{\end{equation}}
\newcommand{\bea}{\begin{eqnarray}}
\newcommand{\eea}{\end{eqnarray}}
\newcommand{\besp}{\begin{equation}\begin{split}}
\newcommand{\eesp}{\end{split}\end{equation}}

\newcommand{\Br}{\text{Br}}
\newcommand{\GeV}{\text{GeV}}
\newcommand{\TeV}{\text{TeV}}
\newcommand{\Det}{\text{Det}}
\newcommand{\tabincell}[2]{\begin{tabular}{@{}#1@{}}#2\end{tabular}}

\newcommand{\Dfbd}{\mathord{\buildrel{\lower3pt\hbox{$\scriptscriptstyle\leftrightarrow$}}\over {D}_{\mu}}}
\newcommand{\dfbd}{\mathord{\buildrel{\lower3pt\hbox{$\scriptscriptstyle\leftrightarrow$}}\over {\partial}_{\mu}}}

\def\barray{\begin{array}{cccccccccc}}
\def\earray{\end{array}}

\def\mG{\mathcal{G}}

\def\mL{\mathcal{L}}

\def\mO{\mathcal{O}}

\def\mT{\mathcal{T}}

\def\1{\textbf{1}}
\def\2{\textbf{2}}
\def\3{\textbf{3}}
\def\4{\textbf{4}}
\def\5{\textbf{5}}
\def\6{\textbf{6}}
\def\7{\textbf{7}}
\def\8{\textbf{8}}
\def\9{\textbf{9}}

\def\iab{ab$^{-1}$}

\newcommand{\Eq}[1]{Eq.~(\ref{#1})}
\newcommand{\Ref}[1]{Ref.~\cite{#1}}
\newcommand{\Fig}[1]{Fig.~\ref{#1}}

\begin{document}

\title{Prospects of searching for composite resonances at the LHC and beyond}

\author{Da Liu}
\email{da.liu@anl.gov}
\affiliation{High Energy Physics Division, Argonne National Laboratory, Argonne, IL 60439, USA}

\author{Lian-Tao Wang}
\email{liantaow@uchicago.edu}
\affiliation{Enrico Fermi Institute, The University of Chicago, 5640 S Ellis Ave, Chicago, IL 60637, USA}
\affiliation{Department of Physics, The University of Chicago, 5640 S Ellis Ave, Chicago, IL 60637, USA}
\affiliation{Kavli Institute for Cosmological Physics, The University of Chicago, 5640 S Ellis Ave, Chicago, IL 60637, USA}

\author{Ke-Pan Xie}
\email{kpxie@snu.ac.kr}
\affiliation{Center for Theoretical Physics, Department of Physics and Astronomy, Seoul National University, Seoul 08826, Korea}
\affiliation{Department of Physics and State Key Laboratory of Nuclear Physics and Technology, Peking University, Beijing 100871, China}

\begin{abstract}
Composite Higgs models predict the existence of resonances. We study in detail the collider phenomenology of both the vector and fermionic resonances, including the possibility of both of them being light and within the reach of the LHC. We present current constraints from di-boson, di-lepton resonance searches and top partner pair searches  on a set of simplified benchmark  models based on the minimal coset $SO(5)/SO(4)$, and make projections for the reach of the HL-LHC. We find that the cascade decay channels for the vector resonances into top partners, or vice versa, can play an important role in the phenomenology of the models. We present a conservative estimate for their reach by using the same-sign di-lepton final states. As a simple extrapolation of our work, we also present the projected reach at the 27 TeV HE-LHC and a 100 TeV $pp$ collider. 

\end{abstract}

\maketitle

\tableofcontents

\section{Introduction}

A promising way of addressing the naturalness problem is to consider the existence of strong dynamics around several to 10 TeV scale. The Higgs boson is a pseudo-Nambu-Goldstone boson, much like the pions in the QCD. This so-called composite Higgs scenario~\cite{Kaplan:1991dc,Kaplan:1983fs,Contino:2003ve} has become a main target for the search of new physics at the Large Hadron Collider (LHC).

A generic prediction of the composite Higgs scenario is the presence of composite resonances. Frequently considered resonances are either  spin 1, analogous to $\rho$-meson in QCD, or spin $1/2$ resonances with quantum numbers similar to those of the top quark, called ``top partners". In this paper, we study in detail the collider phenomenology of both kinds of resonances. We focus on the minimal coset $SO(5)/SO(4)$, denoted as the Minimal Composite Higgs Model (MCHM)~\cite{Agashe:2004rs,Contino:2006qr}. We included several benchmark choices of  both the spin 1 resonance and the top partner:  $\rho_L(\3,\1)$, $\rho_R(\1,\3)$, $\rho_X(\1,\1)$, $\Psi_\4(\2,\2)$ and $\Psi_\1(\1,\1)$. We derive the current constraints, and make projections for the reach of HL-LHC. We also make a simple extrapolation to estimate the prospectives at the 27 TeV HE-LHC~\cite{Zimmermann:2017bbr}  and the 100 TeV $pp$ collider~\cite{Arkani-Hamed:2015vfh,CEPC-SPPCStudyGroup:2015csa,Golling:2016gvc,Contino:2016spe}. Search channels in which the composite resonances are produced via Drell-Yan process and then decay into the Standard Model (SM) final states, such as di-lepton, di-jet, $t \bar{t}$ and di-boson, are well known. We update the limits by including the newest results at the 13 TeV LHC, such as the boosted di-boson jet resonance searches performed  by ATLAS with integrated luminosity  $L=$ 79.8 fb$^{-1}$~\cite{ATLAS-CONF-2018-016}, the di-lepton resonance search at CMS with integrated luminosity $ L = 77.3~{\rm fb}^{-1}$ for the electron channel and $L = 36.3~{\rm fb}^{-1}$ for the muon channel~\cite{CMS-PAS-EXO-18-006}, and the search for the pair production of top quark partners with charge-$5/3$ at CMS with integrated luminosity $L=35.9$ fb$^{-1}$~\cite{Sirunyan:2018yun}.  In addition, we paid close attention to scenarios in which the spin-1 resonances and top partners can be comparable in mass. In this case, cascade decays in which one composite resonance decays into another, can play an important role~\cite{Greco:2014aza,Niehoff:2015iaa,Yepes:2018dlw,Barducci:2012kk,Barducci:2014kxa,Yepes:2017pjr}. In particular, the channels $\rho^+_L \rightarrow t \bar{B}/X_{5/3}\bar{t}$ or $\rho^+_L \rightarrow X_{5/3}\bar{X}_{2/3}$ and $\rho^0_L \rightarrow X_{5/3}\bar{X}_{5/3}$ can have significant branching ratios for models with quartet top partner, if $\rho_L$ is in the intermediate mass region $M_\Psi < M_\rho < 2 M_{\Psi}$ or the high mass region $M_\rho > 2 M_\Psi$, respectively. Such cascade decays can lead to the same-sign di-lepton (SSDL) signals. Since these are relative clean signals, which have already been used for LHC searches, we use them in our recast and estimate the prospective reach on the $M_\rho - M_\Psi$ plane. They are comparable in some regions of the parameter space to the di-boson searches for the spin-1 resonances and the pair-produced top partner searches at the LHC. For the models with a singlet top partner, the cascade decay channel $\widetilde{T}\rightarrow t\rho_X \rightarrow t\bar{t}t$ in the single production channel can play an important role in the mass region $M_{\widetilde{T}} > M_{\rho_X}$. The reach at the LHC is also estimated in the SSDL channels. The projections made based on only the SSDL channel are of course conservative.  Other decay modes of the cascade decay channels mentioned above  can further enhance the reach,  such as the ones including more complicated final states like $1\ell +{\rm jets}$ channels. We leave a detailed exploration of such additional channels for a future work.

The paper is organized as follows. In Section~\ref{sec:pheno}, we summarize the main phenomenological features of the models, including the couplings of the particles  in the mass eigenstates, and the production and the decay of the resonances.   The details of the models are presented in  Appendix~\ref{app:CCWZ} and  Appendix~\ref{app:models}. In Section~\ref{sec:bound}, we show the present bounds from the LHC searches and extrapolate the results to the HL-LHC with an integrated luminosity of  $L = 3$ \iab. An estimate of the reach at the 27 TeV HE-LHC and 100 TeV $pp$ collider is also included. We conclude in Section~\ref{sec:conclusion}.

\section{Phenomenology of the models}
\label{sec:pheno}

We begin with a brief review of the composite Higgs models under consideration.  We will describe the particle content, and give a qualitative discussion of the sizes of various couplings. The details of the models are presented in Appendix~\ref{app:CCWZ} and ~\ref{app:models}.

We will consider models similar to those presented in Ref.~\cite{Greco:2014aza}. The strong dynamics is assumed to have a global symmetry $SO(5)$, which is broken spontaneously to $SO(4) \simeq SU(2)_L \times SU(2)_R$. The resulting Goldstone bosons, parameterizing the coset $SO(5)/SO(4)$, contain the Higgs doublet. This is the minimal setup with a custodial $SU(2)$ symmetry. The composite resonances furnish  complete representations of $SO(4)$.

\begin{table}[h]
\centering
Particle content\\
\vspace{0.5cm}
\begin{tabular}{|c|c|c|c||c|c|c|c|c|}
\hline
&$\rho_L$ & $\rho_R$ & $\rho_X$ & $\Psi_\4$  & $\Psi_\1$ & $q_L^\5$ & $t_R^\5$ & $t_R^{\rm (F)}$\\
\hline
 $SO(4) \simeq SU(2)_L \times SU(2)_R$  & (\3,\1) & (\1,\3) & (\1,\1) & (\2,\2) & (\1,\1) & (\2,\2) & (\1,\1) & (\1,\1) \\
 \hline
\end{tabular} \\
\vspace{0.5cm}
\centering
Models considered\\
\vspace{0.5cm}
\begin{tabular}{|c|c|c|c|c|c|c|c|c|c|c|c|}
 \hline
Interaction& $\rho_L,\Psi_{\bold 4}$ & $\rho_R,\Psi_{\bold 4}$  &  $\rho_X,\Psi_{\bold 4}$  & $\rho_X, \Psi_\1$ \\
\hline
Model & LP(F)$_\4$ & RP(F)$_\4$  & XP(F)$_\4$ & XP(F)$_\1$ \\
\hline
\end{tabular} 
\caption{ Upper table: the particle content considered in this paper and their representations  under the unbroken $SO(4) \simeq SU(2)_L \times SU(2)_R$. The SM left-handed quarks $q_L = (t_L,b_L)^T$ are embedded into an incomplete representation, \5, of $SO(5)$. We consider two possible origins of the right-handed top quark. It can be partially composite, denoted as $t_R^{\rm (P)}$, and it is embedded in an incomplete representation, \5, of $SO(5)$. It can also be a fully composite resonance, denoted as $t_R^{\rm (F)}$, and it is assumed to be  an $SO(4)$ singlet massless bound state. Their representations under the unbroken $SO(4)$ are also presented in the table.  Lower table: the models with different combinations of the composite spin-1 resonances $\rho$ and the fermionic resonances $\Psi$ considered in our paper. P (F) denotes the partially (fully) composite right-handed top quark. }
\label{tab:models}
\end{table}
We summarize the particle content and the models considered in our paper  in Table~\ref{tab:models}.
For the spin-1 resonances $\rho$, we consider three representations under the unbroken $SO(4)\simeq SU(2)_L \times SU(2)_R$: $\rho_L(\3,\1)$, $\rho_R (\1,\3)$, $\rho_X(\1,\1)$, while for the fermionic resonances $\Psi$, we  study the quartet $\Psi_\4(\2,\2)$ and the singlet $\Psi_\1(\1,\1)$. The left handed SM fermions, $q_L = (t_L,b_L)^T$, are assumed to be embedded into (incomplete) \5 representations of $SO(5)$ (see \Eq{qL5_definition})~\cite{Contino:2006qr}. There are two well-studied ways of dealing with the right handed top quark. First, it can be treated as an elementary field, and embedded into a \5 representation of $SO(5)$ (see \Eq{tR5_definition})~\cite{Contino:2006qr}. We call this the partially composite right-handed top quark scenario, and denote right-handed top 
as $ t^{\rm (P)}_R$.  It is also possible that it is a massless bound state of the strong sector and a $SO(4)$ singlet, denoted as $t_R^{\rm (F)}$~\cite{DeSimone:2012fs}. We call this the fully composite right-handed top quark scenario. We will consider both of these cases. In principle, many of the composite resonances can be comparable in their masses in a given model. Rather than getting in the numerous combinations, we consider a set of simplified models in which only one kind of spin-1 resonance(s) and one kind of top partner(s) are light and relevant for collider searches. 
For example, model  LP$_\4$ involves the strong interactions between the $\rho_L$ and the quartet top partner $\Psi_\4$ and the partially composite right-handed top quark. In comparison, model LF$_\4$ is different only in the treatment of the right handed top quark which is assumed to be fully composite.

In the following, we will first discuss all the most relevant interactions and their coupling strengths in Section~\ref{sec:couplings}. The production and decay of the resonances at the LHC are presented in Section~\ref{sec:production_decay}. The mass matrices of different models and their diagonalizations are discussed in Appendix~\ref{app:mm}, where we also list the expressions all the mass eigenvalues.

\subsection{The couplings}
\label{sec:couplings}
Scale $f$, similar to the pion decay constant in QCD, parameterizes the size of global symmetry breaking. The parameter $\xi = v^2/f^2$ measures the hierarchy between the weak scale and the global symmetry breaking scale in the strong sector. It has been well constrained from LEP electroweak precision test (EWPT) and the LHC  Higgs coupling measurements to be $\xi  \lesssim 0.13$~\cite{Sanz:2017tco,deBlas:2018tjm}. In the expressions for the couplings, we will  keep only terms to  the leading order in $\xi$. 

The interactions of the spin-1 resonances in the strong sector are characterized by several couplings, $(g_{\rho_L}, g_{\rho_R}, g_{\rho_X})$, sometimes collectively denoted as $g_\rho$.  Typically, they are assumed to be much larger than the SM gauge couplings, i.e. $g_{\rho} \gg g'$, $g$. We will keep only terms to the leading order in $g/g_\rho$ in the expressions of the couplings~\footnote{Note that the gauge couplings $g',g$ are defined through the leading-order (LO) formulae of the $W,Z$ masses and can be different from the Lagrangian parameters $g_1, g_2$. See Appendix~\ref{app:models} for detail. }. 
Similar to Ref.~\cite{Contino:2011np}, we will also introduce an $\mO(1)$ parameter for each representation of the  spin-1 resonances, defined as
\beq
a_{\rho_{L,R,X}} =\frac{m_{\rho_{L,R,X}}}{g_{\rho_{L,R,X}}f}. 
\eeq
In most of the cases, we will fix $a_\rho$.

The sector of fermionic composite resonances involve another strong coupling, $g_\Psi$, defined as:
\beq
g_{\Psi} = \frac{M_\Psi}{f}, \quad M_\Psi = M_\4,~M_\1.
\eeq
For partially composite SM fermions,  there are mixings between the SM fermions and the top partners before electroweak symmetry breaking (EWSB). For example, the mixing angles between the elementary left (right) handed top and the quartet (singlet) top partners  (defined in \Eq{eq:lmix} and \Eq{eq:rmix}) in models within the partially composite right-handed top quark scenario are:
\beq
\label{eq:lrmix}
s_{\theta_L} \equiv \sin\theta_L = \frac{y_L f}{\sqrt{M_\4^2 + y_L^2 f^2}}, \quad  s_{\theta_R} \equiv \sin\theta_R = \frac{y_R f}{\sqrt{M_\1^2 + y_R^2 f^2}}.
\eeq
and  the same definition applies to $c_{\theta_L}$, $c_{\theta_R}$, $t_{\theta_L}$, $t_{\theta_R}$. The interactions of the spin-1 resonances and the fermions are summarized in Table~\ref{tab:charge} (for the charged sector) and Table~\ref{tab:neutral}, Table~\ref{tab:rhox1} (for the neutral sector). 
\begin{table}[h]
\small\centering
\begin{threeparttable}
\begin{tabular}{|c|c|c|c|c|c|c|c|c|}
\hline
Vertices & $\rho_L^+$  & $\rho_R^+$ &  $W^+$  \\ \hline
\multicolumn{4}{|c|}{Between heavy resonances:}\\
\hline
$\begin{array}{cccc}
\bar{T}_R B_R, \bar{X}_{5/3R} X_{2/3R}\\
 \bar{X}_{5/3L} X_{2/3L}
\end{array} $&$\begin{array}{cc}  \frac{c_1}{\sqrt{2}} g_{\rho_L} \end{array}$ &$\mO(g_{\rho_R} \xi)$ & $\frac{g}{\sqrt{2}} $ \\
 \hline
 $\bar{T}_L B_L$ &$\begin{array}{cc}  \frac{c_1}{\sqrt{2}} g_{\rho_L} c_{\theta_L}^2 \end{array}$ & $\mO(g_{\rho_R} \xi)$ & $\frac{g}{\sqrt{2}} $ \\
 \hline
 $\bar{X}_{2/3R} B_R,\bar{X}_{5/3R} T_R$ &$\mO(g_{\rho_L}\xi)$    & $ c_1\frac{g_{\rho_R}}{\sqrt{2}} \, $ &  $\mO(g\xi)$  \\ 
 \hline
 $\bar{X}_{2/3L} B_L,\bar{X}_{5/3L} T_L$ &$\mO(g_{\rho_L}  \xi)$   & $ c_1\frac{g_{\rho_R}}{\sqrt{2}} \, c_{\theta_L}$ &  $\mO(g\xi)$  \\ \hline
\multicolumn{4}{|c|}{Between heavy resonances and SM fermions:}\\
\hline
 $\bar{T}_L b_L, \bar{t}_L B_L$ &  $ \frac{c_1}{\sqrt{2}}g_{\rho_L}  c_{\theta_L} s_{\theta_L}$ & $\mO(g_{\rho_R} \xi)$ &$ \mO(g\xi)$ \\ \hline
$\bar{X}_{2/3L} b_L,\bar{X}_{5/3L} t_L$ & $\mO( g_{\rho_L}\xi)$  & $c_1\frac{g_{\rho_R}}{\sqrt{2}} \, s_{\theta_L}$ &   $\mO(g\xi)$  \\ \hline
 \multirow{2}{*}{  $\bar{t}_R B_R$} &   $\frac{y_R s_{\theta_L}c_{\theta_L}}{2y_L} c_1g_{\rho_L} \sqrt{\xi}$ (P)&  $-\frac{y_R t_{\theta_L}}{2y_L} c_1g_{\rho_R} \sqrt{\xi}$ (P)&$\frac{y_R s_{\theta_L}c_{\theta_L}}{2y_L} g \sqrt{\xi}$ (P) \\
& $\frac{y_{2L}  s_{\theta_L}^2}{2y_L} c_1g_{\rho_L} \sqrt{\xi}$ (F) & $\mO(\xi^{3/2})$ (F) & $\frac{y_{2L}  s_{\theta_L}^2}{2y_L} g\sqrt{\xi}$ (F)\\ \hline
 \multirow{2}{*}{  $\bar{X}_{5/3R} t_R$} &  $-\frac{y_R t_{\theta_L}}{2y_L} c_1g_{\rho_L} \sqrt{\xi}$ (P)&  $\frac{y_R s_{\theta_L}c_{\theta_L}}{2y_L} c_1g_{\rho_R} \sqrt{\xi}$ (P)& $-\frac{y_R t_{\theta_L}}{2y_L} g \sqrt{\xi}$ (P) \\ 
&  $c_2\frac{g^2}{\sqrt{2}g_{\rho_L}}\sqrt{\xi}$ (F) & $\frac{y_{2L}  s_{\theta_L}^2}{2y_L} c_1g_{\rho_R} \sqrt{\xi}$ (F) &$-c_2\frac{g^2}{\sqrt{2}}\sqrt{\xi}$(F) \\\hline
\multicolumn{4}{|c|}{Between SM particles:}\\
\hline
 $i\phi^-\dfbd \phi^0 $ \tnote{\emph{a}} & $\frac{1}{\sqrt{2}} a_{\rho_L}^2g_{\rho_L}$ & $-\frac{a_{\rho_R}^2g_{\rho_R}}{\sqrt{2}}$  & $\frac{g}{\sqrt{2}} $ \\ \hline 
$\bar{t}_ L b_L $ & $\begin{array}{cc}  \frac{1}{\sqrt{2}}\left(c_1 g_{\rho_L}s_{\theta_L}^2 - \frac{g^2}{g_{\rho_L}}\right)\end{array}$    & $\mO(g_{\rho_R} \xi)$&  $ \frac{g}{\sqrt{2}}$ \\ \hline
$\bar{f}_{\text{el}, L} f^\prime_{\text{el},L}$ & $- \frac{1}{\sqrt{2}} \frac{g^2}{g_{\rho_L}} $ & $\mO\left(\frac{g^2}{g_{\rho_R}} \xi\right)$&$\frac{g}{\sqrt{2}}$\\ \hline
\end{tabular}
\begin{tablenotes}
\item[\emph{a}] For $\rho_R^+$, it should be $i\phi^-\dfbd \phi^{0*}.$
\end{tablenotes}
\end{threeparttable}
\caption{The LO coupling strengths between the charged spin-1 bosons $\rho^\pm_{L,R}$, $W^\pm$ and the fermions in models LP(F)$_\4$, RP(F)$_\4$.  Note that $f_{\text{el}}$ denotes all the SM light fermions, including the first two generation quarks, $b_R$ and all the leptons. Here (P) and (F) mean the partially and fully composite right-handed top quark scenario, respectively.}
\label{tab:charge}
\end{table}

\begin{table}[!h]
\centering
\begin{threeparttable}
\tiny
\begin{tabular}{|c|c|c|c|c|c|c|c|c|}\hline
Vertices & $\rho_L^0$ & $\rho_R^0$ & $\rho_X^0$ & $Z$ \\
 \hline
\multicolumn{5}{|c|}{Between heavy resonances:}\\
\hline
$\barray\bar{X}_{5/3} X_{5/3},  \bar{T}_R T_R\\  \bar{X}_{2/3} X_{2/3}, \bar{B}_R B_R\earray$ &   $\begin{array}{cc}  T^{3_L} c_1 g_{\rho_L}\end{array}$& $\barray T^{3_R}c_1 g_{\rho_R}\earray$ & $\begin{array}{cc}c_1 g_{\rho_X} \end{array}$ & $\begin{array}{ccc}\frac{g}{c_W}\left(T^{3_L}  - Q s_W^2\right) \end{array}$   \\ 
 \hline
$\barray\bar{X}_{2/3L} T_{L},  \bar{T}_L X_{2/3 L}\earray$ &   $\begin{array}{cc} \mO(g_{\rho_L} \xi)\end{array}$& $\barray \mO(g_{\rho_R}\xi)\earray$ & $\begin{array}{cc}\mO(g_{\rho_X}\xi)\end{array}$ & $\begin{array}{ccc}\mO( g\xi) \end{array}$   \\ 
 \hline
$\barray\bar{X}_{2/3R} T_{R},  \bar{T}_R X_{2/3 R}\earray$ &   $\begin{array}{cc} \mO(g_{\rho_L} \xi)\end{array}$& $\barray \mO(g_{\rho_R}\xi)\earray$ & $\begin{array}{cc}\mO(g_{\rho_X}\xi)\end{array}$ & $\begin{array}{ccc}\mO( g\xi) \end{array}$   \\ 
\hline
   $\bar{T}_L T_L,\bar{B}_L B_L $ &    $\begin{array}{cc}  T^{3_L} c_1 g_{\rho_L}c_{\theta_L}^2\end{array}$& $\barray  -\frac12 c_1 g_{\rho_R}c_{\theta_L}^2 \earray$ & $\begin{array}{cc}c_1 g_{\rho_X}  c_{\theta_L}^2\end{array}$ & $\begin{array}{ccc}\frac{g}{c_W}\left(T^{3_L}  - Q s_W^2\right) \end{array}$   \\ 
       \hline
       \multicolumn{5}{|c|}{Between heavy resonances and SM fermions:}\\
       \hline
 $\barray\bar{T}_L t_L,\bar{B}_L b_L \\ \bar{t}_L T_L,\bar{b}_L B_L \earray$ &  $\begin{array}{cc}  T^{3_L} c_1 g_{\rho_L} s_{\theta_L}c_{\theta_L} \end{array}$&  $-\frac12 c_1 g_{\rho_R} s_{\theta_L} c_{\theta_L}$ &  $\begin{array}{cc}c_1g_{\rho_X}s_{\theta_L}c_{\theta_L}\end{array}$ & $\barray \mO(g\xi) \earray$\\
 \hline
 \multirow{2}{*}{  $\bar{T}_R t_R, \bar{t}_R T_R$} &  $\frac{y_R s_{\theta_L}c_{\theta_L}}{2\sqrt{2}y_L} c_1g_{\rho_L} \sqrt{\xi}$ (P)&  $-\frac{y_R s_{\theta_L}c_{\theta_L}}{2\sqrt{2}y_L} c_1g_{\rho_R} \sqrt{\xi}$ (P) &  $\frac{y_R s_{\theta_L}c_{\theta_L}}{\sqrt{2}y_L} c_1g_{\rho_X} \sqrt{\xi}$ (P)&$\frac{y_R s_{\theta_L}c_{\theta_L}}{2\sqrt{2}y_L} \frac{g}{s_w} \sqrt{\xi}$ (P)  \\ 
&  $\frac{y_{2L}  s_{\theta_L}^2}{2\sqrt{2}y_L} c_1g_{\rho_L} \sqrt{\xi}$ (F) & $-\frac{y_{2L}  s_{\theta_L}^2}{2\sqrt{2}y_L} c_1g_{\rho_R} \sqrt{\xi}$ (F) & $\frac{y_{2L}  s_{\theta_L}^2}{\sqrt{2}y_L} (c_1-c_1^\prime)g_{\rho_X} \sqrt{\xi}$ (F) & $\frac{y_{2L}  s_{\theta_L}^2}{2\sqrt{2}y_L} \frac{g}{c_w}\sqrt{\xi}$ (F)  \\  \hline
$\barray\bar{X}_{2/3L} t_{L},  \bar{t}_L X_{2/3 L}\earray$ &   $\begin{array}{cc} \mO(g_{\rho_L} \xi)\end{array}$& $\barray \mO(g_{\rho_R}\xi)\earray$ & $\begin{array}{cc}\mO(g_{\rho_X}\xi)\end{array}$ & $\begin{array}{ccc}\mO( g\xi) \end{array}$   \\ 
 \hline
 \multirow{2}{*}{  $\bar{X}_{2/3R} t_R, \bar{t}_R X_{2/3R}$} &  $\frac{y_R t_{\theta_L}}{2\sqrt{2}y_L} c_1g_{\rho_L} \sqrt{\xi}$ (P)&  $-\frac{y_R t_{\theta_L}}{2\sqrt{2}y_L} c_1g_{\rho_R} \sqrt{\xi}$ (P) & $-\frac{y_R t_{\theta_L}}{\sqrt{2}y_L} c_1g_{\rho_X} \sqrt{\xi}$ (P)&$\frac{y_R t_{\theta_L}}{2\sqrt{2}y_L} \frac{g}{c_w}\sqrt{\xi}$ (P)  \\ 
  &$c_2\frac{g^2}{2g_{\rho_L}}\sqrt{\xi}$ (F)&  $c_2\frac{g'^2}{2g_{\rho_L}}\sqrt{\xi}$ (F)& $\frac{c_2}{2}\frac{g}{c_W}\sqrt{\xi}$ (F) & $\mO(\xi^{3/2})$ (F)  \\ \hline
  \multicolumn{5}{|c|}{Between SM particles:}\\
       \hline
 $i\phi^-\dfbd \phi^+ $ & $\barray\frac{1}{2} a_{\rho_L}^2 g_{\rho_L} \earray$ & $\barray  \frac12 a_{\rho_R}^2g_{\rho_R}\earray$ & $- \frac12 \frac{g^{\prime 2}}{g_{\rho_X}}$  &$\frac{g}{c_W}\left(\frac12  - s_W^2\right) $ \\ 
 \hline
 $ i\phi^{0*}\dfbd \phi^0$ & $\barray -\frac12 a_{\rho_L}^2 g_{\rho_L} \earray$ & $\barray \frac12 a_{\rho_R}^2g_{\rho_R}\earray$ & $-\frac12 \frac{g^{\prime 2}}{g_{\rho_X}}$  &$-\frac{g}{2c_W}$ \\
  \hline
$\bar{t}_L t_L, \bar{b}_L b_L$ &  $\begin{array}{cc}  T^{3_L}\left( c_1 g_{\rho_L}s_{\theta_L}^2- \frac{g^2}{g_{\rho_L}} \right)\end{array}$ &$ -\frac12 c_1 g_{\rho_R} s_{\theta_L}^2 -\frac16 \frac{g^{\prime 2}}{g_{
\rho_R}}$ &  $\begin{array}{cc}c_1g_{\rho_X}s_{\theta_L}^2 - \frac16 \frac{g^{\prime 2}}{g_{\rho_X}} \end{array}$ & $\barray \frac{g}{c_W}\left(T^{3_L}  - Q s_W^2\right) \\+\mO(g\xi) \tnote{\emph{a}} \earray$  \\
 \hline
$\bar t_R t_R$ &  $\mO(g_{\rho_L} \xi) $&$- \frac23\frac{g^{\prime2}}{g_{\rho_R}} + \mO(g_{\rho_R} \xi)$ &$\barray  c_1^\prime g_{\rho_X}  (\text{F})\\ -\frac23\frac{g^{\prime 2}}{g_{\rho_X}} + \mO(g_{\rho_X} \xi) (\text{P}) \earray$ & $-\frac{2gs_W^2}{3c_W} + \mO(g\xi)$    \\ 
\hline
$\bar{f}_{\text{el}} f_{\text{el}}$ & $- T^{3_L} \frac{g^2}{g_{\rho_L}} $ &$-Y\frac{g^{\prime 2}}{g_{\rho_R}} $  & $-Y\frac{g^{\prime 2}}{g_{\rho_X}} $ &$\frac{g}{c_W}\left(T^{3_L}  - Q s_W^2\right)$\\
 \hline 
\end{tabular} 
\begin{tablenotes}
\item[\emph{a}] For model LP(F)$_\4$ and RP(F)$_\4$, it reads  $\frac{g}{4c_W}  \left[- T^{3_L}-\frac12\right] s_{\theta_L}^2 \xi$.
\end{tablenotes}
\end{threeparttable}
\caption{The LO coupling strengths between  the neutral spin-1 bosons, $\rho^0_{L,R,X}$ and $Z$, and the fermions in models LP(F)$_\4$, RP(F)$_\4$, and XP(F)$_\4$.  $f_{\text{el}}$ denotes all the SM elementary fermions including the first two generation quarks, $b_R$ and all the leptons.  Here (P) and (F) in the couplings refer to the partially and fully composite right-handed top quark scenario, respectively. $c_W$ denotes $\cos\theta_W$ with $\theta_W$ being the weak mixing angle.}
\label{tab:neutral}
\end{table}

\begin{table}
\centering
\tiny
\begin{tabular}{|c|c|c|c|c|c|c|c|c|}
\hline
Vertices &$\overline{\widetilde T}_L \widetilde T_L$ & $\overline{\widetilde T}_L  t_L, \bar{t}_L \widetilde T_L$&$\bar{t}_L t_L$ & $\overline{\widetilde T}_R \widetilde T_R$ & $\overline{\widetilde T}_R t_R, \bar{t}_R \widetilde{T}_R$ & $\bar t_R t_R$ & $\bar{b}_L b_L$  \\
 \hline
  $\rho_X^0$ (XP$_\1$)  &$\begin{array}{ccc}c_1 g_{\rho_X} \end{array}$  &    $-\frac{y_L c_{\theta_R} s_{\theta_R}}{\sqrt{2}y_R}c_1 g_{\rho_X} \sqrt{\xi}$& $ - \frac16\frac{g^{\prime2}}{g_{\rho_X}}$ & $c_1 g_{\rho_X} c_{\theta_R}^2$ &  $\begin{array}{ccc}c_1 g_{\rho_X}\,  s_{\theta_R} c_{\theta_R}  \end{array}$ &  $ \barray c_1 g_{\rho_X} s_{\theta_R}^2  \\ -\frac23\frac{g^{\prime 2}}{g_{\rho_X}}\earray$ &$ - \frac16\frac{g^{\prime2}}{g_{\rho_X}}$\\ 
  \hline
    $\rho_X^0$ (XF$_\1$)  &$\begin{array}{ccc}c_1 g_{\rho_X} \end{array}$  &  $- \frac{y_L f}{\sqrt{2} M_1}c_1 g_{\rho_X} \sqrt{\xi}$  &$ - \frac16\frac{g^{\prime 2}}{g_{\rho_X}}$\ & $\begin{array}{ccc}c_1  g_{\rho_X}  \end{array}$ &  $\begin{array}{ccc}c_1^{\prime\prime}g_{\rho_X}\end{array}$  &  $ \barray c_1^\prime g_{\rho_X}\earray$ &$ - \frac16\frac{g^{\prime 2}}{g_{\rho_X}}$\\ \hline
   $Z$   &  $-\frac{2gs_W^2}{3c_W}$&  $\barray \frac{y_L c_{\theta_R} s_{\theta_R}}{\sqrt{2}y_R}\frac{g}{2 c_W}\sqrt{\xi}~(\text{XP}_\1)\\ \frac{y_L f}{\sqrt{2} M_1}  \frac{g}{2 c_W}\sqrt{\xi}~(\text{XF}_\1)\earray$& $\frac{g}{c_W}\left(\frac12 - \frac{2s_W^2}{3}\right)$  &  $-\frac{2gs_W^2}{3c_W}$ & 0  & $-\frac{2gs_W^2}{3c_W}$  &$\frac{g}{c_W}\left(-\frac12 + \frac{s_W^2}{3}\right)$  \\
    \hline
\end{tabular} 
\caption{The LO coupling strengths between  the neutral spin-1 gauge bosons, $\rho^0_{X}$ and $Z$, and the fermions in models XP(F)$_\1$.}
\label{tab:rhox1}
\end{table}
The couplings can be organized into four classes by their typical sizes.
The first class includes the interactions generated directly from the strong dynamics and preserve the non-linearly realized $SO(5)$ symmetry. 
They only  involve  the strong sector resonances $\rho$, $\Psi$, the pseudo-Goldstone bosons $\vec{h}$  and the fully composite right-handed top quark $t_R^{\rm (F)}$. The interaction strengths are of $\mO(g_\rho)$ or $\mO(g_\Psi)$.
Since these  interactions preserve the  unbroken $SO(4)$ symmetry, the interactions between $\rho$ and $\Psi$ are determined by the quantum number of the fermionic resonances under the $SU(2)_L\times SU(2)_R$. The symmetry selection rules permit the following interactions of $\mO(g_{\rho})$:
\beq
\rho_L^+ \bar{T}B, \quad \rho_L^+ \bar{X}_{5/3} X_{2/3}, \quad \rho_R^+ \bar{X}_{2/3} B, \quad \rho_R^+ \bar{X}_{5/3} T, \quad \rho_{L,R,X}^0 \bar{\mT}\mT, \quad \rho_X^0 \overline{\widetilde T} \widetilde T , \quad \rho_X^0 \bar{t}_R^{\rm (F)}t_R^{\rm (F)},
\eeq
where $\mT = T$, $B$, $X_{5/3}$, $X_{2/3}$ denotes the fermionic resonances in the quartet. The last term is for the case of a fully composite right-handed top quark. As will be discussed in the next subsection, these interactions dominate the decay of $\rho$ resonances if the channels are kinematically open. For the interactions involving the $\rho$ and the Higgs doublet $H$, we have (see Appendex~\ref{app:models} for detail):
\beq
\frac{a_{\rho_L}^2 }{2} g_{\rho_L}\rho^{a_L\mu} i H^\dagger \sigma^{a_L} \Dfbd H, \quad  \frac{a_{\rho_R}^2}{2}  g_{\rho_R}\rho^{a_R\mu} J_{\mu}^{a_R} (H),
\eeq
where we have defined the $SU(2)_R$ current
\beq
J_\mu^{ a_R} (H) = \left( -i\left(\widetilde H^\dagger D_\mu H- D_\mu H^\dagger\widetilde H\right) ,-\left(H^\dagger D_\mu\widetilde  H+D_\mu\widetilde H^\dagger H\right),i H^\dagger \Dfbd H\right).
\eeq
The Higgs doublet can be parameterized as
\beq
 H = \begin{pmatrix}
\phi^+ \\ \phi^0 \end{pmatrix} = \begin{pmatrix}
\phi^+ \\
 \frac{h + i \chi}{\sqrt{2}}
\end{pmatrix},
\eeq
with $\phi^\pm$, $\chi$ eaten by the SM $W^\pm$, $Z$ bosons after EWSB. By the Goldstone equivalence theorem, the interactions involve $\phi^\pm$, $\chi$  will determine the couplings of longitudinal modes of $W^\pm$ and $Z$ gauge bosons at high energy,  leading to the following interactions with $\mO(g_\rho)$:
\be\label{eq:RhoVV}\begin{split}
&\frac{a_{\rho_L}^2}{2}g_{\rho_L}\Big[\rho_L^3(\phi^-i\dfbd \phi^+-\phi^{0*}i\dfbd \phi^0)+\sqrt{2}\rho_L^+\phi^-i\dfbd \phi^0+\sqrt{2}\rho_L^-\phi^{0*}i\dfbd \phi^+\Big],\\
&\frac{a_{\rho_R}^2}{2}g_{\rho_R}\Big[\rho_R^3(\phi^-i\dfbd \phi^++\phi^{0*}i\dfbd \phi^0)+\sqrt{2}\rho_R^+\phi^{0*}i\dfbd \phi^-+\sqrt{2}\rho_R^-\phi^+i\dfbd \phi^0\Big].
\end{split}\ee
Hence, $\rho_{L,R}$ will primarily decay into the longitudinal gauge bosons and the Higgs bosons if the other strongly interacting decay channels ($\Psi \Psi$ or $\Psi q$) are not kinematically open. The other type in the first class is the interactions between the resonances $\Psi_{\4}$  and  $t_R^{\rm (F)}$ in \Eq{eq:PF4}:
\be
c_2 \bar\Psi_{\textbf{4}}^i\gamma^\mu id^i_\mu t_R^{\rm (F)} =  \frac{\sqrt{2}c_2}{f} \left(M_{Q_X} \bar Q_{XL}  H t_R^{\rm (F)} - M_Q \bar Q_L \widetilde{H}t_R^{\rm (F)} + \text{h.c.} \right) +\cdots,
\ee
where we have integrated by parts before turning on the Higgs vacuum expectation value (VEV) and focused   only on the trilinear couplings (see Ref.~\cite{DeSimone:2012fs} for detail). $M_{Q_X}$, $M_Q$ are defined in Eq.~(\ref{eq:masstp}).  In the limit $M_\4/f \gg  y_L, y_{2L}$, these are the  dominant interactions between  the top partners and the SM fields. By using Goldstone equivalence theorem, we can easily derive the  well-known approximate decay branching ratios for the top parnters:
 \beq
 \label{eq:decay1}
 \begin{split}
& \Br(T\rightarrow  t h)  \simeq  \Br(T\rightarrow  t Z) \simeq 50\%, \quad   \Br(X_{2/3}\rightarrow  t h)  \simeq  \Br(X_{2/3}\rightarrow  t Z) \simeq 50\%,\\
&\Br(X_{5/3} \rightarrow t W) \simeq 100\%, \quad \Br(B \rightarrow t W) \simeq 100\%.
 \end{split}
 \eeq 
Taking into account the mixing effects, shown in \Eq{eq:lrmix}, will not  modify the conclusions significantly.

The second class of interactions are suppressed either by the left-handed top quark mixing $s_{\theta_L}$ or the right-handed top quark mixing $s_{\theta_R}$ defined in \Eq{eq:lrmix}.  These are the  couplings of $\rho $ to one top partner 
and one SM quark. These interactions preserve SM $SU(2)_L \times U(1)_Y$ gauge symmetries.  Symmetry considerations select the following interactions:
\beq
\rho_L^+ \bar{T}_L b_L, \quad \rho_L^+ \bar{t}_L B_L,  \quad \rho_{L,R,X}^0 \bar{T}_L t_L , \quad \rho_{L,R,X}^0 \bar{B}_L b_L, \quad \rho_X^0 \overline{\widetilde T}_R t_R^{\rm (P)},
\eeq
where the last term  is only present for the partially composite right-handed top quark scenario. The interactions will play an important role in the kinematical region $M_\Psi + M_{t,b} < M_{\rho} < 2 M_\Psi$, if the mixings $s_{\theta_{L,R}}$ are not too small.

The third class of interactions contains the  SM gauge interactions with couplings $g$, $g^\prime$ or SM Yukawa couplings. These include the $W$ and $Z$ interactions with SM elementary fermions (quarks and the leptons) and the fermionic resonances; and the mixed couplings, proportional to $y_{L,R}$, between the top partners $\Psi_{\4,\1}$ and the elementary SM quarks $q_L$, $t_R^{\rm (P)}$. The  gauge interactions are determined by the SM quantum numbers of the fermions. The Yukawa type interactions in \Eq{eq:P4doublet} and \Eq{eq:MXdoublet}  control the decays of the top partners
\beq
y_R \left(\bar Q_L\widetilde H t_R^{\rm (P)}- \bar Q_{XL} H t_R^{\rm (P)} \right), \quad  -y_L \bar{q}_L\widetilde H\widetilde T_R.
\eeq
which leads to the same decay branching ratios as \Eq{eq:decay1} for the quartet. For the singlet top partner, this gives 
 \beq
 \label{eq:decay2}
 \begin{split}
 \Br(\widetilde T\rightarrow   b W^+)  \simeq 2  \Br(\widetilde T\rightarrow  t h)  \simeq 2  \Br(\widetilde T\rightarrow  t Z) \simeq 50\%.
 \end{split}
 \eeq 
The fourth class contains the interactions with coupling strengths suppressed by $g/g_\rho$, $g^\prime/g_\rho$. 
These are the universal couplings between the $\rho$ and the SM fermions, due to the mixings of $\rho$ and SM gauge bosons which are present  before the EWSB. These  interactions include
\beq\label{light_scaling}
\rho_{L,R}^+ \bar{f}_{\text{el}} f^\prime_{\text{el}}, \quad \rho_{L,R,X}^0 \bar{f}_{\text{el}} f_{\text{el}},
\eeq
where $f_{\text{el}}$ denotes all the SM elementary fermions including the first two generation quarks, $b_R$, and all of the leptons.  For the $\rho_L$, the couplings are of $\mO(g^2/g_{\rho_L})$, while for $\rho_{R,X}^0$, they are of $\mO(g^{\prime 2}/g_{\rho_{R,X}})$. For the couplings between $\rho$ and the third generation quarks, there are additional contributions of $\mO(g_\rho s_{\theta_{L,R}}^2)$:
 \beq\label{Rhoqq}
 \rho_L^+ \bar{t}_L b_L, \quad   \rho_{L,R,X}^0 \bar{t}_L t_L, \quad   \rho_{L,R,X}^0 \bar{b}_L b_L, \quad \rho_X^0 \bar{t}_R^{\rm (P)} t_R^{\rm (P)},
 \eeq
with the final term only arises for the partially composite right-handed top quark. All the remaining coupling vertices can only be present after EWSB. Therefore, they are suppressed further by $\xi$ and irrelevant  for the phenomenology of the composite resonances.

\subsection{The production and the decay of the resonances at the LHC}
\label{sec:production_decay}

In this subsection, we discuss the production and decay of the composite resonances. The cross sections are calculated by first implementing the benchmark models into an UFO model file through the \texttt{FeynRules}~\cite{Alloul:2013bka} package and then using \texttt{MadGraph5\_aMC@NLO}~\cite{Alwall:2014hca} to simulate the processes. Most of the calculations are carried out at the LO. The only exception is the QCD pair production of top partners, for which we use the {\tt Top++2.0} package~\cite{Czakon:2011xx,Czakon:2013goa,Czakon:2012pz,Czakon:2012zr,Baernreuther:2012ws,Cacciari:2011hy} to obtain the next-to-next-to-leading-order (NNLO) cross sections. See Appendix~\ref{app:NNLO} for the cross sections at different proton-proton center-of-mass energies. For the decay widths, we have used the analytical formulae calculated by the \texttt{FeynRules}.

\subsubsection{Production at the LHC}
\label{sec:production}

\begin{figure}
\centering
\subfigure[~Charged vector resonances.]{
\label{DY_Rhopm} %% label for first subfigure
\includegraphics[scale=0.4]{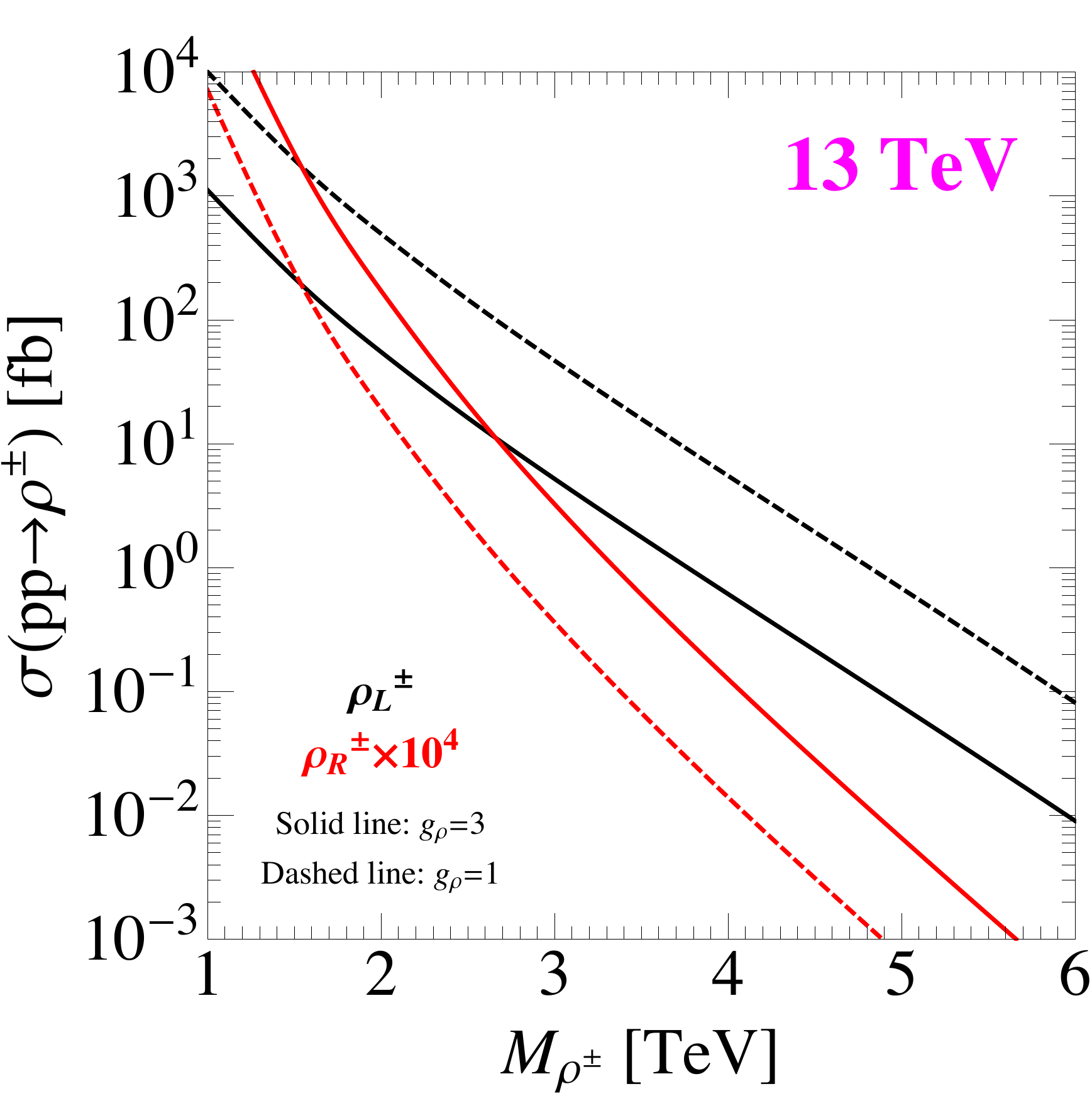}}\quad
\subfigure[~Neutral vector resonances.]{
\label{DY_Rho0} %% label for first subfigure
\includegraphics[scale=0.4]{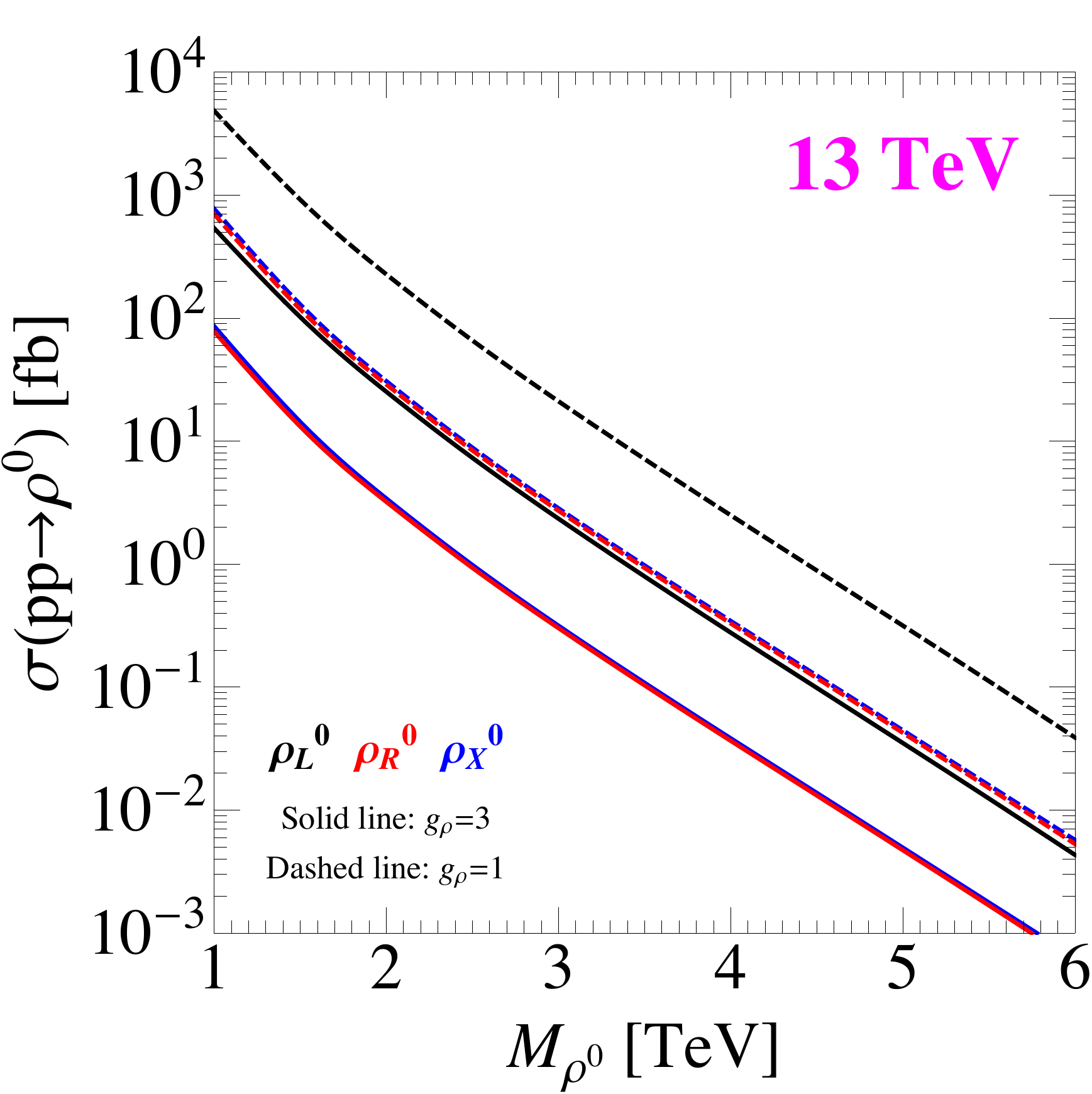}}
\caption{The Drell-Yan cross sections for the vector resonances at the 13 TeV LHC, in the unit of fb. The cross section of $\rho_R^\pm$ is shown as $\sigma(pp\to\rho_R^\pm)\times10^4$. We have fixed the  parameters $a_{\rho_{L,R,X}}^2$ (see \Eq{eq:arhoL}, \Eq{eq:arhoR} and \Eq{eq:arhoX} for the definitions) equal to $1/2$. The rates are calculated within 4-flavor proton scheme, i.e. the $b$ quark is not involved. }
\label{DY_Rho}
\end{figure}
We start from the production of the vector resonances at the LHC. The vector resonances $\rho$ will be dominantly produced via the Drell-Yan processes inspite of their suppressed couplings $\sim g_{\rm SM}^{2}/g_\rho$ to the valence quarks~\cite{Pappadopulo:2014qza,Greco:2014aza}. Although the $\rho$ resonances are strongly interacting with the  longitudinal SM gauge bosons, as shown in \Eq{eq:RhoVV}, the electroweak Vector-Boson-Fusion (VBF) production can barely play an useful role in the phenomenology of the $\rho$ at the LHC~\cite{Pappadopulo:2014qza,Greco:2014aza}. 
For example, for $g_{\rho_L}=3$ and $M_{\rho_L}=3$ TeV, the $W^+W^-\to\rho_L^0$ fusion cross section is two orders of magnitude smaller than that of the Drell Yan process. In \Fig{DY_Rho}, we have shown the $M_\rho$ dependence of the Drell-Yan production cross section for the charged resonances $\rho^\pm_{L,R}$ and neutral resonances $\rho_{L,R,X}^0$, fixing $a_\rho^2 = 1/2$. For the production of the charged resonances, we have summed over the $\rho^+$ and $\rho^-$ contributions.  The cross sections are decreasing functions of the strong coupling $g_\rho$, as expected from the coupling scaling in Tables~\ref{tab:charge} and~\ref{tab:neutral}. The only exception is the production rare of the charged $\rho_R^\pm$, whose couplings to the valence quarks arise after EWSB and are of order $g_{\rho_R} a_{\rho_R}^2M_W^2/M_{\rho_R}^2$. As we are fixing $a_\rho$ in the plot, the cross section is larger  for larger $g_{\rho_R}$, as shown \Fig{DY_Rho}. We also notice that generally, $\rho_L^0$ has one order of magnitude larger production rate than the $\rho_{R,X}^0$ case because of the smallness of $U(1)_Y$ hyper-gauge coupling $g^\prime$ in comparison with $SU(2)_L$ gauge coupling $g$.

In Fig~\ref{DY_Rho}, we have calculated the cross sections using the 4-flavor scheme. The inclusion of bottom parton distribution function (PDF) will increase the cross sections of $\rho_{L,R,X}^0$. As shown in Table~\ref{tab:neutral}, the $\rho_{L,R,X}^0b_L\bar b_L$ couplings in models with quartet top partners  have contributions of  $\mO(g_\rho s_{\theta_L}^2)$ due to the mixing of $b_L$ and $B_L$, which can considerably enhance the cross section in some parameter space. For example, in LP$_\4$, for $y_L=1$ and $M_\4=1$ TeV, $g_{\rho_L}=3$, and $M_{\rho_L}=3$ TeV, the $b\bar b$ fusion can increase $\sigma(pp\to\rho_L^0)$ by $34\%$. In the following section, when we will study the bounds from the searches  at the LHC, we also include the $b\bar{b}$ fusion production.

\begin{figure}
\centering
\subfigure[~Partially composite $t_R^{\rm (P)}$ scenario.]{
\label{Quartet_P} %% label for first subfigure
\includegraphics[scale=0.4]{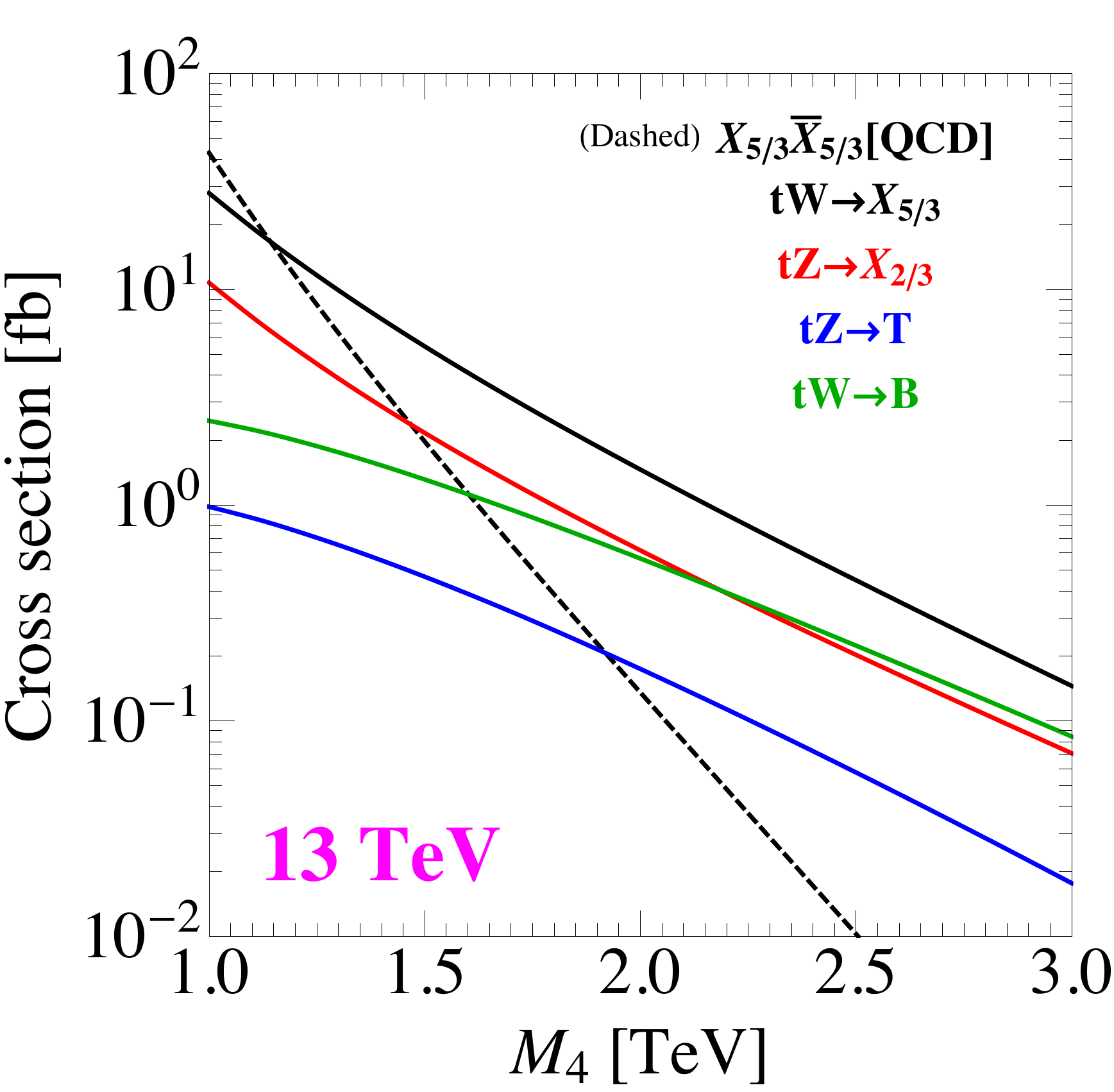}}\quad
\subfigure[~Fully composite $t_R^{\rm (F)}$ scenario.]{
\label{Quartet_F} %% label for first subfigure
\includegraphics[scale=0.4]{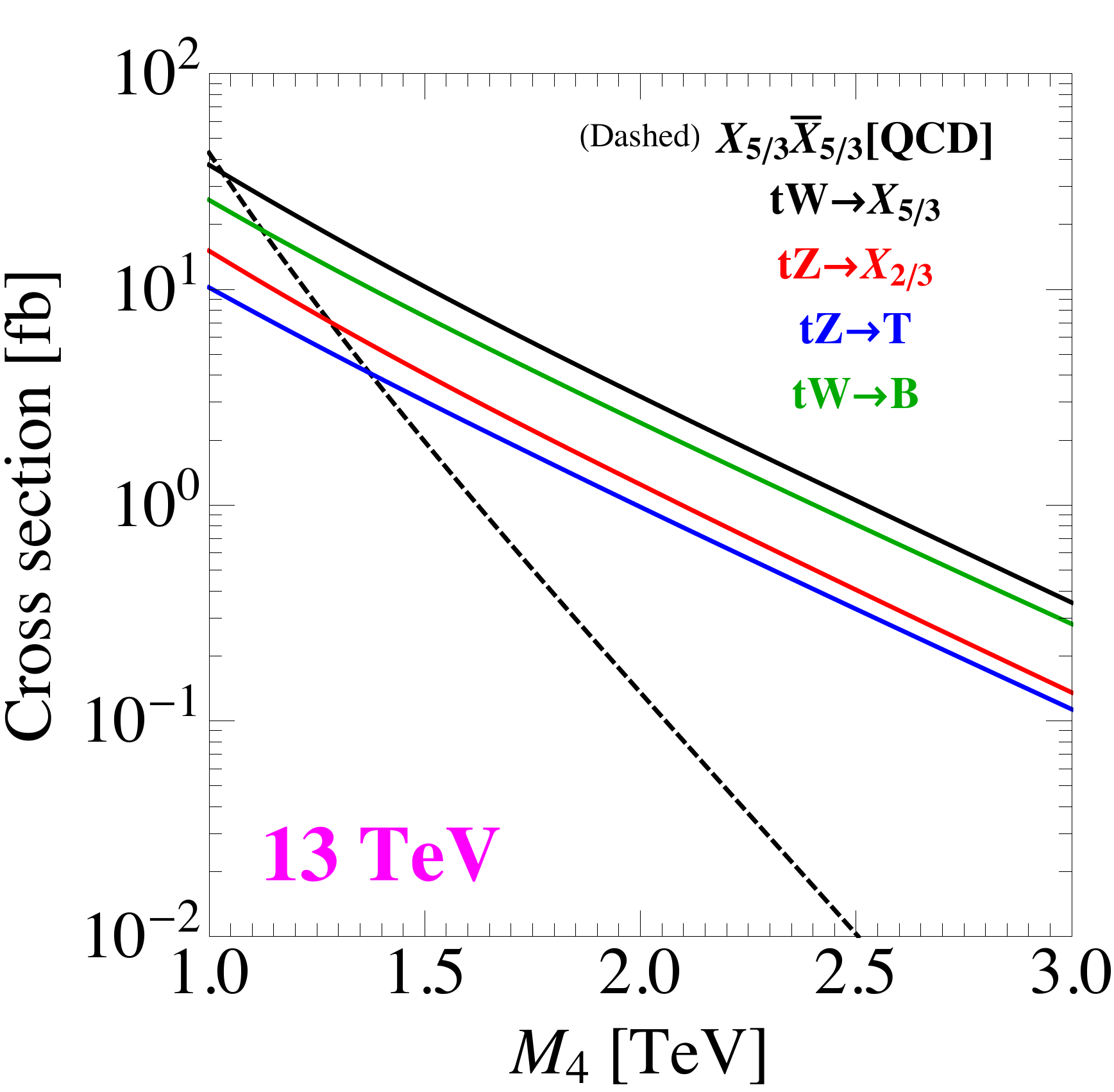}}
\caption{The production cross sections of quartet top partners at the 13 TeV LHC. The parameters are chosen as $f=1$ TeV, $y_L=1$ and $c_2=1$ (for fully composite $t_R^{\rm (F)}$ only), while $y_R$ in partially composite $t_R^{\rm (P)}$ models and  $y_{2L}$ in fully composite $t_R^{\rm (F)}$ models are determined by the top quark mass $M_t = 150$ GeV (see \Eq{eq:mtp4} and \Eq{eq:mtf4}).}
\label{Quartet_production}
\end{figure}
The production of fermion resonances can be categorized into QCD pair production and  electroweak single production processes~(see Ref.~\cite{DeSimone:2012fs} for detail). The QCD production rate depends only on the mass of top partners. Since two heavy fermions are produced, the rate drops rapidly when the resonance's mass increases because of the PDF suppression.  In contrast, the single production channels typically have larger rates in the high mass region, thus it can play an important role in the search for heavier resonance~\cite{Willenbrock:1986cr,DeSimone:2012fs,Li:2013xba,Aguilar-Saavedra:2013qpa,Mrazek:2009yu,Azatov:2013hya,Backovic:2014uma,Matsedonskyi:2014mna,Backovic:2015bca,Gripaios:2014pqa}. This effect can be clearly seen from the Fig.~\ref{Quartet_production}, where we have plotted the cross sections for the resonances in the quartet at the 13 TeV LHC as functions of the Lagrangian parameter $M_\4$. For these plots, we have chosen the following parameters:
\beq
f = 1 \,\TeV, \quad y_L = 1, \quad c_2 = 1~(\text{for $t_R^{\rm (F)}$ only}),
\eeq
where the parameter $y_R$ or $y_{2L}$ is determined by the top mass requirement for the partially composite $t_R^{\rm (P)}$ in~\Eq{eq:mtp4} (the ``P$_\4$ scenario'') or for the fully composite $t_R^{\rm (F)}$ in~\Eq{eq:mtf4} (the ``F$_\4$ scenario''), respectively.
For the single production, we have combined the contribution of the top parters and their anti-particles.
For example, for the charge-$5/3$ resonance $X_{5/3}$ in the quartet case, the $tW$ fusion process  is defined as
\be
\sigma(tW\to X_{5/3})\equiv \sigma(pp\to X_{5/3}\bar tq+\bar X_{5/3}tq).
\ee
The $tW\to B$ and $tZ\to T,~X_{2/3}$ processes are defined in a similar way. Figure~\ref{Quartet_production} shows that, for both P$_\4$ and F$_\4$ scenarios, $tW\to X_{5/3}$ has the largest production rate among the 4 single production channels of the quartet fermionic resonances, and it dominates over the QCD pair production channel for $M_\4\gtrsim1$ TeV. Although the $tW\to X_{5/3}$ rates of those two scenarios are similar under our parameter choice, the rate of $tW\to B$ channel in P$_\4$ scenario is less than that in F$_\4$ scenario. This is because the former is from the composite-elementary Yukawa interaction $-y_R\bar B_L\phi^-t_R^{\rm (P)}$ (see \Eq{eq:P4doublet}) and proportional to $c_{\theta_L}^2$, while the latter is mainly controlled by the strong dynamics term $-(\sqrt{2}c_2/f)\bar B_R \gamma^\mu t_R^{\rm (F)}i\partial_\mu\phi^-$ (see \Eq{eq:F4doublet}) without such suppression. As $c_{\theta_L}$ will increase with $M_\4$, we see the values of the two green lines in Fig.~\ref{Quartet_P} and Fig.~\ref{Quartet_F} become similar at large $M_\4$. By naively using the Goldstone equivalence theorem, we expect 
\be\label{naive_singleF}
\sigma(tW\to X_{5/3})\approx \sigma(tW\to B)>\sigma(tZ\to T)\approx \sigma(tZ\to X_{2/3}),
\ee
if $y_Lf/M_\4\ll1$ and the mass splittings of the top partners become negligible. From the figures we find that in the F$_\4$ scenario it is indeed the case, but in the P$_\4$ scenario it is not. The reasons is that in the P$_\4$ scenario, large $M_\4$ requires large $y_R$ to correctly reproduce the mass of the top quark (see \Eq{eq:mtp4}), which results in a large mixing between the $T$ and $X_{2/3}$ resonances as shown in \Eq{TX23_mixing}. Hence the naive estimate in \Eq{naive_singleF} does not hold. We emphasize that the single production rates are more model-dependent. For example, the $tZ/W$ fusion rates in P$_\4$ scenario increase when $y_L$ decreases. This is because the constraint from observed top quark mass requires a larger $y_R$ as $y_L$ decreases, while the fusion rates are proportional to $(y_R)^2$. But in the F$_\4$ scenario, the cross sections are rather insensitive to $y_L$, since they are mainly determined by  the $c_2$ term. 

\begin{figure}
\centering
\subfigure[~Partially composite $t_R^{\rm (P)}$ scenario.]{
\label{singlet_P} %% label for first subfigure
\includegraphics[scale=0.4]{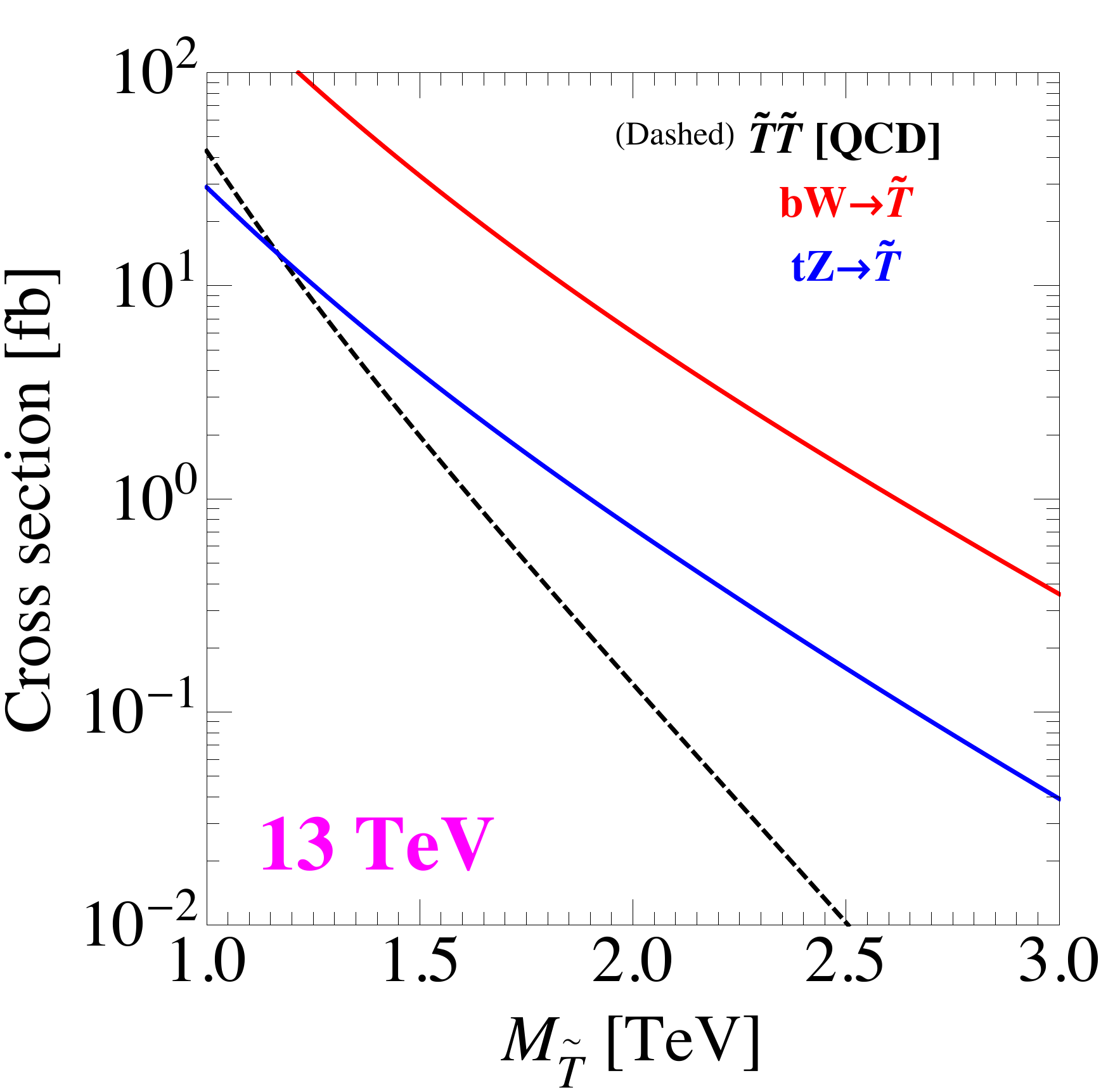}}\quad
\subfigure[~Fully composite $t_R^{\rm (F)}$ scenario.]{
\label{singlet_F} %% label for first subfigure
\includegraphics[scale=0.4]{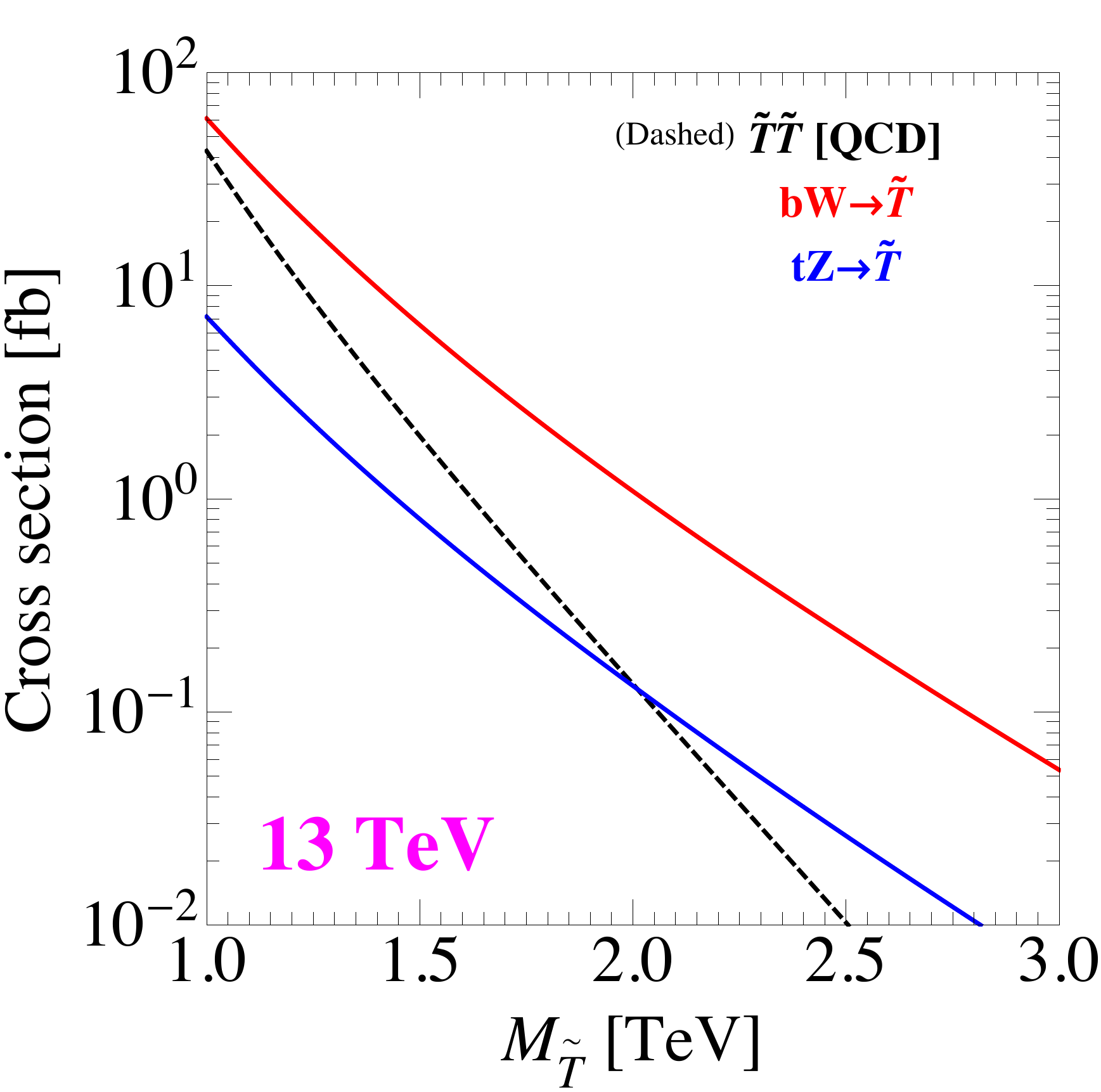}}
\caption{The production cross sections of singlet top partners at the 13 TeV LHC. The parameters used in making these plots are $f=1$ TeV, $y_L=1.5~(1)$ for LP(F)$_\1$, while $y_R$, $y_{2L}$ are determined by the top quark mass $M_t = 150$ GeV (see \Eq{eq:mtp4} and \Eq{eq:mtf4}).  Note that $bW\rightarrow \widetilde T$ process is calculated using bottom PDF.}
\label{singlet_production}
\end{figure}

Similar to the quartet case, the single production mechanism of the singlet top partner $\widetilde T$ dominates over the QCD pair production if it is heavier than  $\mO(1)$ TeV, as shown in Fig.~\ref{singlet_production}. Besides the $tZ\to \widetilde T$ fusion, the singlet can also be produced by $bW$ fusion:
\be
\sigma(bW\to\widetilde T)\equiv\sigma(pp\to \widetilde Tq+\overline{\widetilde T}q).
\ee
In fact, the cross section of this channel is about an order of magnitude larger than the $tZ$ fusion due to the large bottom PDF, as can be seen from the red solid lines in Fig.~\ref{singlet_production}. Note that for the partially composite $t_R^{\rm (P)}$ scenario, we have chosen a somewhat larger value $y_L = 1.5$  in order to correctly reproduce the mass of the top quark  in \Eq{eq:mtp1}.

\subsubsection{Decay of the composite resonances}
\label{sec:decay}

Let's now turn to the decay of the vector resonances \footnote{For the decays of the top partners,  they are mainly determined by the Goldstone Equivalence theorem, as shown in \Eq{eq:decay1} and \Eq{eq:decay2}. See \Ref{DeSimone:2012fs} for the detailed discussion.  For the cascade decays of the top partners into the vector resonances, they can barely play an important role in our interested parameter space. The only exception is that, in the models XP(F)$_\1$,  the decay of $\widetilde T$ into $\rho_X t$ can play an important role as discussed in Sec.~\ref{sec:xpf1}.}. The decay branching ratios into different final states  are determined by both the kinematics and the sizes of the couplings between the vector resonances and the final state particles.  

\begin{figure}[!tb]
\centering
\subfigure[~The branching ratios of $\rho_L^+$ in LP$_\4$.]{
\label{LP4_RhoL_decay} %% label for first subfigure
\includegraphics[scale=0.4]{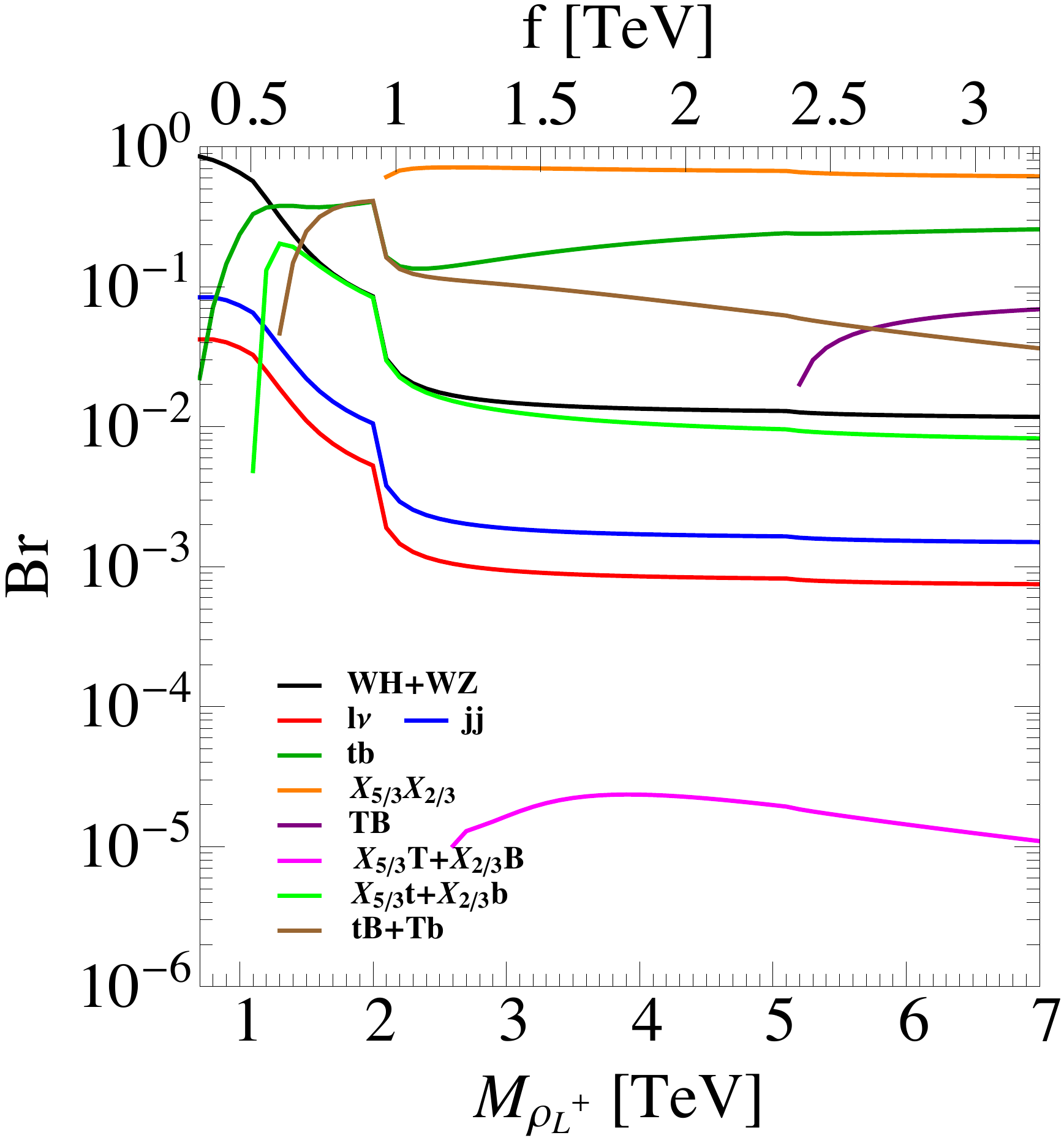}}\quad
\subfigure[~The branching ratios of $\rho_L^0$ in LP$_\4$.]{
\label{LP4_RhoL0_decay} %% label for first subfigure
\includegraphics[scale=0.4]{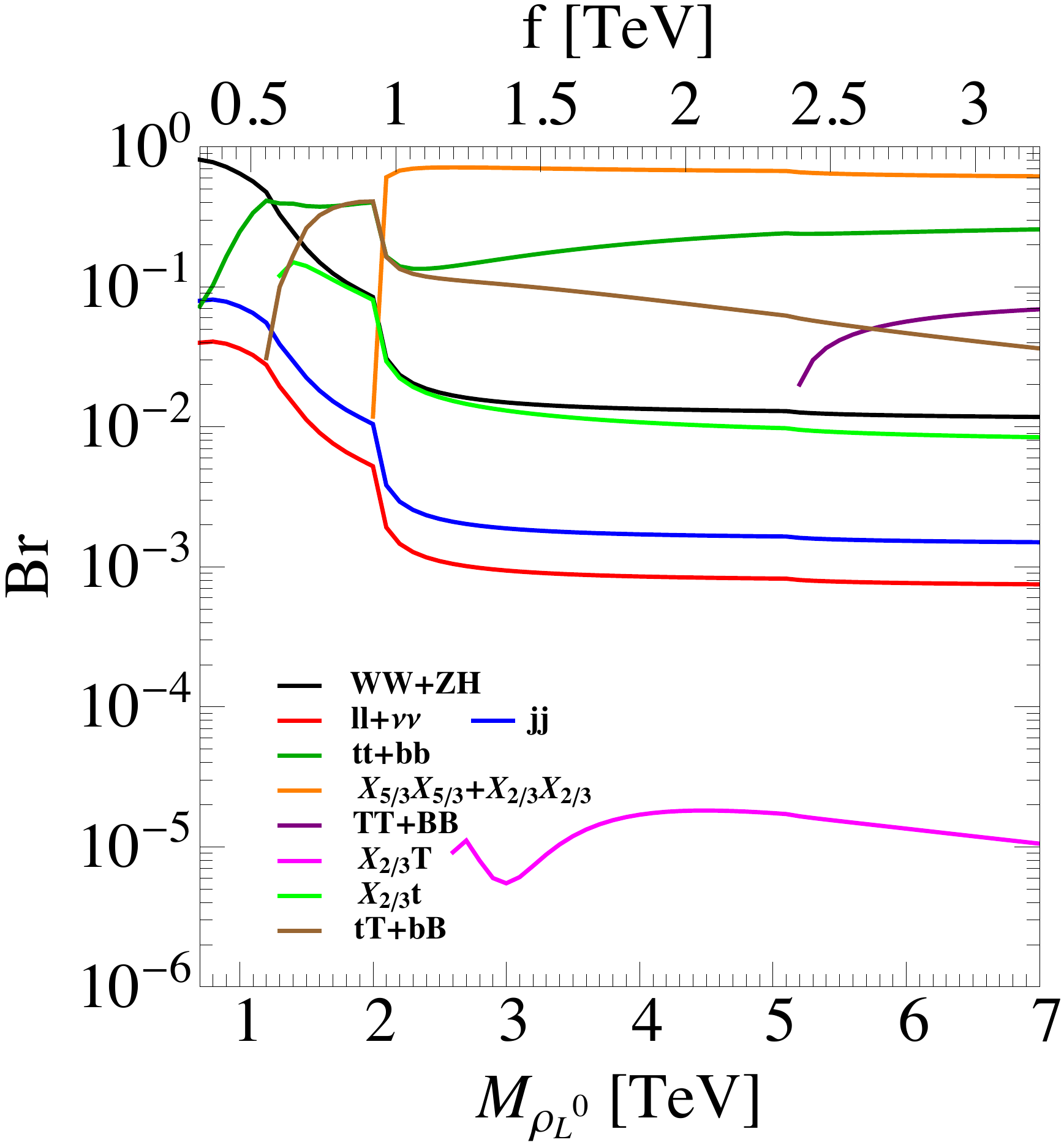}}
\subfigure[~The branching ratios of $\rho_L^+$ in LF$_\4$.]{
\label{LF4_RhoL_decay} %% label for first subfigure
\includegraphics[scale=0.4]{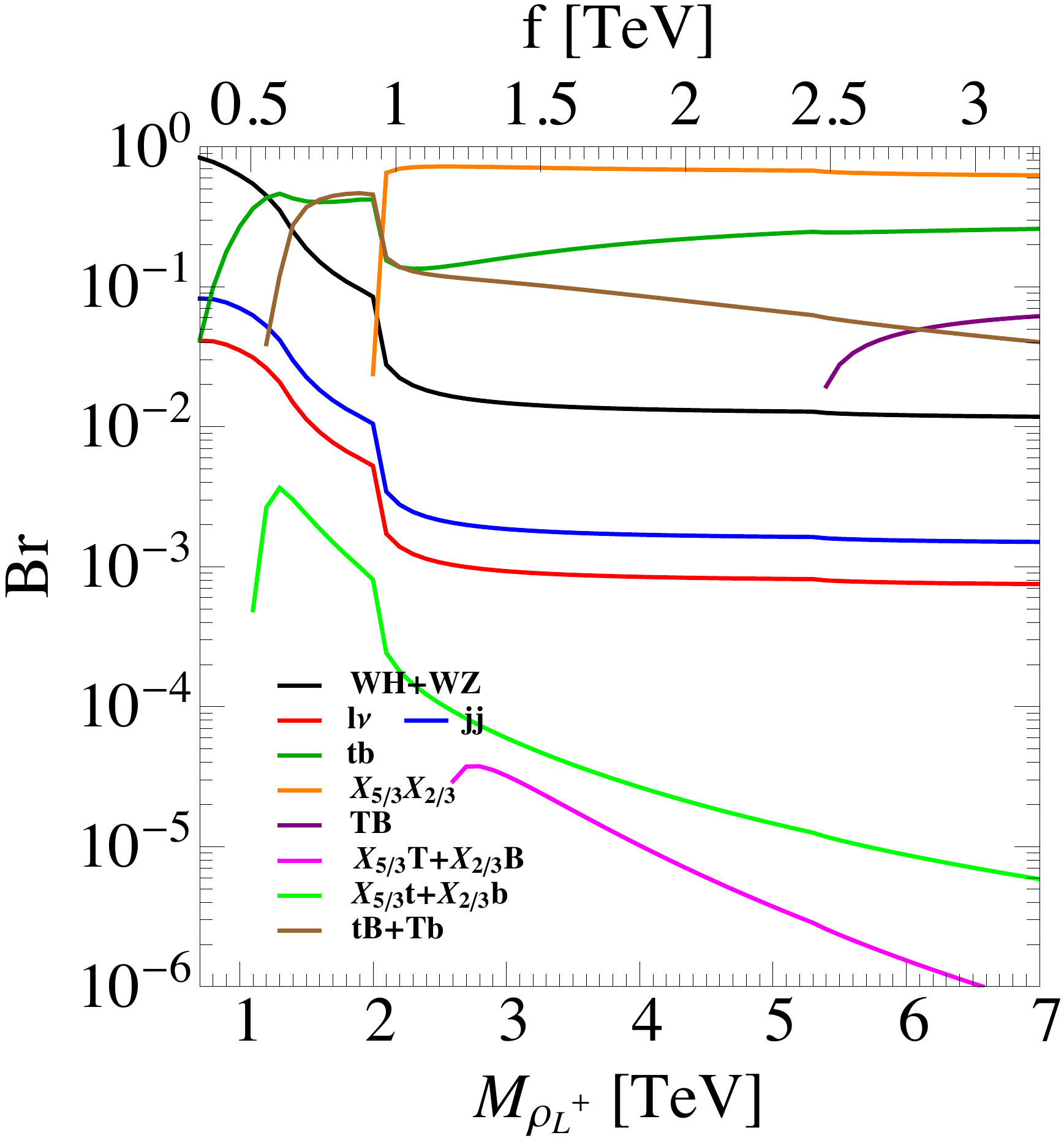}}\quad
\subfigure[~The branching ratios of $\rho_L^0$ in LF$_\4$.]{
\label{LF4_RhoL0_decay} %% label for first subfigure
\includegraphics[scale=0.4]{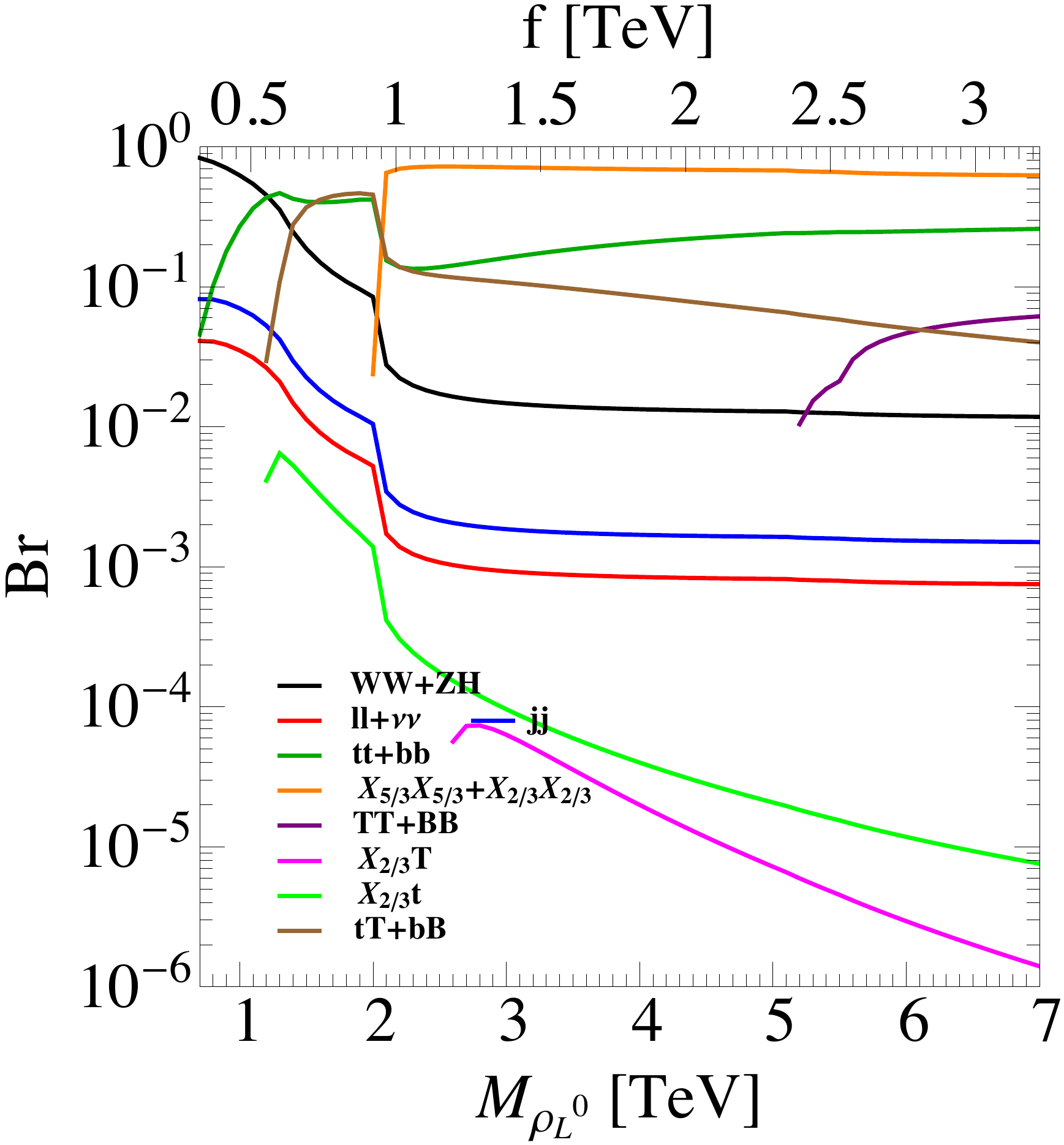}}
\caption{The decay branching ratios of the $\rho_L^{\pm,0}$ resonances. The parameters are $g_{\rho_L}=3$, $a_{\rho_L}^2=1/2$, $M_\4=1$ TeV, $y_L=1$ and $c_1=1$. For LF$_\4$ there is an additional parameter $c_2=1$.}
\label{RhoL_Br}
\end{figure}

Let's start from the $\rho_L(\3,\1)$ resonances in models LP$_\4$ and LF$_\4$.  In Fig.~\ref{RhoL_Br}, we have plotted the decay branching ratios of $\rho_L^{\pm,0}$  as functions of $M_{\rho_L^{\pm,0}}$, choosing the following parameters:
\be
\label{eq:paraL}
g_{\rho_L}=3,\quad a_{\rho_L}^2=\frac{1}{2},\quad M_\4=1~{\rm TeV},\quad y_L=1,\quad c_1=1,\quad c_2=1~\text{(for LF$_\4$ only).} 
\ee
The parameter $f$ is determined by \Eq{eq:arhoL} and the parameters $y_R$ (LP$_\4$), $y_{2L}$ (LF$_\4$) are fixed by reproducing the observed top quark running mass $M_t = 150~\GeV$ at the TeV scale. Several comments are in order.
In the low mass region $M_{\rho_L}<M_\4$, $\rho_L$ can only decay into SM final states. Since we are interested in the mass region $M_\rho \gg M_{W,Z,h}$, we can neglect all the SM masses. Hence, the decaying branching ratios are completely determined by the couplings among $\rho_L$ and SM particles. As discussed above, only $\rho_ L V_L V_L(h)$ $(V = W,Z)$  couplings belong to the first class and are enhanced by the strong coupling $g_{\rho_L}$. Besides this, there are $\rho_L \bar{q}_L q_L$ couplings, where $q_L$ are third generation left-handed quarks. They are of $\mO(g_{\rho_L}s_{\theta_L}^2)$ and can be relevant for the moderate size of $s_{\theta_L}$. Therefore,  the dominant decay channels for this mass region are
\beq
\rho_L^+\rightarrow W^+ Z,~ W^+ h,~ t\bar{b};\quad \rho_L^0\rightarrow W^+ W^-,~Zh, ~t\bar{t}, ~b\bar{b},
\eeq
as shown in Fig.~\ref{RhoL_Br}. There are no significant differences between the two models in this kinematical region.  From the Goldstone equivalence theorem, the decay branching ratio  of $\rho_L^+$ into $W^+Z$ is the same as $W^+ h$ in the limit of $M_{\rho_L} \gg M_{W,Z,h}$ (see \Eq{eq:RhoVV}). We  only plot the sum of the two channels in Fig.~\ref{RhoL_Br}. The same argument applied to the $W^+ W^-$, $Zh$ decay channels of $\rho_L^0$.
We also notice that for the SM light fermion channels,  we have  the accidental relations $\text{Br}(\rho_L^+\rightarrow jj)=2\times\text{Br}(\rho_L^+ \rightarrow \ell^+\nu_\ell)$ and $\text{Br}(\rho_L^0 \rightarrow jj)=2\times\text{Br}(\rho_L^0 \rightarrow \ell^+\ell^-+\nu_\ell\bar\nu_\ell)$ as illustrated by \Ref{Brooijmans:2014eja,Greco:2014aza}.

For the intermediate mass region, i.e. $M_\4<M_{\rho_L}<2M_\4$, the decay channels with one third generation quark and one top partner (the ``heavy-light'' channels)  are  open kinematically. For  the charged resonance $\rho_L^+$, we have plotted  the  sum of branching ratios of the decay channels $t\bar{B}$ and $T\bar{b}$ and  the sum of the decay channels $ X_{5/3}\bar t $ and $X_{2/3}\bar b$.
For the neutral resonances $\rho_L^0$, we have combined the channels $t\bar{T}$ and $b\bar{B}$ and their charge conjugate processes. Let's start the discussion from the model LP$_\4$. The branching ratios of such channels grow  quickly once they are kinematically open.  This rapid increase is due to the strong coupling  enhancement.  At the same time, there is also a difference between  the $t\bar B+T\bar b$  channels  and the $X_{5/3}\bar t+X_{2/3}\bar b$ channels. The branching ratio for the former increases as $M_{\rho_L^+}$ becomes larger, while the branching  ratio of the latter increases at the beginning then decreases as the mass of $\rho_L^+$ increase. We first note that the couplings $\rho_L^+X_{5/3}\bar t$, $\rho_L^+  X_{2/3}\bar b$ are suppressed by the fine-tuning parameter $\xi = v^2/f^2$ (see Table~\ref{tab:charge}). Since $g_{\rho_L}$ and $a_{\rho_L}$ are fixed,  increasing mass $M_{\rho_L^+}$ will result in an increasing of the decay constant $f$ and a smaller $\xi$ parameter. The same behavior is also observed in the neutral resonance $\rho_L^0$  decay channels of $t\bar T+b\bar B$ and $X_{2/3}\bar t$ and their charge conjugates due to similar reasons.  There is a difference here between the two models LP(F)$_\4$.  For the partially composite $t_R^{\rm (P)}$ scenario, the decay channels $\rho_L^+\rightarrow X_{5/3}\bar t+X_{2/3}\bar b$  and $\rho_L^0\rightarrow X_{2/3}\bar t +  \bar X_{2/3}t $ can become sizable $\sim 10\%$. However, for the fully composite $t_R^{\rm (F)}$, their branching ratios are below $1\%$. This is due to the fact that the couplings $\rho_L^+\bar{X}_{5/3R} t_R$, $\rho_L^0 \bar X_{2/3R} t_R$ arise from $\mO(\sqrt{\xi})$ in model LP$_\4$ and $\mO(\xi)$ in model LF$_\4$, as can be seen clearly from  Table~\ref{tab:charge} and Table~\ref{tab:neutral}. We also notice that $\rho_L^+\rightarrow t\bar{b}$, $\rho_L^0\rightarrow t\bar{t}+ b\bar{b}$ decay channels are always sizable  even in the  intermediate mass region and the high mass region $M_{\rho_L}>2M_\4$. This is due to the fact that we are fixing $y_L$ and $M_\4$. Hence,  increasing  $M_{\rho_L}$ will also increase $f$. As a result, the left-handed mixing angle $s_{\theta_L}$ becomes larger. The branching ratio ranges from $20\%$ to $40\%$  in the intermediate mass region and above $10\%$ in the high mass region. 

For the mass region of $M_{\rho_L}>2M_\4$, the pure strong dynamics channels are kinematically allowed. Since their couplings are of $\mO(g_\rho)$ and we expect that they will dominate.
Among those channels, the $\rho_L^+\to X_{5/3}\bar X_{2/3}$ channel has the largest branching ratios (above $60\%$), because they are the first and second lightest top partners. Note that the decaying channel into $T\bar{B}$ opens very slowly. In the parameter space under consideration, its branching ratio is always below $10\%$ and smaller than those of the decay channels $t\bar{b}$ and $t\bar{B}+T\bar{b}$. This behavior is due to the particular choice of our parameters in \Eq{eq:paraL}. In particular, the masses of $T,B$ are roughly given by
\beq
M_{T,B} \sim \sqrt{M_\4^2 + y_L^2 M_{\rho_L}^2/(a_{\rho_L}^2 g_{\rho_L}^2)} \sim   \sqrt{M_\4^2 + 2 M_{\rho_L}^2/9}.
\eeq
Even for large $M_{\rho_L}$, the masses of $T,B$ are $\sim0.47\times M_{\rho_L}$ and the decay into
$T\bar B$ suffers from phase space suppression. We also expect that other choices of the parameters (for example smaller value of $y_L$)  will make this channel more relevant.  Things are similar in the case of $\rho_L^0$, where the decay channels into $\bar{X}_{5/3}X_{5/3}$, $\bar{X}_{2/3} X_{2/3}$ are dominant ($>60\%$) and $\bar{T}T+ \bar{B}B$ decaying channels are below 10$\%$. 

\begin{figure}
\centering
\subfigure[~The branching ratios of $\rho_R^+$ in RP$_\4$.]{
\label{RP4_RhoR_decay} %% label for first subfigure
\includegraphics[scale=0.4]{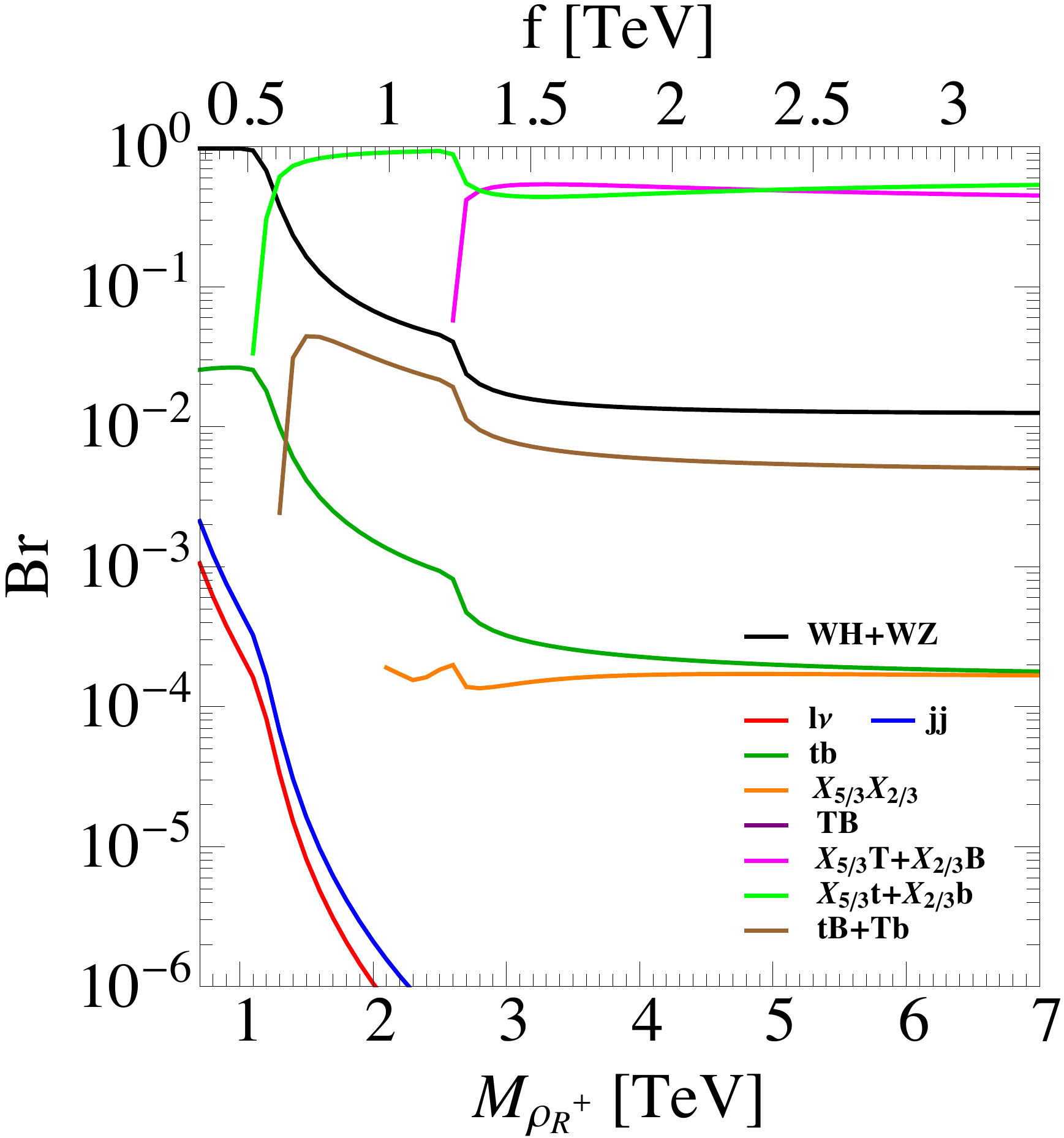}}\quad
\subfigure[~The branching ratios of $\rho_R^0$ in RP$_\4$.]{
\label{RP4_RhoR0_decay} %% label for first subfigure
\includegraphics[scale=0.4]{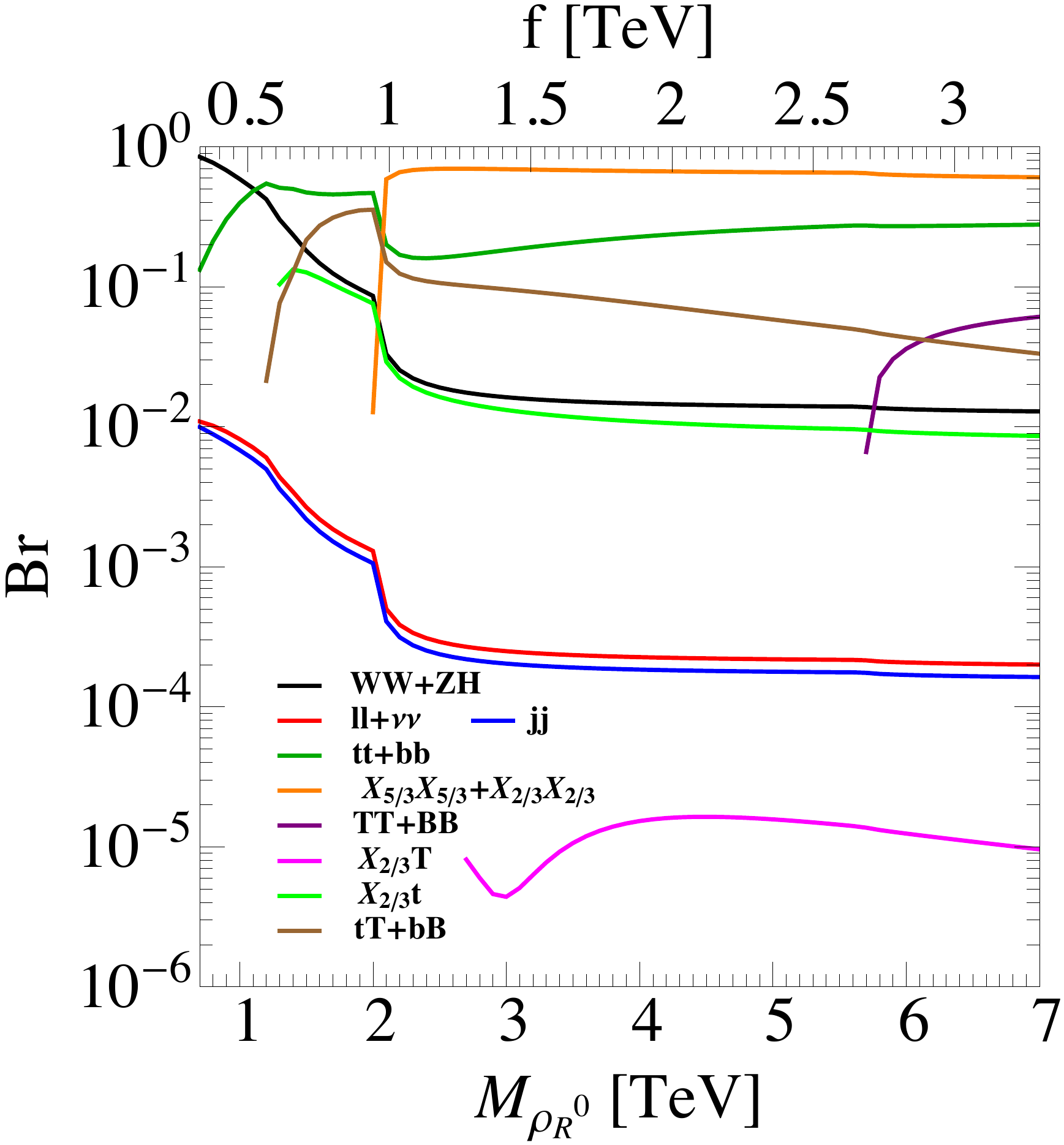}}
\subfigure[~The branching ratios of $\rho_R^+$ in RF$_\4$.]{
\label{RF4_RhoR_decay} %% label for first subfigure
\includegraphics[scale=0.4]{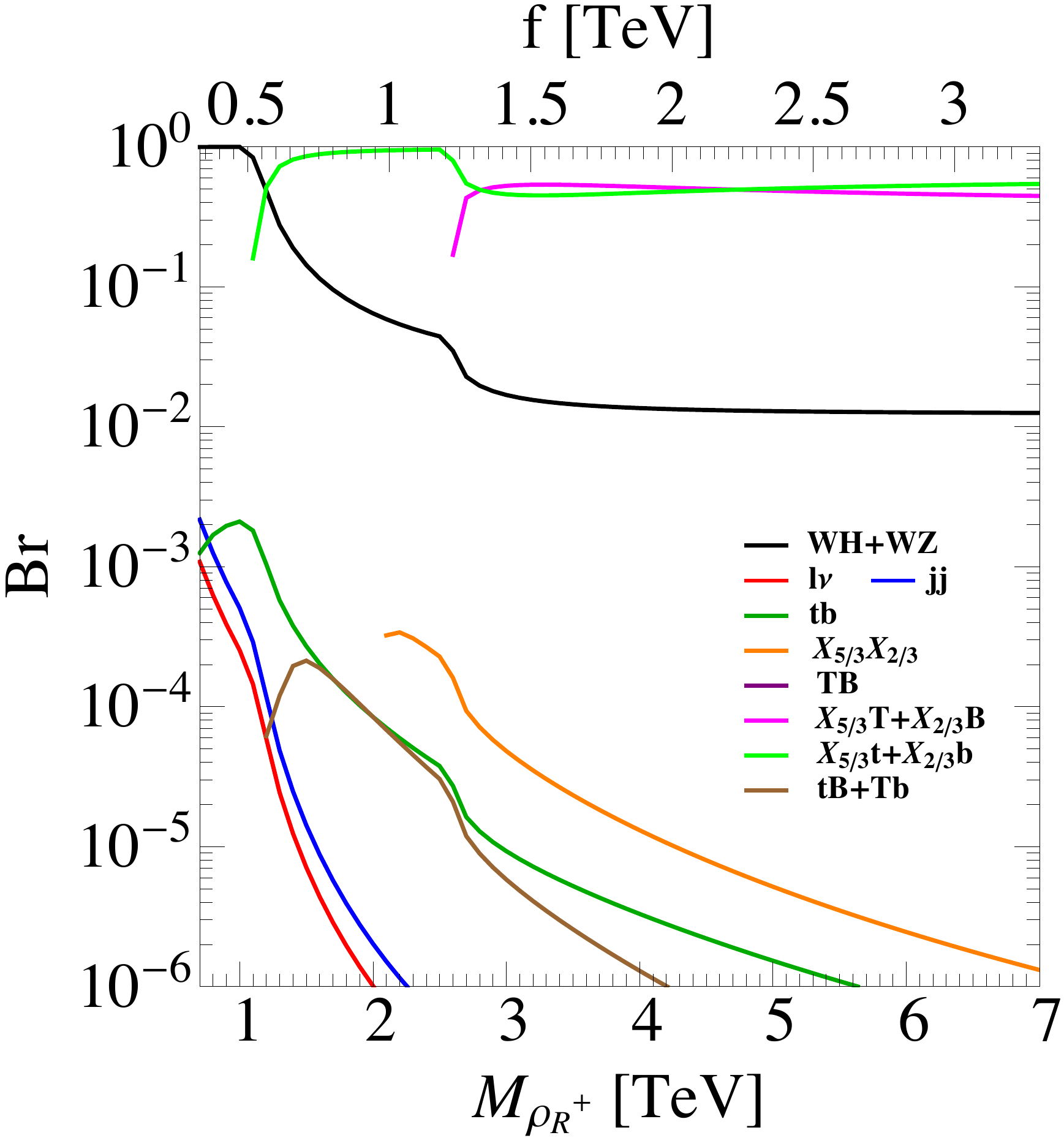}}\quad
\subfigure[~The branching ratios of $\rho_R^0$ in RF$_\4$.]{
\label{RF4_RhoR0_decay} %% label for first subfigure
\includegraphics[scale=0.4]{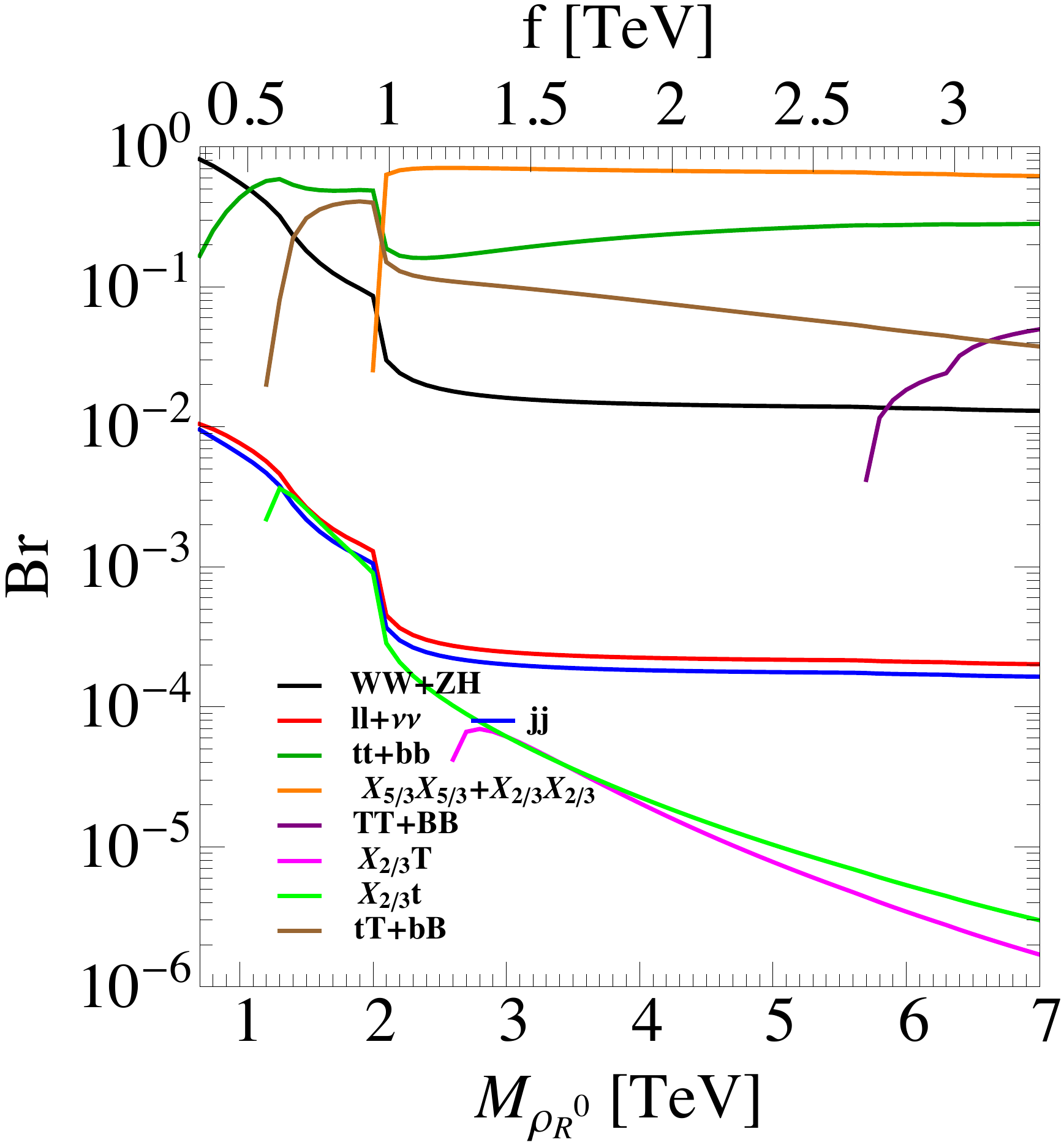}}
\caption{The decay branching ratios of the $\rho_R^{\pm,0}$ resonances. The parameters are $g_{\rho_R}=3$, $a_{\rho_R}^2=1/2$, $M_\4=1$ TeV, $y_L=1$ and $c_1=1$. For RF$_\4$ there is an additional parameter $c_2=1$. The $\text{Br}(T\bar B)$ in subfigures (a) and (c) are less than $10^{-7}$, thus not shown in the figures.}
\label{RhoR_Br}
\end{figure}

Next we turn to the $(\1,\3)$ resonances $\rho_R^{\pm,0}$. The benchmark point is the same as that in the $\rho_L^{\pm,0}$ case, with the replacement $\rho_L \to \rho_R$. Unlike $\rho_L^+$, the $\rho_R^+$ does not mix with SM gauge bosons before EWSB because of its quantum number. Consequently, its decay branching ratios to SM light fermions are  tiny. For example, it is less than $10^{-3}$ for the parameter space shown in Fig.~\ref{RP4_RhoR_decay} and \ref{RF4_RhoR_decay}. The decaying branching ratio into $t\bar b$ is also suppressed because the corresponding coupling arises after EWSB and is of order $\mO(g_{\rho_R}\xi)$. As a consequence, the $\rho_R^+$ mainly decays into di-boson channels $W^+h+W^+Z$. In the intermediate mass region, the decaying into $X_{5/3}\bar{t} + X_{2/3}\bar{b}$  channels dominate over all the other channels with branching ratio larger than $90\%$ in both model  LP$_\4$ and  LF$_\4$, as  their left-handed couplings arise before EWSB.  The decay channels into $t\bar{B}+T\bar{b}$ are very small ($2\%-4\%$) for model LP$_\4$ and below $10^{-3}$ for model LF$_\4$. In the high-mass region, the dominant decaying channels are $X_{5/3}\bar{T} + X_{2/3} \bar{B}$ and $X_{5/3}\bar{t} + X_{2/3}\bar{b}$ with similar branching ratios. It is interesting to see that the heavy-light decay channel is still  sizable in the high-mass region, as the mixing angle $s_{\theta_L}$ becomes larger for larger $\rho_R^+$ mass and the mass of $T$, $B$ increase with $M_{\rho_R}$ as discussed before. The neutral resonance $\rho_R^0$ mixes with the SM Hypercharge  gauge boson before EWSB, resulting in the relation $\text{Br}(jj)=22/27\times\text{Br}(\ell^+\ell^-+\nu_\ell\bar\nu_\ell)$~\cite{Brooijmans:2014eja}, as shown in Fig.~\ref{RP4_RhoR0_decay} and \ref{RF4_RhoR0_decay}. The branching ratios of the other decay channels of $\rho_R^0$  are very similar to those of $\rho_L^0$, and we will not discuss them further.

\begin{figure}[t]
\centering
\subfigure[~The branching ratios of $\rho_X^0$ in XP$_\4$.]{
\label{XP4_RhoX0_decay} %% label for first subfigure
\includegraphics[scale=0.4]{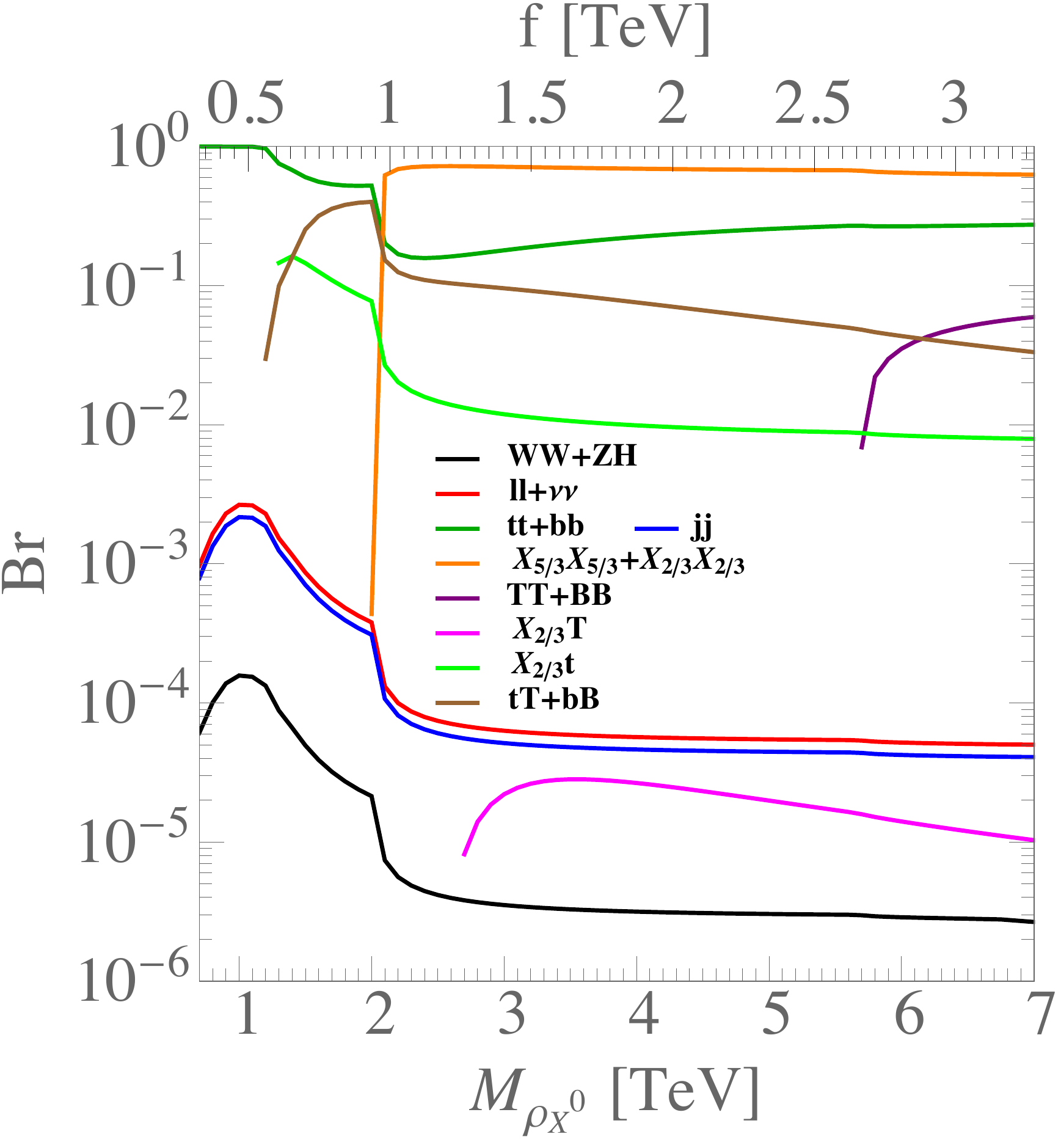}}\quad
\subfigure[~The branching ratios of $\rho_X^0$ in XF$_\4$.]{
\label{XF4_RhoX0_decay} %% label for first subfigure
\includegraphics[scale=0.4]{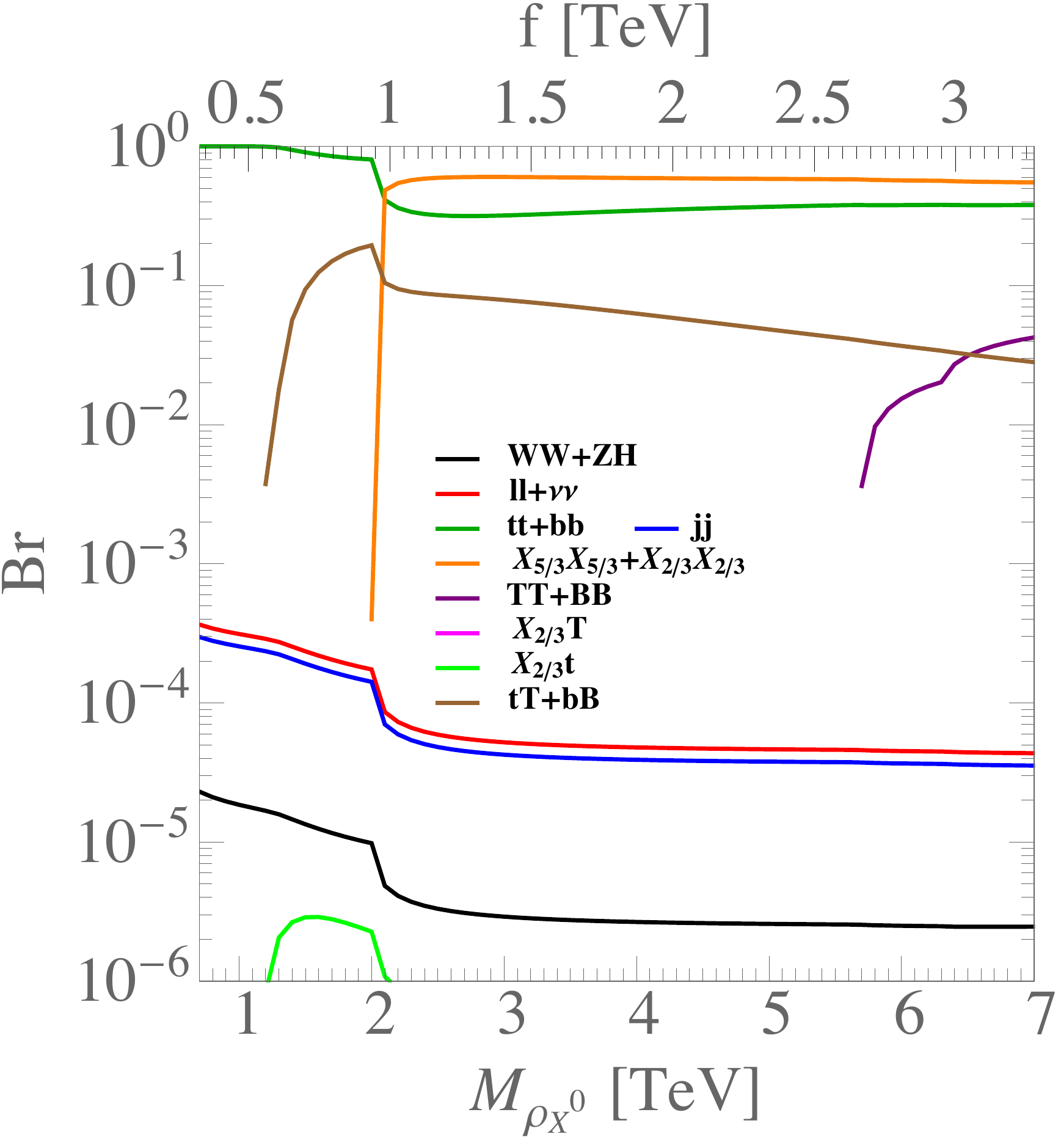}}
\subfigure[~The branching ratios of $\rho_X^0$ in XP$_\1$.]{
\label{XP1_RhoX0_decay} %% label for first subfigure
\includegraphics[scale=0.4]{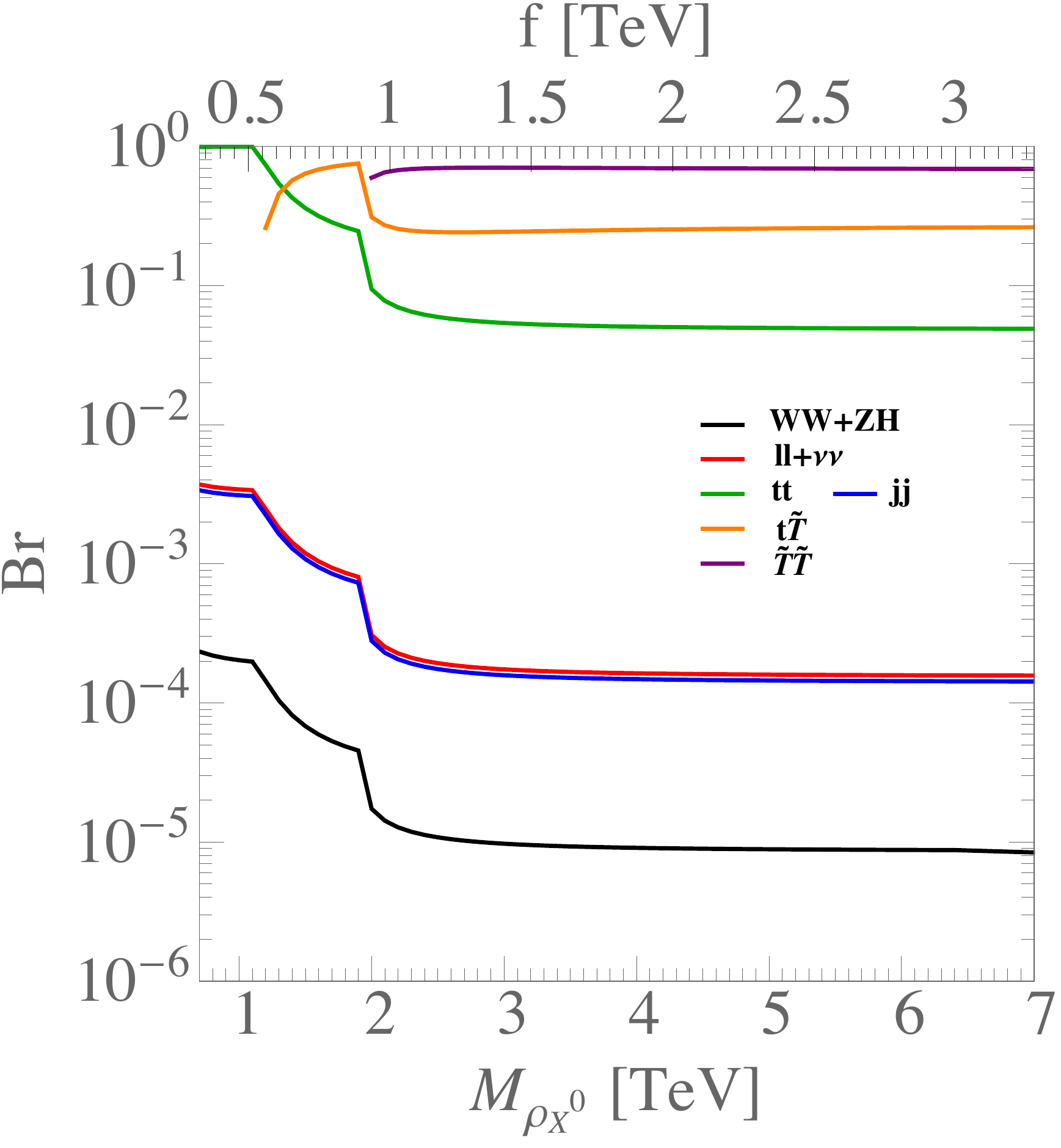}}\quad
\subfigure[~The branching ratios of $\rho_X^0$ in XF$_\1$.]{
\label{XF1_RhoX0_decay} %% label for first subfigure
\includegraphics[scale=0.4]{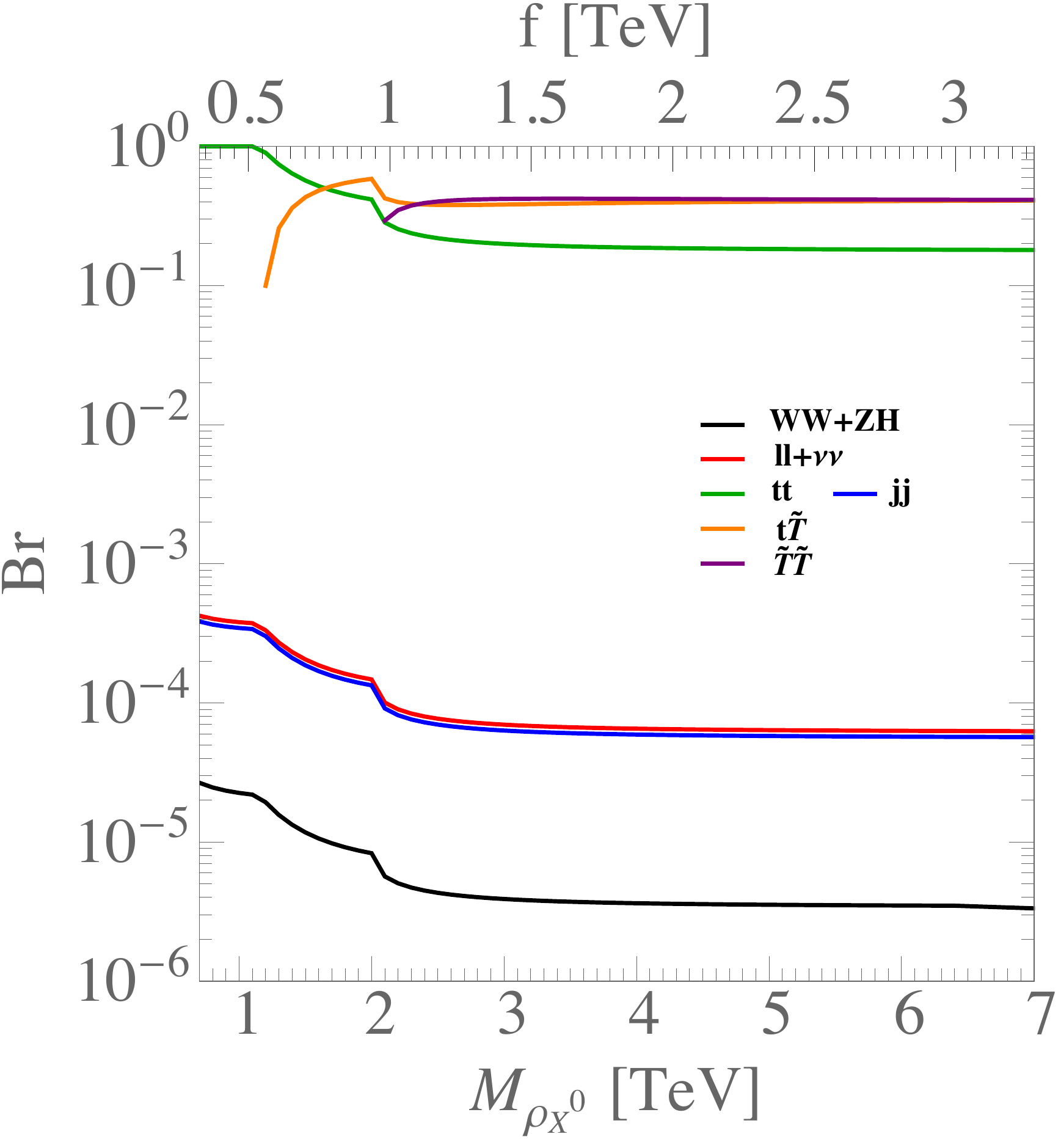}}
\caption{The decay branching ratios of the $\rho_X^0$ resonances. The parameters are  chosen as $g_{\rho_X}=3$, $a_{\rho_X}^2=1/2$, $M_{\4,\1}=1$ TeV and $c_1=1$. For the XP$_\1$ model  we have chosen $y_L=1.5$, while for other three models we set $y_L=1$. For the XF$_\4$ model, we set $c'_1=1$, and for XF$_\1$, we set $c'_1=c''_1=1$.}
\label{RhoX_Br}
\end{figure}

Finally, we study the $(\1,\1)$ resonance $\rho_X^0$. As an $SO(4)$ singlet, the $\rho_X^0$ can couple either to quartet $\Psi_\4$ or to the singlet $\Psi_\1$, and the corresponding models are XP(F)$_\4$ and XP(F)$_\1$, respectively.  In our plots, the parameters chosen are very similar to the benchmark point of $\rho_L^{\pm,0}$, except for XP$_\4$ where we choose $y_L=1.5$. For the XF$_{\4,\1}$ model, there is another parameter $c_1'$ describing the direct interaction between the fully composite $t_R^{\rm (F)}$ and the $\rho_X^0$ resonance, and it is set to be 1. For the XF$_\1$ model,  we further set $c''_1$ (the parameter describing the interaction between $t_R^{\rm (F)}$ and the $\rho_X^0$, $\Psi_\1$ resonances) to be 1. Since the $U(1)_X$ has no direct connection to the dynamical symmetry breaking $SO(5)\to SO(4)$, its corresponding spin-1 resonance $\rho_X^0$ does not couple to the Goldstone boson $H$ before EWSB. Consequently, the decaying branching ratios into SM di-bosons $W^+W^-+Zh$ are very small ($< 10^{-4}$). The di-fermion decay channels of XP(F)$_\4$ are very analogous to those of $\rho_R^0$ in RP(F)$_\4$. The most relevant channels are  $\rho_X^0\rightarrow t\bar{t}+b\bar{b}$ in the low-mass region, $ \rho_X^0\rightarrow t\bar T+\bar t T+b\bar B+\bar bB$ in the intermediate mass region, and $\rho_X^0 \rightarrow \bar{X}_{5/3} X_{5/3} + \bar{X}_{2/3} X_{2/3}$ in the high-mass region .  In models with singlet top partner XP(F)$_\1$, since the $b$ quark does not mix with the resonance, we classify as one of the ``SM light fermions''. Therefore, we have $\text{Br}(jj)=\text{Br}(\ell^+\ell^-+\nu_\ell\bar\nu_\ell)$, as shown in the bottom panel of Fig.~\ref{RhoX_Br}. In model XP$_\1$,  the dominant decaying channels are $\rho_X^0\rightarrow t\bar{t}$ in the low-mass region, $\rho_X^0\rightarrow \bar{t}\widetilde{T}+\overline{\widetilde T} t$ ($\sim$ 70\%) in the intermediate mass region and $\rho_X^0\rightarrow \overline{\widetilde T} \widetilde T (\sim 70\%)$ in the high mass region. The situation is similar in the model XF$_\1$ except that in the high-mass region, the $\rho_X^0\rightarrow t\bar{t}$ and  $\rho_X^0\rightarrow \bar{t}\widetilde{T}+\overline{\widetilde T} t$ decaying channels are also relevant. Their branching ratios are around $20\%$ and $40\%$, respectively.

\section{The present limits and prospective reaches at the LHC}
\label{sec:bound}

In this section, we present the current limits and prospective reaches for the simplified models at the LHC. 

\subsection{Making projections}\label{sec:lhcsearch}

For the projections at the high luminosity or high energy LHC, we  extrapolate from the current LHC searches using  a similar method as in \Ref{Thamm:2015zwa}.  We described the method in detail in Appendix~\ref{app:method}.  

There have been a number of searches for beyond the SM (BSM) resonances at the LHC, providing constraints to the composite Higgs models. To use a more generic and uniform notation in describing the searches, we denote the spin-1 resonances as $\rho$ and the spin-$1/2$ resonances as $F_Q$, where $Q$ is the electric charge. The results at the 13 TeV LHC can be classified into two main groups. The first group is the Drell-Yan production and two-body decay of $\rho$, its various final states can be summarized as follows,
\begin{enumerate}
\item SM di-fermion final states, including di-lepton, di-jet, and the third generation quark-involved channels. We list the relevant measurements in Table~\ref{table:difermion}.

\begin{table}[h!]\footnotesize
\begin{tabular}{|c|c|} \hline
Channel & Collaboration and corresponding integrated luminosity \\ \hline
$q\bar q\to\rho^0\to\ell^+\ell^-$ & \tabincell{c}{ATLAS at 36 fb$^{-1}$~\cite{Aaboud:2017buh};\\
CMS at 77.3 fb$^{-1}$ (for $e$ channel) and 36.3 fb$^{-1}$ (for $\mu$ channel)~\cite{CMS-PAS-EXO-18-006}.} \\ \hline
$q\bar q'\to\rho^\pm\to\ell^+\nu/\ell^-\bar\nu$ & ATLAS at 79.8 fb$^{-1}$~\cite{ATLAS-CONF-2018-017}; CMS at 35.9 fb$^{-1}$~\cite{Sirunyan:2018mpc,CMS-PAS-EXO-17-008}.\\ \hline
$q\bar q^{(\prime)}\to \rho^{\pm,0}\to jj$ & ATLAS at 37 fb$^{-1}$~\cite{Aaboud:2017yvp}; CMS at 77.8 fb$^{-1}$~\cite{CMS-PAS-EXO-17-026}.\\ \hline
$q\bar q^{(\prime)}\to \rho^{\pm,0}\to jj$ (with $b$-tagging) & ATLAS at 36.1 fb$^{-1}$~\cite{Aaboud:2018tqo}.\\ \hline
$q\bar q\to \rho^0\to t\bar t$ & ATLAS at 36.1 fb$^{-1}$~\cite{Aaboud:2018mjh}; CMS at 35.9 fb$^{-1}$~\cite{Sirunyan:2018ryr}.\\ \hline
$q\bar q'\to\rho^\pm\to t\bar b/\bar tb$ & ATLAS at 36.1 fb$^{-1}$~\cite{Aaboud:2018juj,Aaboud:2018jux}; CMS at 35.9 fb$^{-1}$~\cite{Sirunyan:2017vkm}.\\ \hline
\end{tabular}
\caption{The present searches for BSM spin-1 resonances in di-fermion final states at the 13 TeV LHC.}\label{table:difermion}
\end{table}

\item SM di-boson final states. The topology is $q\bar q^{(\prime)}\to\rho^{\pm,0}\to W^\pm Z/W^+W^-/W^\pm h/Zh$, and different final states are summarized in Table~\ref{table:di-boson}. Note that we have no $hh$ or $ZZ$ final states  from the decay of spin-1 resonances in composite Higgs models .

\begin{table}[h!]\footnotesize
\begin{tabular}{|c|c|c|} \hline
& $V\to qq$ & $W\to\ell\nu$ or $Z\to\ell\ell/\nu\bar\nu$ \\ \hline
$V\to qq$ & ATLAS at 79.8 fb$^{-1}$~\cite{ATLAS-CONF-2018-016}; CMS at 35.9 fb$^{-1}$~\cite{Sirunyan:2017acf}. & $-$ \\ \hline
$W\to\ell\nu$ & ATLAS at 36.1 fb$^{-1}$~\cite{Aaboud:2017fgj}; CMS at 35.9 fb$^{-1}$~\cite{Sirunyan:2018iff}; & ATLAS at 36.1 fb$^{-1}$~\cite{Aaboud:2017gsl}. \\ \hline
$Z\to\ell\ell$ & ATLAS at 36.1 fb$^{-1}$~\cite{Aaboud:2017itg}; CMS at 35.9 fb$^{-1}$~\cite{Sirunyan:2018hsl}; & ATLAS at 36.1 fb$^{-1}$~\cite{Aaboud:2018ohp}. \\ \hline
$Z\to\nu\nu$ & ATLAS at 36.1 fb$^{-1}$~\cite{Aaboud:2017itg}; CMS at 35.9 fb$^{-1}$~\cite{Sirunyan:2018ivv}. & $-$  \\ \hline
$h\to bb$ & ATLAS at 36.1 fb$^{-1}$~\cite{Aaboud:2017ahz}; CMS at 35.9 fb$^{-1}$~\cite{Sirunyan:2017wto}; & ATLAS at 36.1 fb$^{-1}$~\cite{Aaboud:2017cxo}; CMS at 35.9 fb$^{-1}$~\cite{Sirunyan:2018qob}.\\ \hline
$h\to\tau\tau$ & CMS at 35.9 fb$^{-1}$~\cite{Sirunyan:2018fuh}. & $-$ \\ \hline
\end{tabular}
\caption{The searches for BSM spin-1 resonances in di-boson final states at the 13 TeV LHC, where $V$ denotes $W^\pm$, $Z$. The search in Ref.~\cite{Aaboud:2018ohp} is also sensitive to the VBF production of $\rho^\pm$, but the constraint is quite weak ($\lesssim1$ TeV). For a summary of the di-boson results from ATLAS at $\sim36~{\rm fb}^{-1}$, we refer the readers to Ref.~\cite{Aaboud:2018bun}.}
\label{table:di-boson}
\end{table}

\item The $tF_Q$ final state, i.e. the heavy resonance-SM fermion channel with one top quark and one top partner. The newest result is $q\bar q\to\rho^0\to t\bar F_{2/3}\to t\bar tZ$, measured by CMS at 35.9 fb$^{-1}$~\cite{Sirunyan:2017ynj,CMS-PAS-B2G-17-015}. Hereafter, for simplicity the charge conjugate of the particle decay final state is always implied; for example, $t\bar F_{2/3}$ denotes both $t\bar F_{2/3}$ and $\bar t F_{2/3}$.
\end{enumerate}

The second group of limits are from the search of the fermionic resonances $F_Q$. Their production mechanisms can be categorized into QCD pair production and single production processes. For the QCD pair production,  $pp \to F_Q \bar{F}_Q$, the experimental collaborations have searched for resonances with different charges in various decay channels. The results relevant to our models are the searches for $F_{2/3}\to tZ/th/bW$, $F_{-1/3}\to tW^-$ and $F_{5/3}\to tW^+$ in various final states by the ATLAS and CMS collaborations at an integrated luminosities around 36 fb$^{-1}$~\cite{Aaboud:2017qpr,Aaboud:2018xuw,Aaboud:2017zfn,Sirunyan:2017pks,Aaboud:2018uek,CMS-PAS-B2G-16-019,Sirunyan:2018yun,ATLAS-CONF-2018-032,Aaboud:2018saj,Sirunyan:2018omb,Aaboud:2018xpj,Aaboud:2018wxv}. A summary of the QCD pair produced top partner searches by the ATLAS Collaboration can be found in Ref.~\cite{Aaboud:2018pii}.

In addition, there have been searches for singly produced top partners. Such channels typically have larger rates than the QCD pair production. However, they are also more model-dependent. Currently, the $bW\to F_{2/3}\to bW$ channel is explored by ATLAS at 3.2 fb$^{-1}$~\cite{ATLAS-CONF-2016-072} and CMS at 2.3 fb$^{-1}$~\cite{Sirunyan:2017tfc}; while the singly produced $F_{2/3}\to tZ$ channel has been searched  with 36 fb$^{-1}$ by CMS~\cite{Sirunyan:2017ynj}    ($tZ$ fusion) and ATLAS~\cite{Aaboud:2018saj} ($bW$ fusion). The $tW$ fusion production of $F_{5/3}$ (or $F_{-1/3}$) has been searched by CMS at 35.9 fb$^{-1}$~\cite{Sirunyan:2018ncp} and by ATLAS at 36.1 fb$^{-1}$~\cite{Aaboud:2018xpj}.

For the new channels we propose in this paper, especially the cascade decays of the $\rho$  resonances to the heavy fermionic resonances, there have been no dedicated searches. We estimate their exclusion by recasting existing searches using the SSDL final states $\ell^\pm\ell^\pm{\rm +jets}$. In Table~\ref{table:same-sign}, we have listed the existing searches  for the resonances at the LHC using $\ell^\pm\ell^\pm{\rm +jets}$ final state. The upper limit on the cross section of the highest mass points considered in the searches and the corresponding number of events before any kinematic cuts are reported in the table. Motivated by these results, we assume that a limit can be set for $N(\ell^\pm\ell^\pm+{\rm jets})=20$ before any kinematical cuts.

\begin{table}[!h]\footnotesize
\begin{tabular}{|c|c|c|} \hline
Experimental searches & Cross section upper limit of the tail & Event number before cut\\ \hline
CMS at 35.9 fb$^{-1}$~\cite{CMS-PAS-B2G-16-019} & $\sigma(pp\to X_{5/3}\bar X_{5/3}\to tW\bar tW)=16$ fb at 1.5 TeV & $N(\ell^\pm\ell^\pm+{\rm jets})=23$ \\ \hline
CMS at 35.9 fb$^{-1}$~\cite{Sirunyan:2017uyt} & \tabincell{c}{$\sigma(pp\to t\bar t,tW,tq+X) {\rm Br} (X\to t\bar t)=30$ fb at 0.55 TeV\\ X denotes a new scalar or pseudo-scalaer} & $N(\ell^\pm\ell^\pm+{\rm jets})=42$\\ \hline
ATLAS at 36.1 fb$^{-1}$~\cite{Aaboud:2017dmy} & \tabincell{c}{$\sigma(pp\to \widetilde b_R \widetilde b_R\to \bar t\bar s\bar t\bar s)=14$ fb at 1.6 TeV\\ $\widetilde  b_R^\dagger\widetilde  b_R^\dagger$ is not included because PDF $f_{\bar d}\ll f_{d}$.} & $N(\ell^-\ell^-+{\rm jets})=20$\\ \hline
\end{tabular}
\caption{The experimental results of the SSDL final states at the 13 TeV LHC.}
\label{table:same-sign}
\end{table}

Next, we present the results for models LP(F)$_\4$, RP(F)$_\4$, XP(F)$_\4$ and XP(F)$_\1$ in subsequent subsections. 

\subsection{The results of LP$_\4$ and LF$_\4$}
\begin{figure}
\centering
\subfigure[~The results of LP$_\4$.]{
\label{LP4_results} %% label for first subfigure
\includegraphics[scale=0.4]{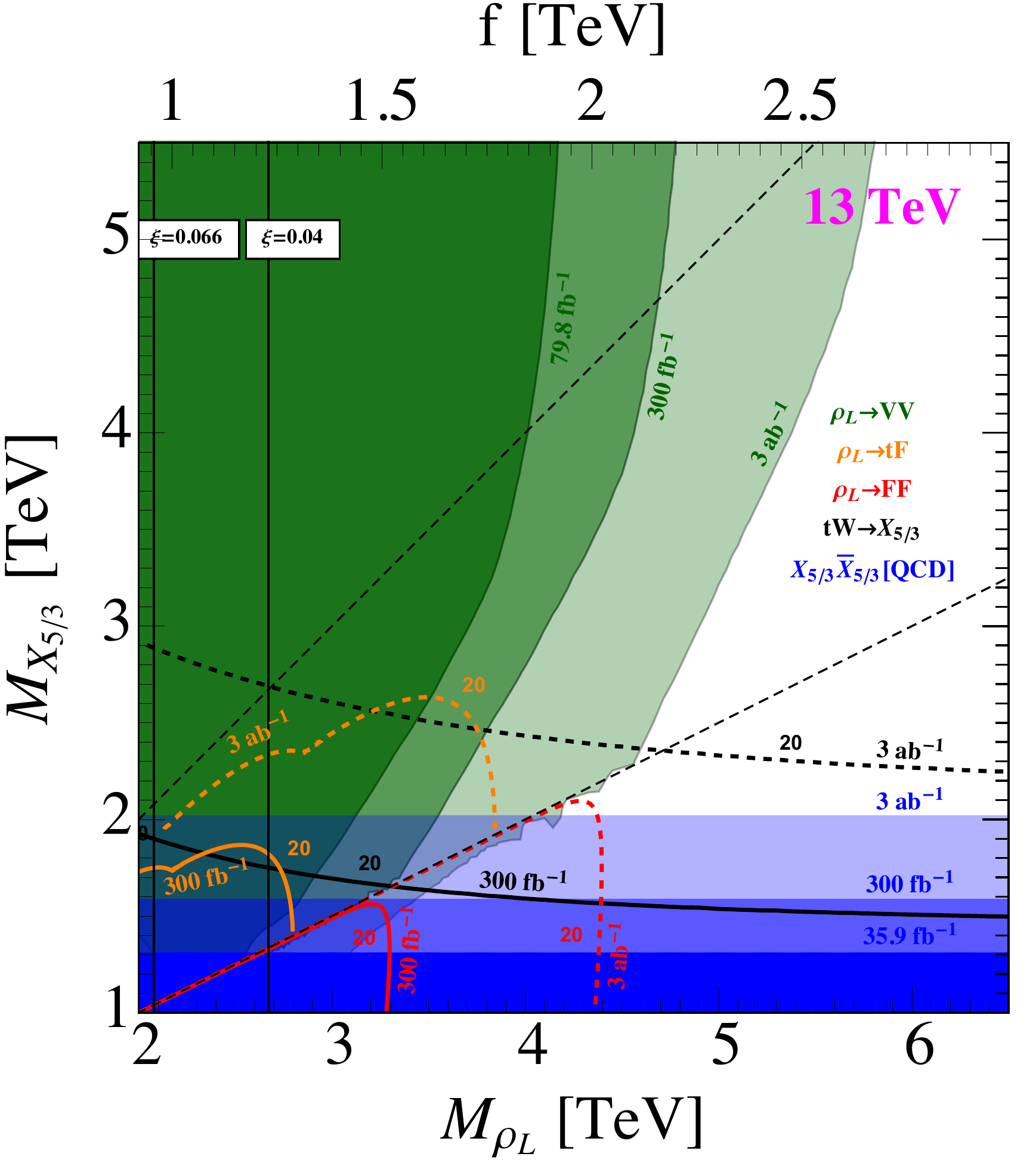}}\quad
\subfigure[~The results of LF$_\4$.]{
\label{LF4_results} %% label for first subfigure
\includegraphics[scale=0.4]{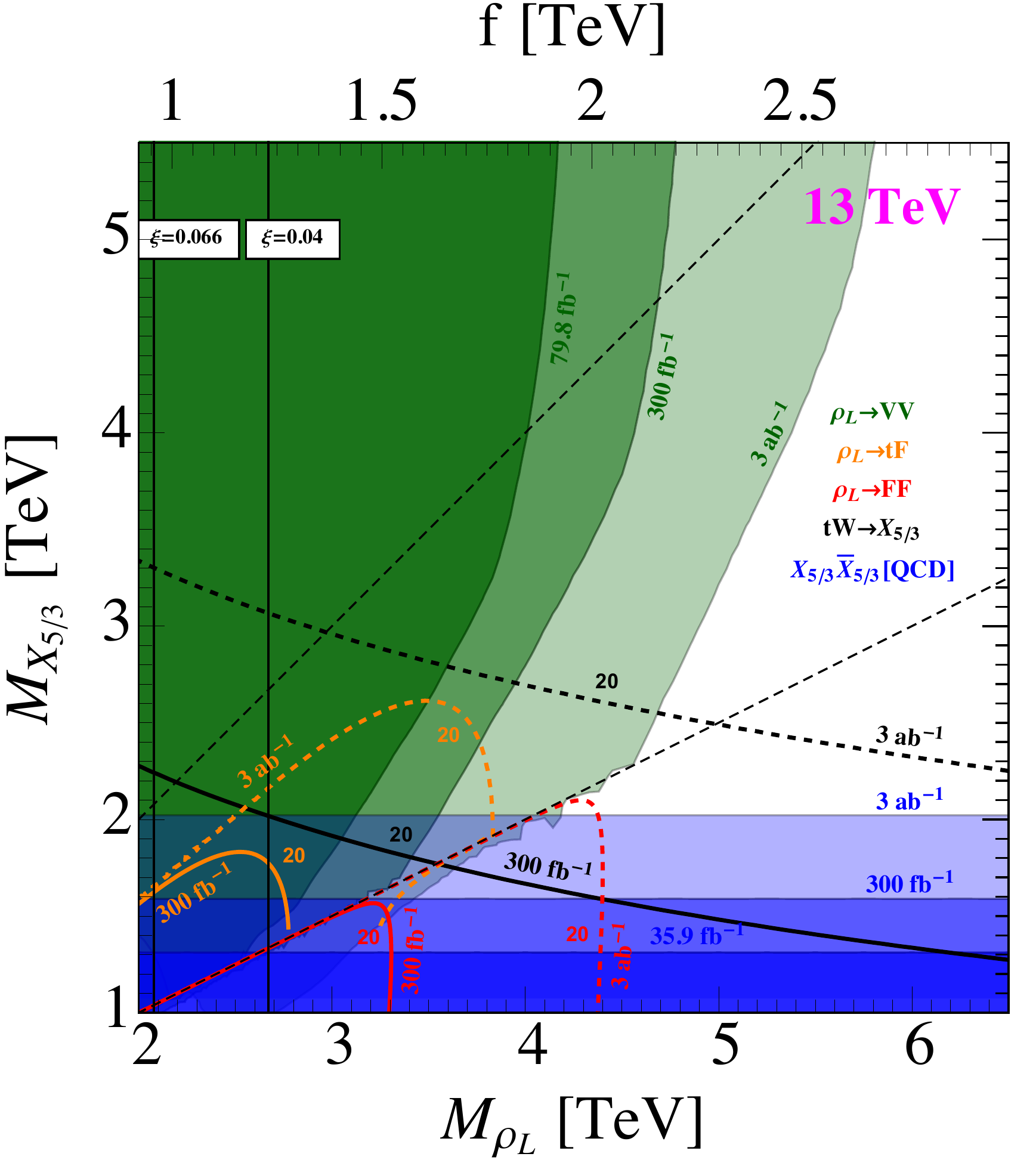}}
\caption{The current limits and prospective reaches on model LP$_\4$ (left plot) and model LF$_\4$ (right plot). The parameters are chosen as \Eq{eq:benchrhoL}. The existing limits are shown as the darkest shaded regions, while the projections for 300 (3000) fb$^{-1}$ are plotted the lighter shaded regions. The production channels include Drell-Yan $\rho_L^{\pm,0}\to W^\pm Z/W^+W^-$~\cite{ATLAS-CONF-2018-016} (denoted as $\rho_L\to VV$) and QCD pair production of $X_{5/3}\bar X_{5/3}$~\cite{Sirunyan:2018yun}. The event number contours for $N(\ell^\pm\ell^\pm+\text{jets})=20$ are plotted in solid (dashed) lines for 300 (3000) fb$^{-1}$, to set a prospective limit for the proposed channels, including $\rho_L^{\pm}\to t\bar B/X_{5/3}\bar t$ (denoted as $\rho_L\to tF$), $\rho_L^{\pm,0}\to X_{5/3}\bar X_{2/3}/X_{5/3}\bar X_{5/3}$ (denoted as $\rho_L\to FF$) and $tW\to X_{5/3}$. See the main text for more details.}
\label{ML4_results}
\end{figure}

In this subsection, we investigate the current limits and prospective reaches on the models LP(F)$_\4$ at the 13 TeV LHC.
In the Lagrangian of LP$_\4$ in \Eq{LP4_lll}, there are 10 parameters: 
$$\{g_2,\;g_1,\;v,\;g_{\rho_L},\;a_{\rho_L},\;f,\;M_\4,\;y_L,\;y_R,\;c_1\}.$$
The electroweak input parameters $\{\alpha,M_Z,G_F,M_t\}$ provide 4 constraints, leaving us 6 free parameters, which we have chosen to be $\{g_{\rho_L}, a_{\rho_L}, M_{\rho_L^0},M_\4,y_L,c_1\}$. Because $M_{\rho_L^0}$ is nearly degenerate with $M_{\rho_L^\pm}$, we denote them with the same variable $M_{\rho_L}$. Note that the Lagrangian mass $M_\4$ is also the exact physical mass of  $X_{5/3}$ after EWSB, since none of the SM  particles  can mix with it. For LF$_\4$, the situation is almost the same, expect there is an additional $\mO(1)$ strong dynamics parameter $c_2$. To better demonstrate the interplay of the spin-1 and spin-$1/2$ resonances, we scan $(M_{\rho_L},M_{X_{5/3}})$ while fixing the other parameters to a benchmark point\footnote{Smaller value of $g_{\rho_L}$ will make Drell-Yan di-lepton resonance search more relevant. Large $g_{\rho_L}$ will make the production cross section too small to have  relevant effects at the LHC.}
\be
\label{eq:benchrhoL}
g_{\rho_L}=3,\quad a_{\rho_L}^2=\frac{1}{2},\quad y_L=1,\quad c_1=1,\quad c_2=1~(\text{for LF$_\4$ only}).
\ee
The results of LP$_\4$ and LF$_\4$ are shown in Fig.~\ref{LP4_results} and \ref{LF4_results}, respectively. Since $g_{\rho_L}$ is fixed, $f$ is determined by $M_{\rho_L}$, and we use its value to label the top horizontal axis.

We plot the existing bounds from LHC searches and their extrapolations at 300 (3000) fb$^{-1}$ in colored shaded regions. Besides the direct searches for resonances, the measurement of and $\xi$ parameter can provide an indirect constraint. Currently LHC results imply $\xi\lesssim0.13$~\cite{Sanz:2017tco,deBlas:2018tjm}, while the further constraints are expected to be as good as $0.066$ ($0.04$) with 300 (3000) fb$^{-1} $ of data~\cite{CMS-NOTE-2012-006,ATL-PHYS-PUB-2013-014,Dawson:2013bba}. We also plot the constraints on $\xi$ in the figures as vertical  black thin lines.

Putting all the constraints and projections together, we  see that the future data at the LHC will explore the parameter space of LP(F)$_\4$ extensively~\footnote{The left hand side of the figures start from  $M_{\rho_L} = 2$ TeV. The $\hat S$-parameter constraint roughly gives us $M_{\rho_L} > 1.9$ TeV in our parameter choice and we didn't show it~\cite{Barbieri:2004qk,Giudice:2007fh}.}. The constraints are similar in the two models LP(F)$_\4$ . For a relatively large value of $g_{\rho_L}$ (for example, $g_{\rho_L}=3$ in our benchmark point), the most sensitive channel in the $M_{\rho_L}<2M_{X_{5/3}}$ region is the $W^\pm Z/W^+W^-$ search with boosted di-jet channel  performed by the ATLAS Collaboration with integrated luminosity $L=$ 79.8 fb$^{-1}$  in Ref.~\cite{ATLAS-CONF-2018-016}. In such a mass region, the $\Gamma_{\rho_L}/M_{\rho_L}$ ratio is $\sim0.8\%$ for our chosen parameters, thus the narrow width approximation works very well. If $g_{\rho_L}\lesssim2$, the di-lepton $\ell^+\ell^-$ channel by CMS with $L = 77.3~{\rm fb}^{-1}(e^+e^-)+36.3~{\rm fb}^{-1}(\mu^+\mu^-)$~\cite{CMS-PAS-EXO-18-006} gives the strongest limit. Because of the large experimental uncertainty, the $\rho \to t\bar t$, $b\bar b$ and $t\bar b$ channels are not able to give competitive limits, although they have significant branching ratios. In Fig.~\ref{ML4_results}, we only show the present limits and prospective reaches from ATLAS di-boson boosted jet channels in Ref.~\cite{ATLAS-CONF-2018-016}. It is clear from the figure that the interactions with light top-partner has affected the phenomenology of $\rho_L$ significantly. In particular, the present bound is relaxed from 4.2 TeV to 2.6 TeV for our benchmark parameters in \Eq{eq:benchrhoL} as the mass of top partner changes from $M_{X_{5/3}} \gg M_{\rho_L}$ to  $M_{\rho_L} \gtrsim 2 M_{X_{5/3}}$. Once the decays into pair of top partners are kinematically open, the bound becomes very weak. At the same time, very light top partners have been excluded by the direct searches for the top partner.

In the mass region of $M_{X_{5/3}}<M_{\rho_L}\leqslant 2M_{X_{5/3}}$,  the decays of $\rho_L$ into one top partner and one SM particles are kinematically allowed. The width of the $\rho_L$ resonance is enhanced by the existence of those new channels, but still within the narrow width range. For example, in our benchmark \Eq{eq:benchrhoL}, $\Gamma_{\rho_L}/M_{\rho_L}$ is $4.8\%-1.2\%$ for $M_{X_{5/3}}=0.6\times M_{\rho_L}-0.9\times M_{\rho_L}$. The $t\bar tZ$ final state from the decay channel $\rho_L^{0}\to  t\bar T$ has been studied both experimentally~\cite{Sirunyan:2017ynj, CMS-PAS-B2G-16-001} and theoretically~\cite{Backovic:2016ogz}, but current experimental results are still too weak to be visible in Fig.~\ref{ML4_results}.  The $\rho_L^\pm\to T\bar b\to t\bar bZ$ channel is studied phenomenologically in Ref.~\cite{Vignaroli:2014bpa}. In this work, we propose that the $t\bar tW^\pm\to\ell^\pm\ell^\pm+{\rm jets}$ final state from $\rho_L^{\pm}\to  t\bar B/t\bar X_{5/3}$ can also be a good channel to probe such a heavy-light decay. In Fig.~\ref{ML4_results}, we have plotted the contours for the constant number ($=20$) of SSDL events summing all these decay channels at 300 fb$^{-1}$ and 3 ab$^{-1}$ LHC. These channels have sensitivity to the parameter space up to $M_{\rho_L} = 3.8$ TeV at 3 ab$^{-1}$ LHC, but it still can't compete with the di-boson jet searches. This is due to the fact that the branching ratios into the heavy-light channels are not significantly larger than the di-boson channel and the decaying branching ratios to the SSDL are very small. It is interesting to explore other more complicated final states like $1\ell + {\rm jets}$ and we leave this for future possible work.

In the mass region of $M_{\rho_L}>2M_{X_{5/3}}$, the spin-1 resonances will decay dominantly  into pairs of top partners, as discussed in detail in Sec.~\ref{sec:decay}. We focus here on the decay channels resulting in the SSDL final states: $\rho_L^{\pm,0}\to X_{5/3}\bar X_{2/3}/X_{5/3}\bar X_{5/3}$ (see also Refs.~\cite{Barducci:2015vyf,Vignaroli:2014bpa} for the study of these channels). We plot the contours with 20  SSDL events, summing over all the above decay channels for 300 fb$^{-1}$ and 3 ab$^{-1}$ LHC. The prospective for the cascade decay channels are very promising and comparable with direct searches for the pair produced $X_{5/3}$. If the top partner is around 1 TeV, these channels can be promising to discover the heavy spin-1 resonance~\footnote{If the first generation light quarks have some degrees of compositeness as studied in Ref.~\cite{Low:2015uha}, the cascade decay channels are more important as the Drell-Yan cross sections of $\rho_L$ are enhanced by the extra piece of coupling of $\mO(g_\rho s_{\theta_{1q}})$. Here $\theta_{1q}$ is the mixing angle between the first generation quark and the corresponding partners.}. Note that in such region the $\Gamma_{\rho_L}/M_{\rho_L}$ can be large. For example, for our benchmark point \Eq{eq:benchrhoL}, $\Gamma_{\rho_L}/M_{\rho_L}$ varies from 56\% to 37\% when $M_{X_{5/3}}$ varies from $0.1\times M_{\rho_L}$ to $0.4\times M_{\rho_L}$. It is interesting to study the effects of large decay width on the resonance searches and we leave this for a future work. Here we just estimate the bounds by an event-counting method based on the SSDL final state, which does not require the reconstruction of a resonance peak. We expect such an estimate has less dependence on the width of $\rho_L$.

We have shown the present bounds and the prospectives of  the searches for  QCD pair produced $X_{5/3}\bar X_{5/3}$ in the $1\ell +{\rm jets}$ final state by CMS~\cite{Sirunyan:2018yun}~\footnote{See also Refs.~\cite{Dennis:2007tv,Contino:2008hi} for the phenomenological study of these channels.}. The single top partner production may play an important role in the relatively high top parter mass region as discussed in Sec.~\ref{sec:production}. Currently, the $tZ\to T/X_{2/3}$ channel has been searched by CMS at 35.9 fb$^{-1}$~\cite{Sirunyan:2017ynj}, and the $tW\to X_{5/3}$ channel has been searched by CMS at 35.9 fb$^{-1}$~\cite{Sirunyan:2018ncp} in $1\ell+{\rm jets}$ final state and by ATLAS at 36.1 fb$^{-1}$~\cite{Aaboud:2018xpj} in SSDL final state. However, the mass reaches of all those searches are still too low to be visible in our figures. Instead, in Fig.~\ref{ML4_results} we present the contours with constant number of events (= 20) in the $tW\rightarrow X_{5/3}\rightarrow \ell^\pm\ell^\pm +{\rm jets}$ final states as a projection for the future run of the LHC. The reach in model LP$_\4$ range from 1.5 TeV to 2 TeV at the 300 fb$^{-1}$ LHC and from 2.3 TeV to 3.1 TeV at the 3 ab$^{-1}$ HL-LHC which is better than the QCD pair searches (1.3 TeV at 300 fb$^{-1}$ and 2.0 TeV at 3 ab$^{-1}$).

\subsection{The results of RP$_\4$ and RF$_\4$}

\begin{figure}
\centering
\subfigure[~The results of RP$_\4$.]{
\label{RP4_results} %% label for first subfigure
\includegraphics[scale=0.4]{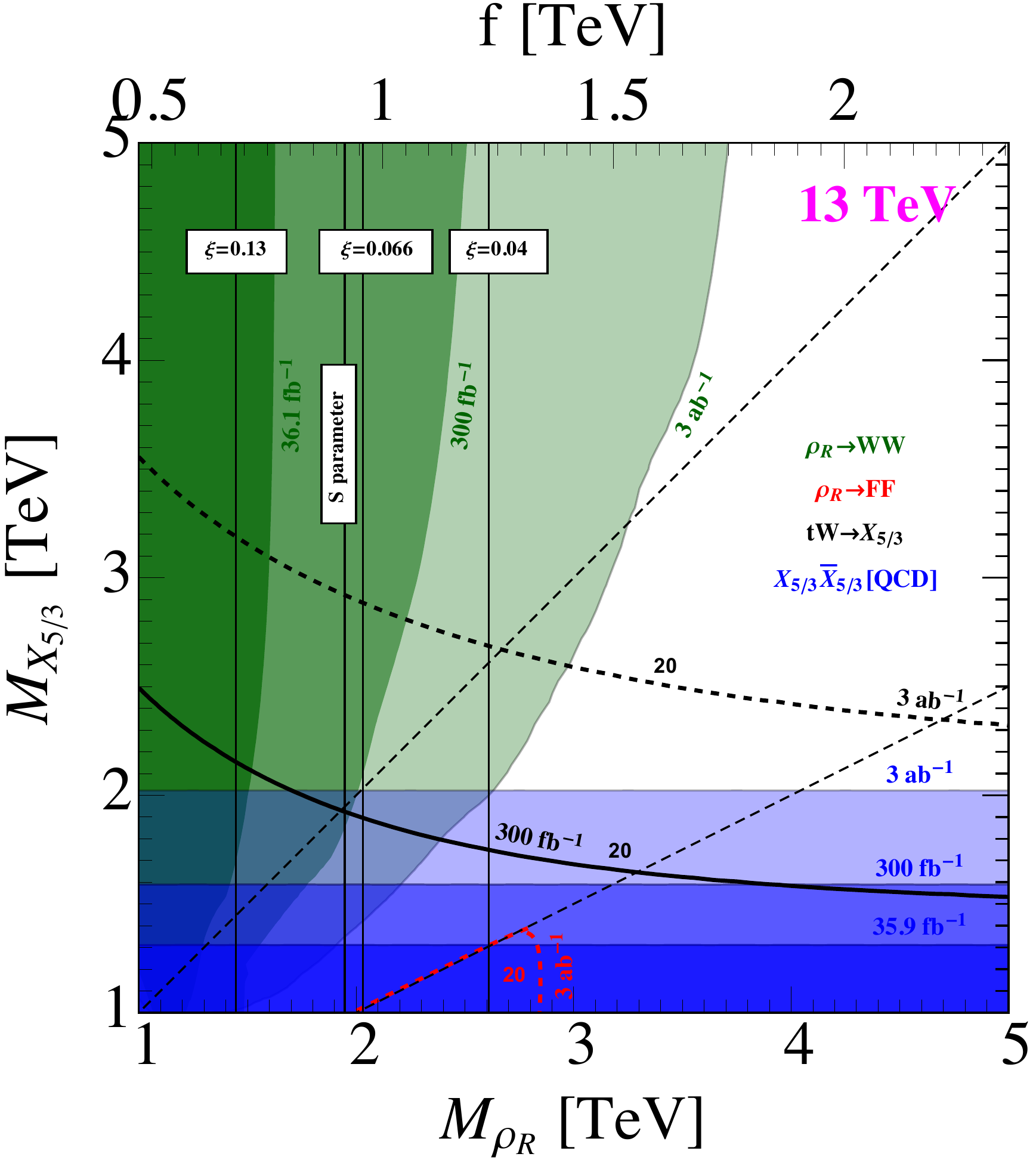}}\quad
\subfigure[~The results of RF$_\4$.]{
\label{RF4_results} %% label for first subfigure
\includegraphics[scale=0.4]{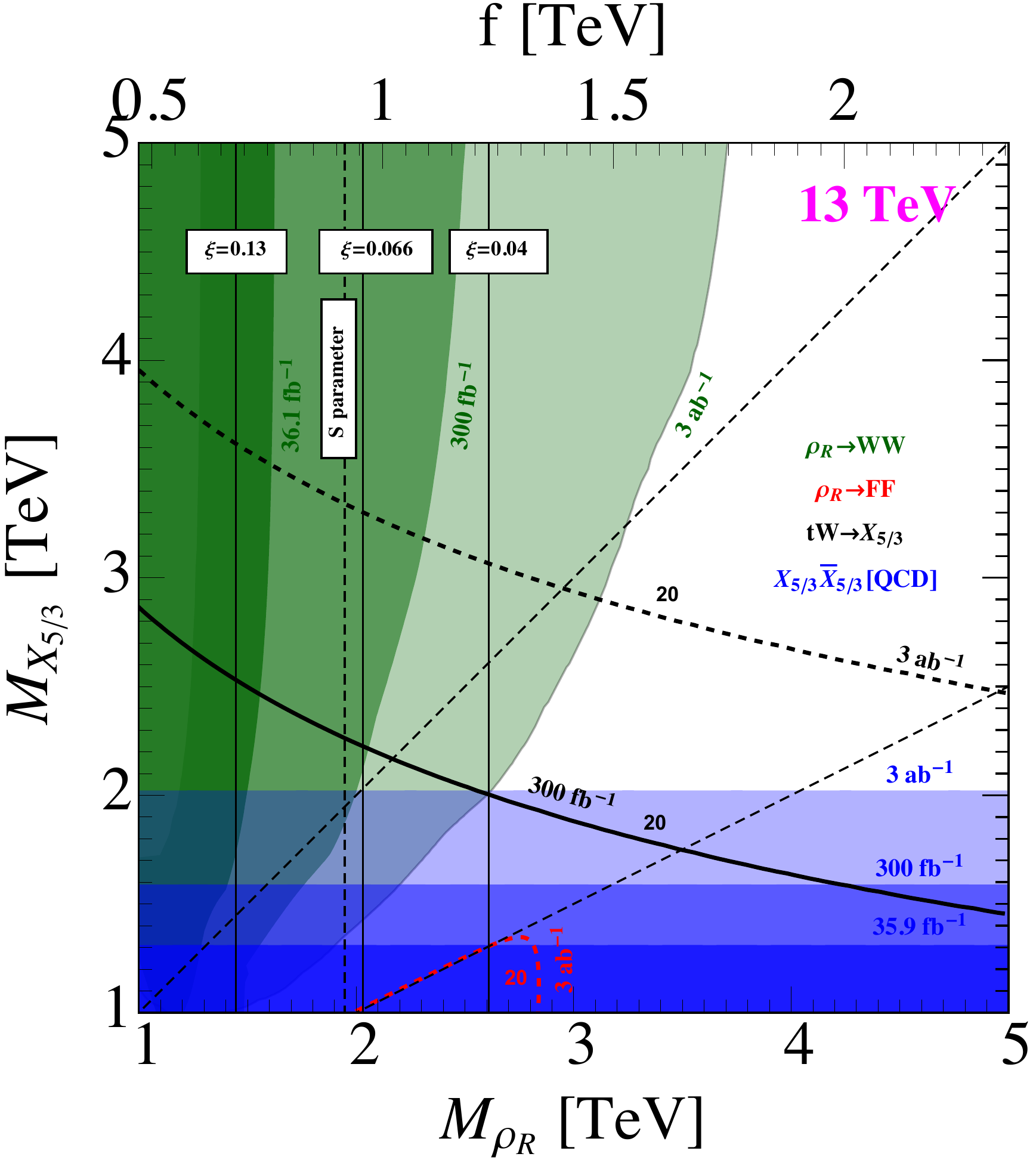}}
\caption{The current limits and prospective reach on model RP$_\4$ (left plot) and model RF$_\4$ (right plot). The parameters are chosen as \Eq{eq:benchrhoR}. The existing limits from Drell-Yan $\rho_R^0\to W^+W^-$~\cite{Aaboud:2017fgj} and $X_{5/3}\bar X_{5/3}$ [QCD]~\cite{Sirunyan:2018yun} are shown as the darkest shaded regions, while the projections for 300 (3000) fb$^{-1}$ are shown as lighter shaded regions. The event number contours for $N(\ell^\pm\ell^\pm+\text{jets})=20$ are drawn in solid (dashed) lines for 300 (3000) fb$^{-1}$, as a prospective limit for the $\rho_R^0\to X_{5/3}\bar X_{5/3}$ (denoted as $\rho_R\to FF$) and the $tW\to X_{5/3}$ channels. }
\label{MR4_results}
\end{figure}

We now turn to discuss the models RP(F)$_\4$. Similar to the cases of LP(F)$_\4$, we have set the following parameters as
\be
\label{eq:benchrhoR}
g_{\rho_R}=3,\quad a_{\rho_R}^2=\frac{1}{2},\quad y_L=1,\quad c_1=1,\quad c_2=1~\text{(for RF$_\4$ only)},
\ee
and scanned over $(M_{\rho_R},M_{X_{5/3}})$. The results are plotted in Fig.~\ref{MR4_results}. The meanings of the shaded regions and contour lines are similar to those in Fig.~\ref{ML4_results}. Note that we have started from $M_{\rho_R}$ from 1 TeV. Because the production cross sections of charged $\rho_R^\pm$ resonances are very small, we only use  the searches for the Drell-Yan production of $\rho_R^0$ at the LHC. Similar to the search for the $\rho_L$ resonances, the di-boson channel provides the strongest constraints in the region of $M_{\rho_R}<2M_\4$. Among the existing limits, we found that  the diboson  resonance searches  by ATLAS in the semi-leptonic channel~\cite{Aaboud:2017fgj} and in the fully hadronic channel in~\cite{ATLAS-CONF-2018-016} give the strongest constraints, and their results are  similar. Here we show the limits from results of \Ref{Aaboud:2017fgj}. As expected, due to the smallness of hypercharge gauge coupling,
the bound is weaker than the $\rho_L$ resonances. The present bound is around 1.6 TeV and will reach 3.8 TeV at the HL-LHC. 
 In the mass region of $M_{X_{5/3}}<M_{\rho_R}\leqslant 2M_{X_{5/3}}$,  the $\rho_R^0\to t\bar X_{2/3}\to t\bar tZ$ may be relevant, but the current search in Ref.~\cite{Sirunyan:2017ynj} is still not possible to put any relevant constraint in our parameter space. Thus, it is  not shown in the figure. In the mass region of  $M_{\rho_R}> 2M_{X_{5/3}}$, the cascade decay channel $\rho_R^0\to X_{5/3}\bar X_{5/3}$ in the SSDL final state is not comparable with the searches for the QCD pair $X_{5/3}$ production, due to the smallness of the production cross section.  We can also read from the figure that the electroweak precision $\hat S$-parameter measured by LEP~\cite{Tanabashi:2018oca} sets a strong constraint on the models with $\rho_R$, requiring $M_{\rho_R}\gtrsim1.95$ TeV, which is heavier than current experimental reach. However,  the reach of LHC with an integrated luminosity of  300 fb$^{-1}$ could surpass this constraint. The bounds for the top parters are the same as models LP(F)$_\4$ and not discussed here anymore.

\subsection{The results of XP$_\4$ and XF$_\4$}

\begin{figure}
\centering
\subfigure[~The results of XP$_\4$.]{
\label{XP4_results} %% label for first subfigure
\includegraphics[scale=0.4]{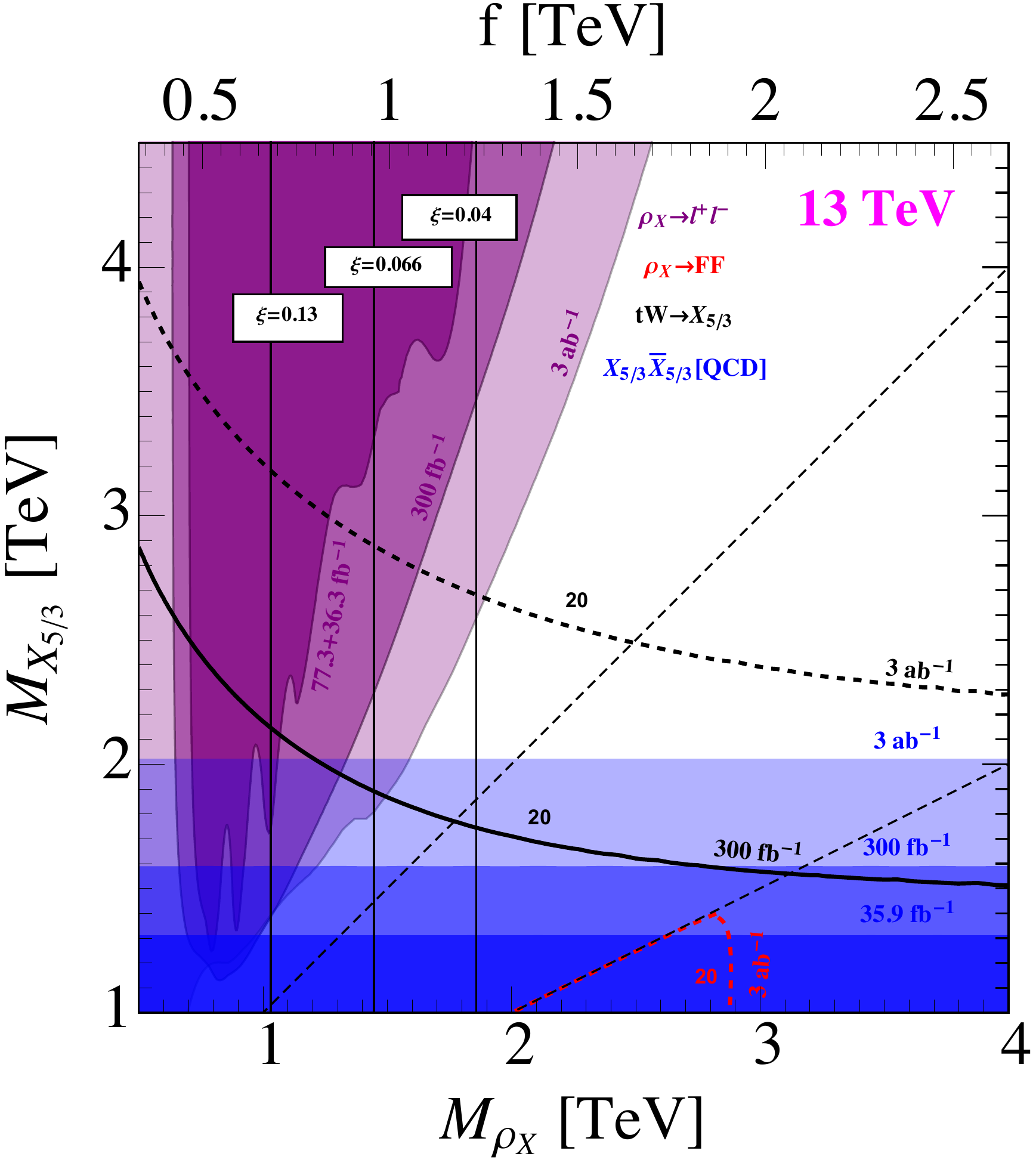}}
\subfigure[~The results of XF$_\4$.]{
\label{XF4_results} %% label for first subfigure
\includegraphics[scale=0.4]{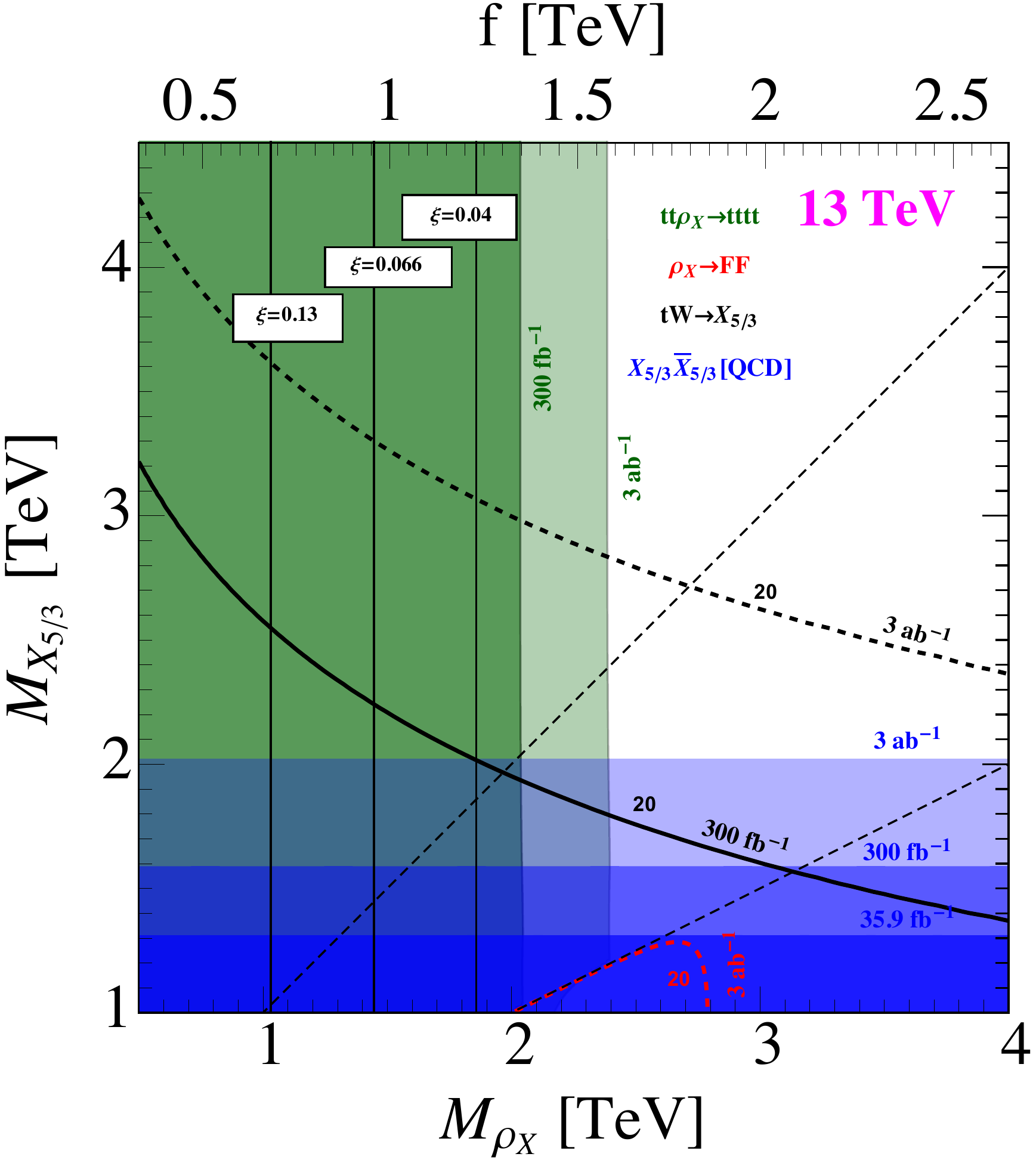}}
\caption{The current limits and prospective reach on model XP$_\4$ (left plot) and model XF$_\4$ (right plot). The parameters are chosen as \Eq{eq:benchrhox4}. The existing limits and projections from $X_{5/3}\bar X_{5/3}$ [QCD]~\cite{Sirunyan:2018yun} are plotted as shaded regions. The green regions  come from $t\bar t\rho_X^0$ associated production, by the phenomenological study of Ref.~\cite{Liu:2015hxi}. The purple regions represent the limit from the $\ell^+\ell^-$ search~\cite{CMS-PAS-EXO-18-006} and its extrapolations. The contours for $N(\ell^\pm\ell^\pm+\text{jets})=20$ are drawn with solid (dashed) lines for 300 (3000) fb$^{-1}$, as a prospective reach for the $\rho_X^0\to X_{5/3}\bar X_{5/3}$ (denoted as $\rho_X\to FF$) and the $tW\to X_{5/3}$ channels. See the text for more details.}
\label{MX4_results}
\end{figure}

We now turn to the models with a singlet vector resonance $\rho_{X}^0$. In this subsection we will discuss its interactions with the quartet top partner in models XP(F)$_\4$, while in the next subsection we will investigate its interactions with the singlet top partner XP(F)$_\1$.  As discussed in \Ref{Greco:2014aza}, $\rho_X$ only contributes to the $Y$-parameter of the electroweak precision test (see also \Eq{rho_X-Y}). Due to  the $(g'/g_{\rho_X})^2$ suppression, the indirect constraint on the $\rho_X$ is weak. As a result, $\rho_X$ could be very light especially in the case of large $g_{\rho_X}$. We choose the benchmark  values for the parameters as
\be
\label{eq:benchrhox4}
g_{\rho_X}=3,\quad a_{\rho_X}^2 = \frac14,\quad y_L=1,\quad c_1=1,\quad c_1'=c_2=1~\text{(for XF$_\4$ only)},
\ee
and scan over $(M_{\rho_X},M_{X_{5/3}})$ in Fig.~\ref{MX4_results}. Note that we have chosen a slightly smaller value of $a_{\rho_X}$ in order to relax the bound from $\xi$ measurement. Here we can see a difference between the partially composite $t_R^{\rm (P)}$ and the fully composite $t_R^{\rm (F)}$ scenario. While the di-lepton channel~\cite{CMS-PAS-EXO-18-006} can play an important role in model XP$_\4$ in the large $M_{X_{5/3}}$ region (i.e. $M_{X_{5/3}}>M_{\rho_X}$), it won't put any significant constraint on the model XF$_\4$. This is due to the fact that the branching ratio of di-lepton in the model  XP$_\4$ scales like $\left[g'/(g_{\rho_X}s_{\theta_L})\right]^4$, while in model XF$_\4$, it scales like $(g'/g_{\rho_X})^4$. As we fix $y_L$, larger value of $M_{X_{5/3}}$ will induce smaller value of $s_{\theta_L}$ and an enhancement of the di-lepton branching ratio in model XP$_\4$. Note that in the region $M_{\rho_X}\leqslant M_{X_{5/3}}$ where $\rho_X^0$ only decays to SM particles, the $t\bar t$ and $b\bar b$ channels dominate. The sensitivity in these  channels at the 13 TeV LHC is roughly three order of magnitude worse than the di-lepton channel, assuming the same branching ratios. Thus they can only play a role in the large $g_{\rho_X}$ region. However, large $g_{\rho_X}$ will lead to small  Drell-Yan production cross section  and make $t\bar{t}$, $b\bar{b}$ channels not relevant in our parameter space. In contrast, the authors of Ref.~\cite{Liu:2015hxi} have pointed out that the $pp\to t\bar t\rho_X^0\to t\bar tt\bar t$ channel  with the SSDL final states can probe the fully composite $t_R^{\rm (F)}$ scenario very well, as the production cross section scales like $g^2_{\rho_X}$. In Fig.~\ref{MX4_results}, we have reinterpreted the results of Ref.~\cite{Liu:2015hxi} in our parameter space  in model XF$_\4$.  We see that  $\rho_X^0$ with mass below 2 (2.4) TeV can be probed at 300 (3000) fb$^{-1}$  LHC  with our choice of $g_{\rho_X}=3$ in model  XF$_\4$. While for model XP$_\4$, the bound (not shown in the figure) is weaker ($\sim 1.0$ TeV at 3 ab$^{-1}$) due to the suppression of $\rho_X t\bar{t}$ couplings either by the $t_L-T_L$ mixing or the $B_\mu -\rho_{X\mu}$ mixing. We can also see that the limits from $t\bar{t}\rho_X$ channel become stronger in the low $M_{X_{5/3}}$ region in model XP$_\4$, as the left-handed top quark mixing angle $s_{\theta_L}$ becomes large. We also noticed that the cascade decays to top partner can barely play an important role, as the cross section of $\rho_X^0$ is small.  The bounds on the quartet top partners are the same as models LP(F)$_\4$.

\subsection{The results of XP$_\1$ and XF$_\1$}
\label{sec:xpf1}

Finally we come to the models containing a singlet top partner,  XP$_\1$ and XF$_\1$. While the scanning over $(M_{\rho_X},M_{\widetilde T})$, the other parameters are chosen as 
\be
\label{eq:benchrhox1}
g_{\rho_X}=3,\quad a_{\rho_X}^2=\frac{1}{4},\quad c_1=1,\quad y_L = 1.5~(1.0)~\text{for XP$_\1$ (XF$_\1$)}, \quad c_1'=c_1''=1~(\text{XF$_\1$)}.
\ee
where we have chosen a slightly larger value of $y_L$ in model XP$_\1$ in order to reproduce the observed value of top quark mass.  Note that in  model XP$_\1$, the top quark mass is approximately given by \Eq{eq:mtp1}
$$
M_t =\frac{y_L v s_{\theta_R}}{\sqrt{2}},
$$
and the choice for $y_L$ in \Eq{eq:benchrhox1} has fixed $s_{\theta_R} \sim 0.6$. This means that the couplings of the interactions $\rho_X \bar{t}_R t_R$, $\rho_X \bar{t}_R \widetilde{T}_R$ are roughly constants with varying  mass of the top partner (see Table~\ref{tab:rhox1}).
In both models, the Drell-Yan production of the $\rho_X$ can't play an important role in our interested parameter space, because of the lack of the sensitivity to the dominant decay channel $t\bar{t}$ and the suppression of the decay branching ratio into the di-lepton final state. 
\begin{figure}
\centering
\subfigure[~The results of XP$_\1$.]{
\label{XP1_results} %% label for first subfigure
\includegraphics[scale=0.4]{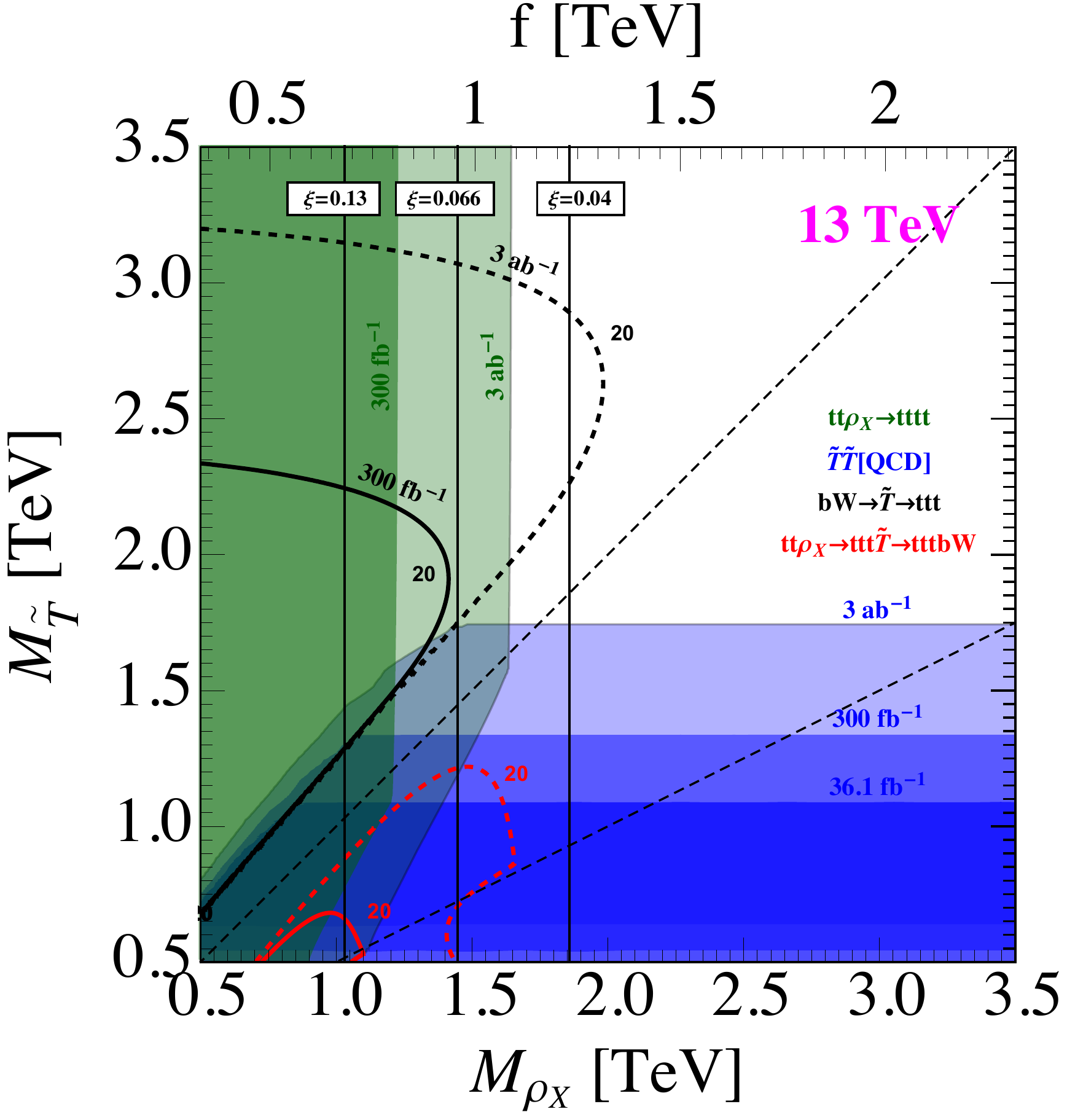}}\quad
\subfigure[~The results of XF$_\1$.]{
\label{XF1_results} %% label for first subfigure
\includegraphics[scale=0.4]{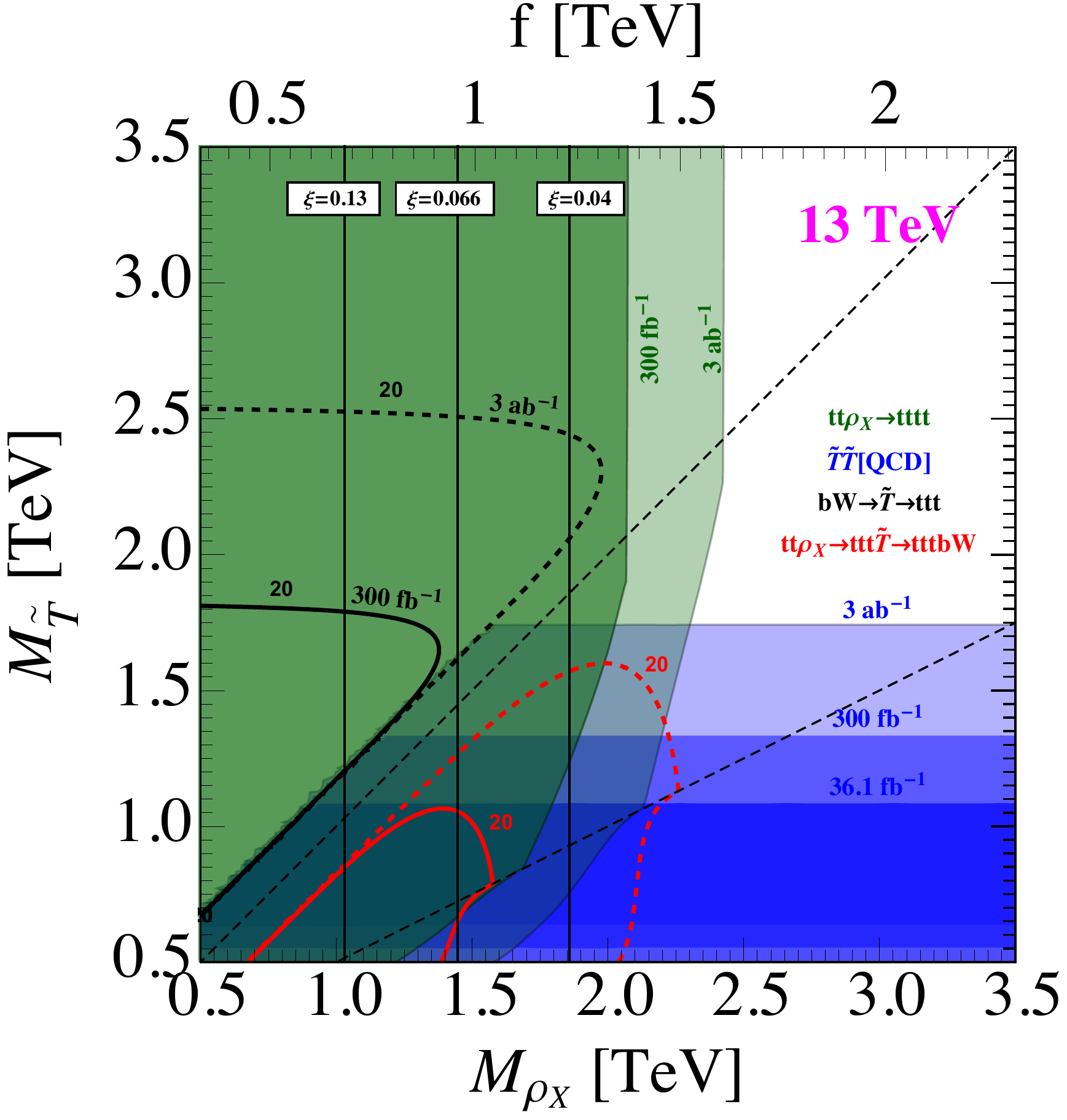}}
\caption{The current limits and prospective reach on model XP$_\1$ (left plot) and model XF$_\1$ (right plot). The parameters are chosen as \Eq{eq:benchrhox1}. The existing bound and the projections are shown as shaded regions, using $\widetilde T\overline{\widetilde T}$ [QCD] in $bW^+\bar bW^-$ channel~\cite{Aaboud:2017zfn}. The dark green regions come from $t\bar t\rho_X^0$ associated production, based on the phenomenological study of Ref.~\cite{Liu:2015hxi}. The contours for $N(\ell^\pm\ell^\pm+\text{jets})=20$ are drawn in solid (dashed) lines for 300 (3000) fb$^{-1}$, as a prospective limit for the $bW\to\widetilde T\to t\rho_X^0(t \bar t)$ channel (in black) and $t\bar{t}\rho_X \rightarrow t\bar{t} t\widetilde{T}(bW)$ channel (in red). See  the main text for more details.}
\label{MX1_results}
\end{figure}
 In Fig.~\ref{MX1_results}, we have shown the reach from the $t\bar{t} \rho_X$ production with the SSDL channel,  including the analysis of \Ref{Liu:2015hxi} in the four top final state and the 
 cascade decay of $\rho_X$ into $\bar{t}\widetilde{T}(bW)$. We see that the SSDL in the four top final state at the 3 ab$^{-1}$ HL-LHC can probe the $\rho_X$ up to 1.6 TeV in model XP$_\1$ and up to 2.4 TeV in model XF$_\1$. The cascade decay channel of $\rho_X\rightarrow t\widetilde{T}$ plays a more important role in model XF$_\1$ than in model XP$_\1$, due to the strong interaction in the fully composite $t_R^{\rm (F)}$ scenario ($c_1^{\prime \prime}$ term in \Eq{eq:rhox1}). 
 For the top partner, we present the current limits and prospective reaches coming from the ATLAS searches for the QCD pair production of the top partner with the $bW^+\bar bW^-( 1\ell +{\rm jets})$ final states~\cite{Aaboud:2017zfn}. Note that the single top partner searche performed by ATLAS in \Ref{ATLAS-CONF-2016-072} with integrated luminosity $L = 3.2$ fb$^{-1}$ using the $bW(\to\ell \nu)$  decay channel is not sensitive to our parameter space yet~\footnote{For the theoretical studies of $bW\to \widetilde T\to bW/tZ/th$ channels, see Refs.~\cite{Ortiz:2014iza,Liu:2017sdg,Liu:2016jho,Backovic:2015bca,Backovic:2015lfa,Reuter:2014iya,Li:2013xba}.}.
 Instead, we find that the cascade decay of the  top partner $\widetilde{T}$ into $\rho_X t$  with $\rho_X$ decaying into top pair in the single production channel can become relevant in the mass region of $M_{\widetilde T} > M_{\rho_X}$. For example, for $M_{\widetilde T}=2$ TeV and $M_{\rho_X}=1$ TeV, the branching ratio can reach 65.8\% (93.8\%) for XP(F)$_\1$ in our parameter choice, due to the large coupling of $\rho_X t_R \widetilde{T}_R$ in both models.
 Moreover,  it will lead to the SSDL signature. In Fig.~\ref{MX1_results}, we have estimated the reach of this channel  with SSDL searches at the LHC with integrated luminosities  $300~{\rm fb}^{-1}$ and $3~{\rm ab}^{-1}$. This channel is very promising,  and can become comparable with the four top final states in both models, especially in XP$_\1$.  This is due to the fact that in model XP$_\1$,  the branching ratio of this cascade decay channel is further enhanced by the  $s^2_{\theta_R}$ suppression of $\bar{t}_R^{\rm (P)} t_R^{\rm (P)} \rho_X^0$ coupling, as can be seen from Table~\ref{tab:rhox1}.

\subsection{Summary}

In summary, focusing on the coupling regime $g_\rho \sim 3$, we  have investigated the present limits and prospective reaches  in the $M_{\rho}-M_{X_{5/3}}$ space for models LP(F)$_\4$, RP(F)$_\4$, XP(F)$_\4$, and  in the $M_{\rho}-M_{\widetilde T}$ space in models XP(F)$_\1$.  For the spin-1 resonances in non-trivial representation of $SO(4)$, such as the $\rho_L(\3,\1)$ of LP(F)$_\4$ and the $\rho_R(\1,\3)$ of RP(F)$_\4$, the Drell-Yan production followed by decaying into the di-boson final state in the fully hadronic channel provide the best probe in the $M_\rho\leqslant M_\4$ region, where the spin-1 resonances can only decay to pure SM final states. For LP(F)$_\4$, the  mass region of $M_\rho > M_\4$  can also be explored by Drell-Yan production followed by decaying into the heavy-light final state $t\bar B/X_{5/3}\bar t$ and the pure strong dynamics final state $X_{5/3}\bar X_{2/3}/X_{5/3}\bar X_{5/3}$ in the SSDL channel. For the $SO(4)$ singlet resonance $\rho_X(\1,\1)$, the sensitivity to the dominant $t\bar t$ final state from Drell-Yan production is still limited by the experimental uncertainty. Instead, the $\ell^+\ell^-$ channel is useful for XP$_\4$, while the $t\bar t\rho_X^0$ associated production is useful for XF$_\4$ and XF(P)$_\1$, as the cross section scales like $g_{\rho_X}^2(g_{\rho_X}^2 s_{\theta_R}^4)$  and it  can lead to four top final states with SSDL signature.  We  have recasted the analysis of \Ref{Liu:2015hxi} in this SSDL channels  in our parameter space. The cascade decaying channels (heavy-light and heavy-heavy) in models XP(F)$_\4$ can rarely play an important role because the cross section is small in the high mass region, and the very light top partners have already been excluded by the present experiments. In models XP(F)$_\1$, we find that the  SSDL final states from the $bW\to \widetilde T\to t\rho_X^0$ process can be very important in the $ M_{\rho_X} < M_{\widetilde{T}}  $ region, while the SSDL channel of $t\bar{t}\rho_X\to t\bar{t}\bar t \widetilde T$ can be relevant in intermediate mass region. Finally, the QCD pair production of top partners offers a robust probe for the models. At the same time, the singly produced channels have a much higher mass reach. For example, for the models  with quartet top partners, the QCD pair channel and $tW\to X_{5/3}$ channel could probe the parameter $M_{X_{5/3}}$ up to $\sim2$ TeV  and $\sim2.5-4$ TeV (depends on the $f$ parameter) at the HL-LHC, respectively. The limits and reaches of the mass scale from present and future searches at he LHC  are summarized in Fig.~\ref{fig:rhoLR} (for models LP(F)$_\4$ and RP(F)$_\4$) and  in Fig.~\ref{fig:rhoX}  (for models XP(F)$_\4$ and XP(F)$_\1$).

\subsection{Future colliders}

\begin{figure}[!t]
\centering
\subfigure[~The results of LF$_\4$.]{
\label{LF4_results_27TeV} %% label for first subfigure
\includegraphics[scale=0.4]{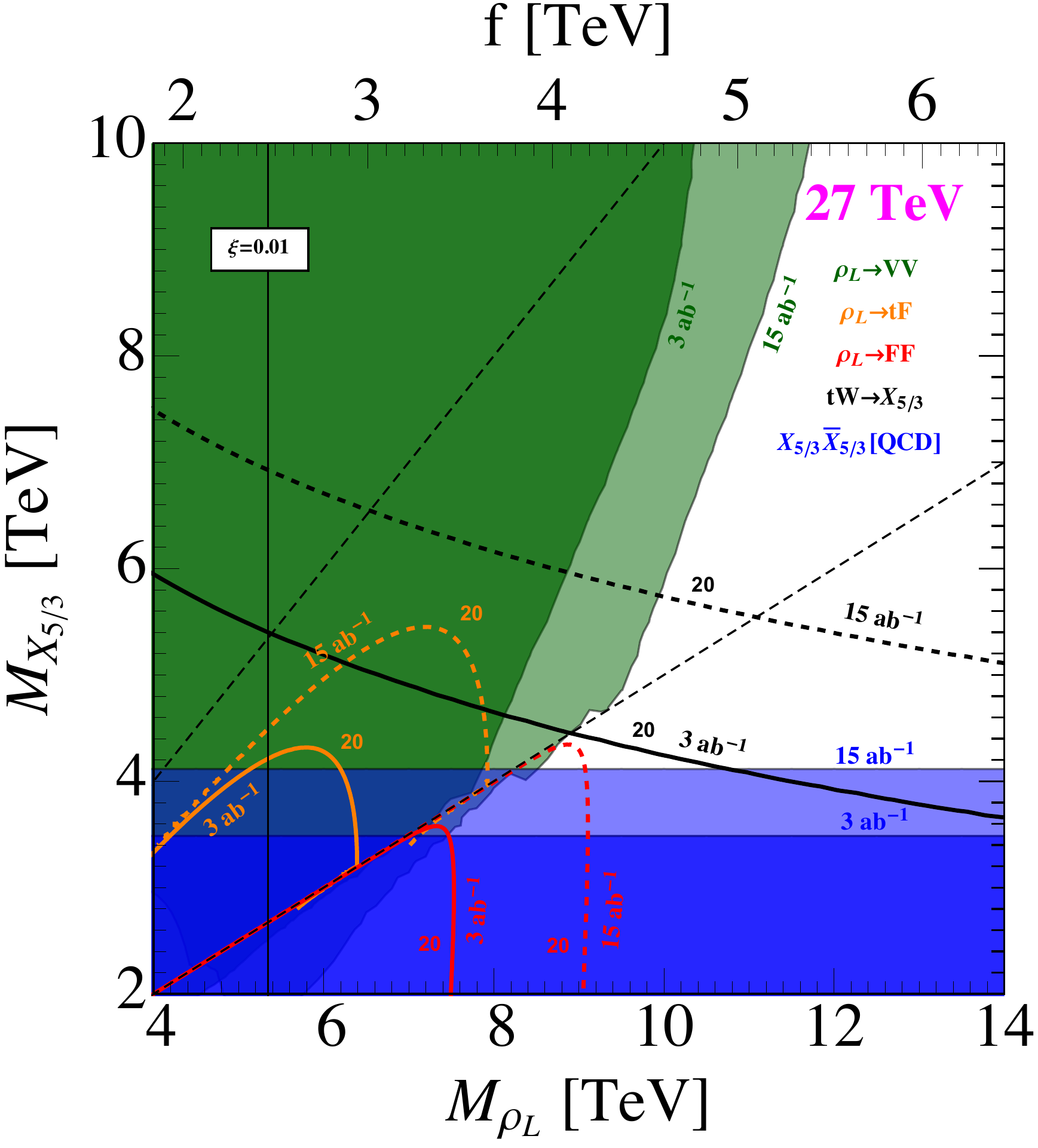}}
\subfigure[~The results of LF$_\4$.]{
\label{LF4_results_100TeV} %% label for first subfigure
\includegraphics[scale=0.4]{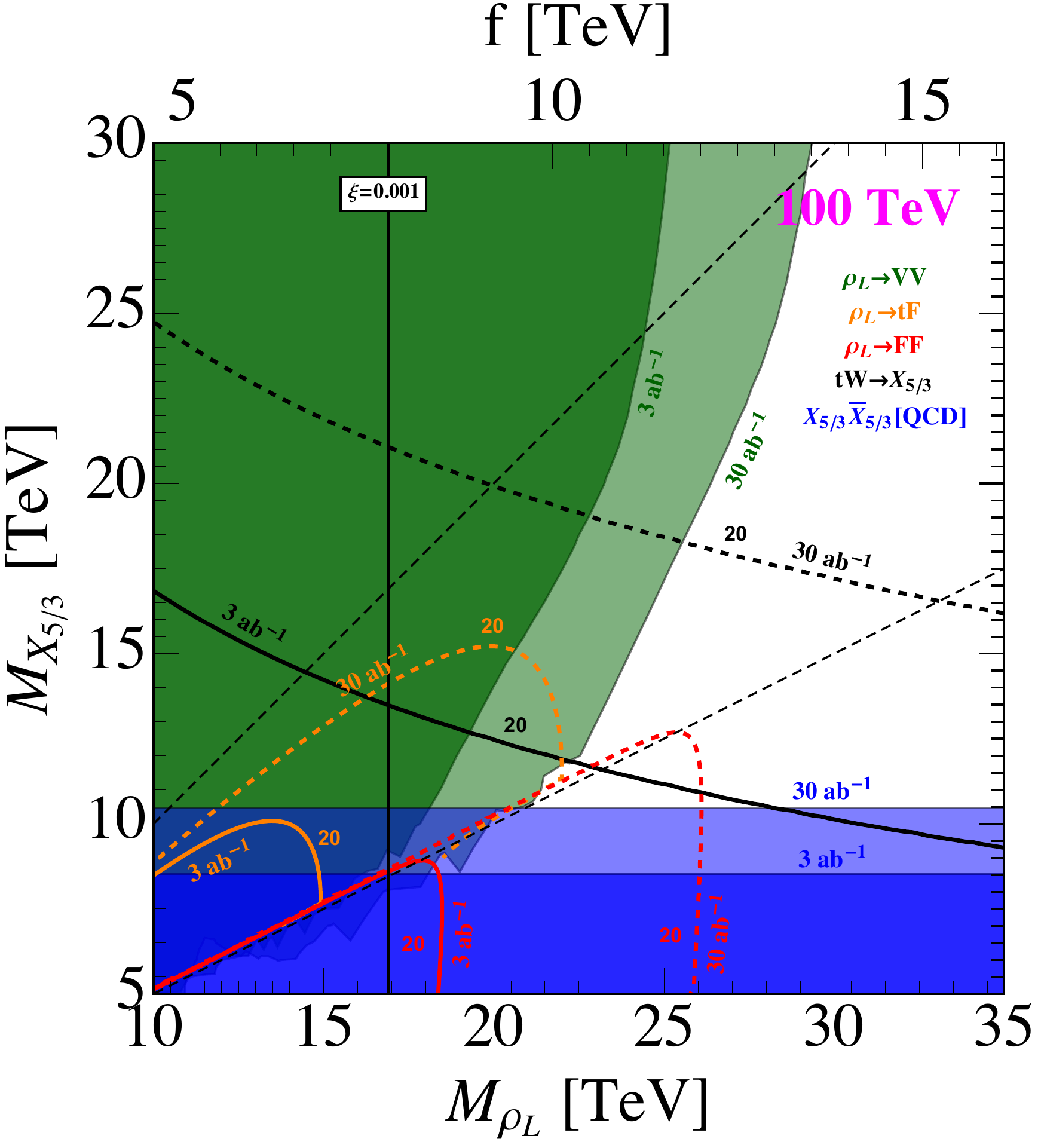}}
\caption{Left: the prospective reaches at the 27 TeV HE-LHC with integrated luminosities 3 ab$^{-1}$ and 15 ab$^{-1}$; Right: the prospective reaches at the 100 TeV $pp$ collider with integrated luminosities 3 ab$^{-1}$ and 30 ab$^{-1}$. The parameters and conventions are the same as Fig.~\ref{ML4_results}.}
\label{fig:HE}
\end{figure}

Before we conclude our study, we make some estimates of the prospective reaches on the mass scales in our models at the 27 TeV HE-LHC and 100 TeV $pp$ collider. In Fig.~\ref{fig:HE}, we  have used the method described in Appendix~\ref{app:method} to extrapolate, based one the  di-boson boosted-jet resonance searches  at ATLAS~\cite{ATLAS-CONF-2018-016} and the pair top partner searches in the $1\ell+{\rm jets}$ channel at CMS~\cite{Sirunyan:2018yun} in model LF$_\4$. We present the results with the integrated luminosities of 3 ab$^{-1}$ and 15 ab$^{-1}$  for the HE-LHC. For the 100 TeV collider, we show the results with 3 ab$^{-1}$ and 30 ab$^{-1}$ integrated luminosities. Compared with HL-LHC, we approximately gain a factor of 2 for the reach of the mass scales at the 15 ab$^{-1}$ HE-LHC and  a factor  of  5 at the 30 ab$^{-1}$ 100 TeV collider.  The SSDL channels (including $tW\rightarrow X_{5/3}$, $\rho \rightarrow tF$ and $\rho\rightarrow FF$) have slightly better reach at the 100 TeV collider with  a factor of 5.5 gained with an integrated luminosity of 30 ab$^{-1}$.

\section{Conclusion}
\label{sec:conclusion}

In this paper, we have studied the phenomenology of the vector resonances and the fermionic resonances in several classes of benchmark simplified models in the minimal coset $SO(5)/SO(4)$, with some emphasis on the importance of the interplay of the phenomenology of the composite resonances. We have considered three irreducible representations under the unbroken $SO(4)$ for the spin-1 resonances: $\rho_L(\3,\1)$, $\rho_R(\1,\3)$, $\rho_X(\1,\1)$ and two irreducible representations for the spin-1/2 resonances: $\Psi_\4(\2,\2)$, $\Psi_\1(\1,\1)$. In addition, we have also studied the two scenarios depending on whether   the  right-handed top quark is elementary or fully composite.
\begin{figure}[!t]
\centering
\includegraphics[scale=0.65]{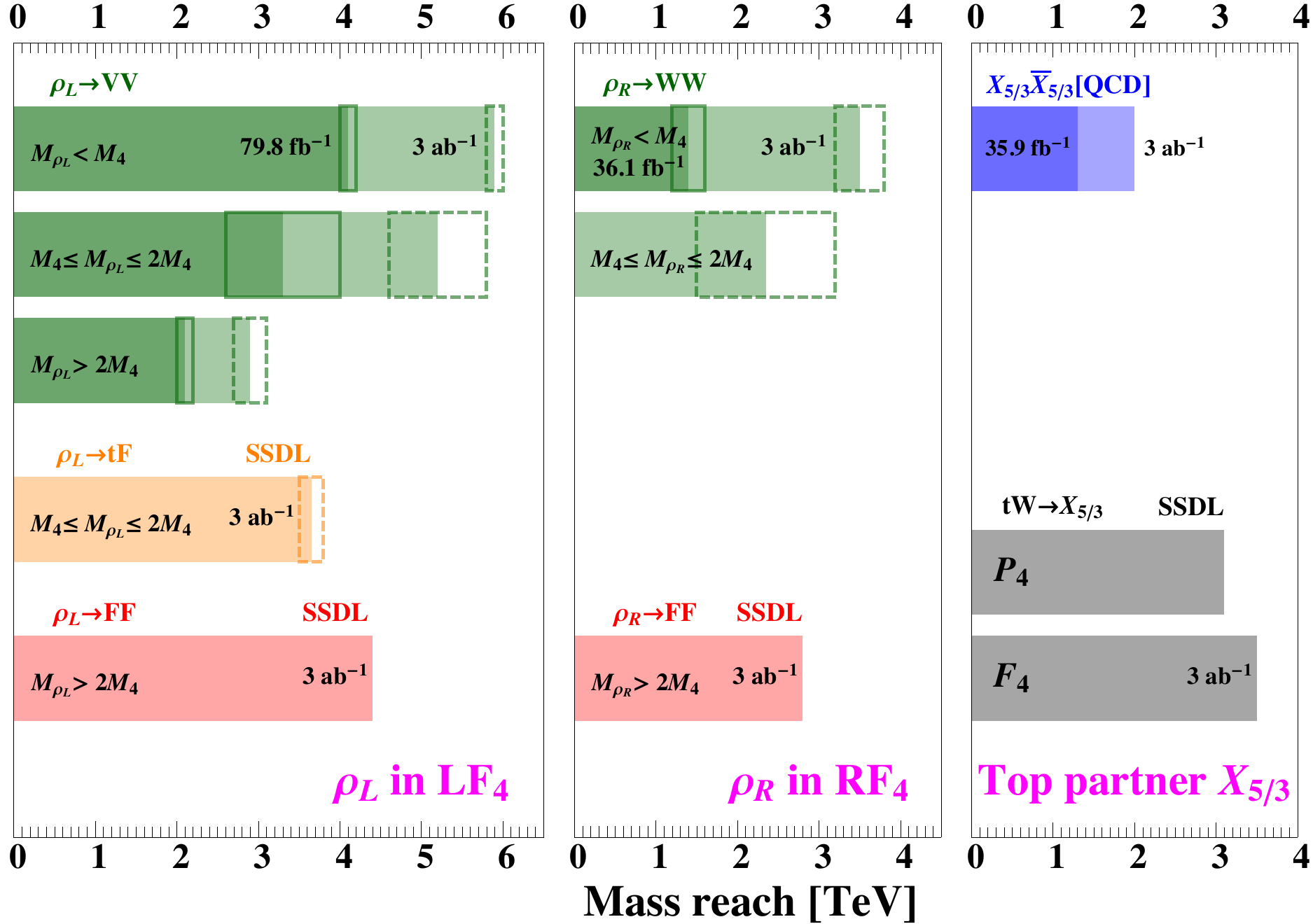}
\caption{Summary of the present limits and prospective reach on the mass scales in models LF$_\4$ and RF$_\4$ for the benchmark parameters in \Eq{eq:benchrhoL} and \Eq{eq:benchrhoR}. The bounds in models with partially composite $t_R$ are similar and not shown here except that  for the  single production of $X_{5/3}$, we have shown the bounds in both P$_\4$ and F$_\4$ scenarios. The bounds on $M_{\rho}$ are shown in three kinematical region ($M_\rho <  M_\4$, $M_\4 <M_\rho < 2 M_\4$, $M_\rho >  2 M_\4$). The rectangles for the existing searches $\rho_L \rightarrow VV$, $\rho_R \rightarrow WW$ indicate the ranges of the bound when varying the parameter $M_\4$. The bounds from single production of $X_{5/3}$ are obtained by choosing $\xi=0.1$.}
\label{fig:rhoLR}
\end{figure}

\begin{figure}[!t]
\centering
\includegraphics[scale=0.55]{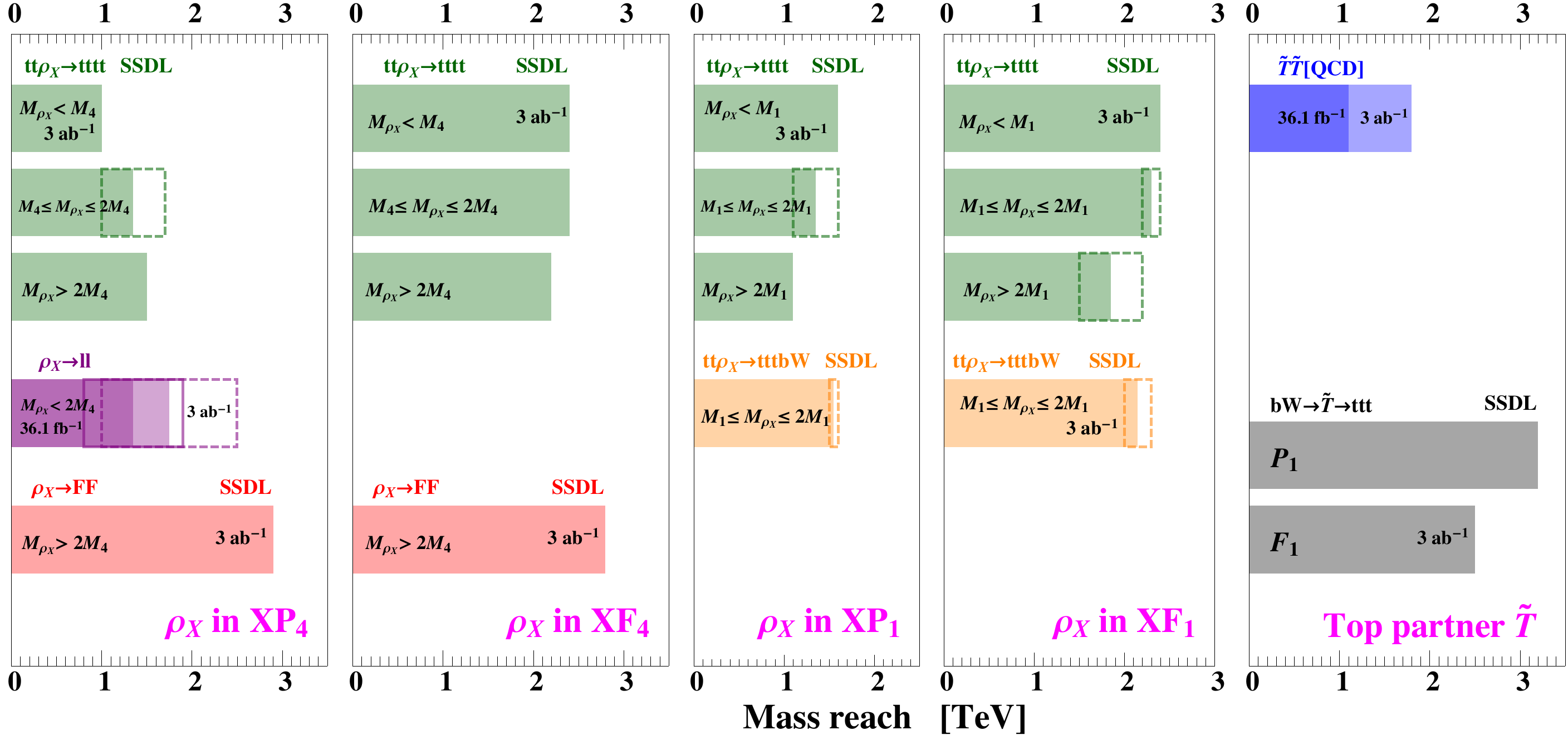}
\caption{Summary of the present limits and prospective reach on the mass scales in models involving the singlet spin-1 resonance XP(F)$_{\4,\1}$ for the benchmark parameters in \Eq{eq:benchrhox4} and \Eq{eq:benchrhox1}. The bounds on $M_{\rho}$ are shown in three kinematical region ($M_\rho <  M_\4$, $M_\4 <M_\rho < 2 M_\4$, $M_\rho >  2 M_\4$). The rectangles  indicate the ranges of the bound when varying the parameter $M_{\4,\1}$. The bounds from single production of $\tilde{T}$ are obtained by choosing $\xi=0.1$.}
\label{fig:rhoX}
\end{figure}

 We have categorized the couplings of the composite resonances into four classes according to their expected sizes,  $\mO(g_\rho)$, $\mO(g_\rho s_{\theta_L}$, $g_{\rho}s_{\theta_R})$, $ \mO(g_{\rm SM})$, and $\mO(g_{\rm SM}^2/g_\rho)$, where $s_{\theta_{L,R}}$ are the elementary-composite mixing angles $s_{\theta_L}, s_{\theta_R}$, and $g_{\rm SM}$ is of the size of  the Standard Model gauge and Yukawa couplings. The results are summarized in Table~\ref{tab:charge}, Table~\ref{tab:neutral} and Table~\ref{tab:rhox1}. Based on the discussion of the couplings, we have studied different production and decay channels for the composite resonances, paying special attention to the relevance of the cascade decay channels between the composite resonances. We have shown the present and future prospective bounds on our parameter space in the $M_{\rho} - M_{\Psi}$ plane in different models, focusing on the moderate large  coupling $g_\rho = 3$.  We found that the cascade decay channels into one top partner and one top quark $t\Psi$ or two top partners  $\Psi\Psi$ strongly affect the phenomenology of the $\rho$ if they are kinematically open. Their presence significantly weakens the reach of the channels with only SM particles, such as the di-boson channel. 
In addition, the decay channels $\rho^+_L \rightarrow t \bar{B}/X_{5/3}\bar{t}$ and $\rho^+_L \rightarrow X_{5/3}\bar{X}_{2/3}$, $ \rho^0_{L,R,X} \rightarrow X_{5/3}\bar{X}_{5/3}$  can lead to the SSDL final states, which are used as an estimate of the reach on the $M_{\rho} - M_\Psi$ plane. We found that they are comparable in some regions of the parameter space to the di-boson searches or the top partner searches at the HL-LHC, especially for the $\rho_L$ models LP(F)$_\4$. For the $\rho_{R,X}$ models RP(F)$_\4$, XP(F)$_\4$, because the Drell-Yan production is suppressed by the smallness of the hypercharge gauge coupling, the cascade decay channels play less important roles. We also find that the SSDL channels in the single production of the charge-$5/3$ top partner $X_{5/3}$ can always play an important role in our parameter spaces. In the models involving the singlet spin-1 resonance XP(F)$_\4$ and XP(F)$_\1$, the associated production of top pair and the $\rho_X$ with the four top final states can play an important role, as the coupling between $\rho_X$ and $\bar{t}t$ is of $\mO(g_{\rho})$ for the fully composite $t_R^{\rm (F)}$ models and $\mO(g_\rho s^2_{\theta_L})$ or $\mO(g_\rho s^2_{\theta_R})$ for the partially composite $t_R^{\rm (P)}$ models.  
We have recast the analysis in the SSDL channel by  Ref.~\cite{Liu:2015hxi} in our parameter space. In models XP(F)$_\1$, the single production of the top partner $\widetilde T$, followed by cascade decaying into $t\rho_X( t\bar{t})$ can be important  in the region $M_{\widetilde T} > M_{\rho_X} $, and we have explored its sensitivity in the SSDL channel. It can be better than the $t\bar{t}\rho_X(t\bar{t})$ SSDL channel in model XP$_\1$. In the mass region $M_{\widetilde T} < M_{\rho_X} < 2 M_{\widetilde T} $,  the $t\bar{t}$ fusion production of $\rho_X$, which decays into $t\overline{\widetilde{T}}$, can lead to the $t\bar{t}tbW^+$ final state with SSDL signature. We have used this to explore its sensitivity. In Fig.~\ref{fig:rhoLR} and Fig.~\ref{fig:rhoX}, we have summarized the prospective reach on the mass scale $M_{\rho}$ and $M_{\Psi}$ by the different existing searches at the LHC and by various SSDL channels from the cascade decays.

Several directions should be explored further. Among the various cascade decay channels, we have only considered the SSDL final state. The reach obtained this way is conservative.  Other decay final states, such as $1\ell{\rm +jets}$, should also be studied in detail. The final kinematical variables are usually very complicated, and new techniques such as machine learning may be useful to enhance the sensitiy. We hope to address the issues in a future work.

\section{Acknowledgement}

We would like to thank  Andrea Tesi for the  collaboration in the early stage of this work. KPX thanks Andrea Wulzer for useful discussions. KPX would like to thank the hospitality of the University of Chicago where part of this work was performed. LTW is supported by the DOE grant DE-SC0013642. DL is supported in part by the U.S. Department of Energy under Contract No. DE-AC02-06CH11357. KPX is supported in part by the National Science Foundation of China under Grant Nos. 11275009, 11675002, 11635001 and 11725520, and the National Research Foundation of Korea under grant 2017R1D1A1B03030820.

\appendix

%%%%%%%%%%%%%%%%%%%%%%%%%%%%%%%%%%%%%%%%%%%
\section{ CCWZ for SO(5)/SO(4) and its matching to BSM EFT}
\label{app:CCWZ}
%%%%%%%%%%%%%%%%%%%%%%%%%%%%%%%%%%%%%%%%%%%

\subsection{The CCWZ operators}

We first present the $SO(5)$ generators as follows~\cite{Panico:2015jxa}:
\be\label{app:SO(5)}\begin{split}
T^{\hat{i}}_{IJ} =&- \frac{i}{\sqrt{2}}\left(\delta^{\hat{i} I} \delta^{5J} - \delta^{\hat{i}J} \delta^{5I}\right), \\ 
T^{a_{L/R}}_{IJ} =&-\frac{i}{2} \left( \frac12 \epsilon^{abc}(\delta^{bI}\delta^{cJ} -\delta^{bJ}\delta^{cI} )\pm (\delta^{aI}\delta^{4J} - \delta^{aJ} \delta^{4I})\right),
\end{split}\ee
where $\hat{i} = 1,\cdots, 4$, while $a, b,c = 1,2,3 $ and $I,J = 1, \cdots, 5$. Here $T^{a_L}$ and $T^{a_R}$ correspond to the unbroken $SO(4)\simeq SU(2)_L \times SU(2)_R$ generators and they are in the form of 
\beq\label{eq:44}
T^{a_{L/R}} = \begin{pmatrix}
t^{a_{L/R}} & 0 \\
0  & 0 \end{pmatrix},
\eeq
with $t^{a_{L/R}}$ is the $4\times 4 $ matrix, which will be useful in the following discussion. 

The standard Callan-Coleman-Wess-Zumino (CCWZ) framwork~\cite{Coleman:1969sm,Callan:1969sn} is used to describe the general interactions in our models. The Goldstone quartet $\vec{h} =(h_1,h_2,h_3,h_4)^T$ lives in the coset space $SO(5)/SO(4)$. The Goldstone matrix is defined as
\be
U[\vec{h}]=e^{i\frac{\sqrt{2}}{f}h_iT^{\hat i}}=\begin{pmatrix}\delta_{ij}-(1-\cos\frac{h}{f})\frac{h_ih_j}{h^2}&\frac{h_i}{f}\sin\frac{h}{f}\\
-\frac{h_j}{f}\sin\frac{h}{f}&\cos\frac{h}{f}\end{pmatrix}.
\ee
Under the non-linearized $\mG \in SO(5)$, it transforms as $U[\vec{h}]\to\mathcal{G}U[\vec{h}]\mathcal{H}^{-1}[\vec{h}; \mG]$, where  $\mathcal{H} \in SO(4)$ is  a function of the Goldstone fields and the global group element $\mG$. We use the Maurer-Cartan form to define the covariant objects $d_\mu$ and $e_\mu$  as follows:
\be
U^\dagger (A_\mu+i\partial_\mu)U=d_\mu^iT^{\hat i}+e_\mu^aT^a,
\ee
where $A_\mu\equiv A_\mu^aT^a$ are the gauge fields corresponding to the unbroken generators. The $d_\mu$ and $e_\mu$ objects will transform under the non-linearized $SO(5)$ group as:
\be\label{eq:detrans}
d_\mu\to\mathcal{H}d_\mu\mathcal{H}^{-1},\quad e_\mu\to\mathcal{H}(e_\mu+i\partial_\mu)\mathcal{H}^{-1}.
\ee
In MCHMs, only the subgroup $SU(2)_L\times U(1)_Y\in SO(4)\times U(1)_X$ is gauged, i.e.
\be
A_\mu^{a_L}=g_2W_\mu^{a_L},\quad A_\mu^{1_R,2_R}=0,\quad A_\mu^{3_R}=g_1B_\mu,\quad X_\mu=g_1B_\mu.
\ee
The last gauge field $X_\mu$, corresponding to the $U(1)_X$ group, is introduced to give correct hypercharge for the fermions, and the Goldstone bosons are neutral under this symmetry.

The full formulae of $d_\mu$ and $e_\mu$ symbols can be obtained as follows~\cite{Panico:2015jxa}
\be\label{d_and_e_decompose}\begin{split}
d_\mu^i=&\sqrt{2}\left(\frac{1}{|\vec{h}|}\sin\frac{|\vec{h}|}{f }-\frac{1}{f }\right)\frac{\vec{h}^TD_\mu\vec{h}}{|\vec{h}|^2}\vec{h}_i-\frac{\sqrt{2}}{|\vec{h}|}\sin\frac{|\vec{h}|}{f }D_\mu\vec{h}_i,\\
e_\mu^{a_L}=&g_2 W^{a_L}_\mu-\frac{4}{|\vec{h}|^2}\sin^2\frac{|\vec{h}|}{2f }\vec{h}^Ti t^{a_L}D_\mu\vec{h},\\
e_\mu^{a_R}=& \delta^{a 3}g_1 B_\mu-\frac{4}{|\vec{h}|^2}\sin^2\frac{|\vec{h}|}{2f }\vec{h}^Ti t^{a_R}D_\mu\vec{h},
\end{split}\ee
where the covariant derivative is given by:
\be
D_\mu\vec{h}=(\partial_\mu-ig_2t^{a_L}W_\mu^{a_L}-ig_1t^{3_R}B_\mu)\vec{h},
\ee
and the matrices $t^{a_{L/R}}$ are defined in Eq.~(\ref{eq:44}). Because of \Eq{eq:detrans}, the leading Lagrangian of the Goldstone fields is simply
\be\label{d2}
\mL^{d_2}=\frac{f^2}{4}d_\mu^id^{i\mu}.
\ee
For the fermionic heavy resonances, they fall into the irreducible representations of  the unbroken group $SO(4)\times U(1)_X \simeq SU(2)_L \times SU(2)_R\times U(1)_X$. We will consider two irreducible representations: the quartet  $\4_{2/3}$ and the singlet $\1_{2/3}$ as the lightest top partners. They are parametrized as follows:
\beq
\Psi_\4 = \frac{1}{\sqrt{2}}\left(
\begin{array}{c}
i B - i X_{5/3} \\
 B + X_{5/3} \\
i T + i X_{2/3} \\
 -T + X_{2/3} 
\end{array}
\right)_{2/3}, \quad \Psi_\1 =\begin{pmatrix} \widetilde{T}\end{pmatrix}_{2/3},
\eeq
and transform as $\Psi\to\mathcal{H}_{\textbf{r}_\Psi}\otimes\mG_X\Psi$, where $\textbf{r}_\Psi$ is the $SO(4)$ representation of $\Psi$, and $\mG_X$ denotes the group element of $U(1)_X$. From the transformation rules in Eq.~(\ref{eq:detrans}), we can construct a covariant derivative acting on the composite fermionic fields $\Psi$:
\beq
\nabla_\mu=\partial_\mu-ie_\mu^aT^a_{\textbf{r}_\Psi}.
\eeq
Taking into account of the $U(1)_X$ group, the covariant derivative becomes $(\nabla_\mu - i g_1 X B_\mu)$. For the spin-1 resonances, we consider three irreducible representations under the unbroken $SO(4)$: $\rho_L (\3,\1)$, $\rho_R (\1,\3)$ and $\rho_X(\1,\1)$.

\subsection{The matching to the Higgs doublet notation}

The CCWZ operators and the effective Lagrangians for the composite resonances can be written in terms of the fields that have the definite quantum number under the SM gauge group $SU(2)_L\times U(1)_Y$ . To see this, we first notice that the SM Higgs doublet with hypercharge $Y=1/2$  can  be written as follows:
\be
H=\begin{pmatrix}\frac{h_2+ih_1}{\sqrt{2}}\\ \frac{h_4-ih_3}{\sqrt{2}}
\end{pmatrix}, \quad \widetilde H=i\sigma^2H^*.
\ee
It is related with the quartet notation $\vec{h}$ by an unitary matrix $P$ with determinant -1:
\be
\label{eq:matrixp}
\vec{h}=P\begin{pmatrix}
H\\ -\widetilde{H}
\end{pmatrix},\quad  P=\begin{pmatrix}-\frac{i}{\sqrt{2}}&0&0&\frac{i}{\sqrt{2}}\\ \frac{1}{\sqrt{2}}&0&0&\frac{1}{\sqrt{2}}\\ 0&\frac{i}{\sqrt{2}}&\frac{i}{\sqrt{2}}&0\\0&\frac{1}{\sqrt{2}}&-\frac{1}{\sqrt{2}}&0\end{pmatrix}, \quad P^\dagger P =\mathbb{I}_{4}, \quad \Det \,P = -1.
\ee
The $SO(4)$ generators can be converted to  the doublet notation by using $P$: 
\be
\begin{split}
P^\dagger t^{a_L}P &=\begin{pmatrix}
\frac{1}{2}\sigma^{a_L}&\\&\frac{1}{2}\sigma^{a_L}
\end{pmatrix},\quad P^\dagger t^{3_R}P=\begin{pmatrix}
\frac{1}{2}\mathbb{I}_{2\times2}&\\&-\frac{1}{2}\mathbb{I}_{2\times2}
\end{pmatrix},  \\
P^\dagger t^{1_R}P &=\begin{pmatrix}
&\frac{1}{2}\mathbb{I}_{2\times2}\\ \frac{1}{2}\mathbb{I}_{2\times2}&
\end{pmatrix},\quad P^\dagger t^{2_R}P=\begin{pmatrix}
&-\frac{i}{2}\mathbb{I}_{2\times2}\\ \frac{i}{2}\mathbb{I}_{2\times2}&
\end{pmatrix}.
\end{split}
\ee
Consequently, the $\vec{h}$ covariant derivative term can be rewritten as:
\beq
D_\mu \vec{h} =P \begin{pmatrix}
D_\mu H  \\ - D_\mu \widetilde H
\end{pmatrix},  
\eeq
where the $D_\mu$ in the right-hand side of the equation is the normal SM covariant derivative:
\be
D_\mu =\partial_\mu-ig_2\frac{\sigma^{a_L}}{2}W_\mu^{a_L}-ig_1YB_\mu.
\ee
where hypercharge $Y$ is given by $Y = T^{3_R} + X$.
Using above results, we can easily rewrite the leading Lagrangian in \Eq{d2} in the doublet notation:
\be
\frac{f ^2}{4}d_\mu^id^{i\mu}=\frac{f ^2}{2|H|^2}\sin^2\frac{\sqrt{2}|H|}{f }D_\mu H^\dagger D^\mu H+\frac{f ^2}{8|H|^4}\left(\frac{2|H|^2}{f ^2}-\sin^2\frac{\sqrt{2}|H|}{f }\right)(\partial_\mu|H|^2)^2,
\ee
with $|H| = \sqrt{H^\dagger H}$. For further convenience, we list the following useful identities:
\be\label{h2H}\begin{split}
&\vec{h}^T D_\mu\vec{h}=  \partial_\mu (H^\dagger H), \quad \vec{h}^T t^{a_L}D_\mu\vec{h}  = \frac{1}{2} H^\dagger\sigma^{a_L} \Dfbd H,\quad \vec{h}^Tt^{3_R}D_\mu\vec{h}= \frac{1}{2} H^\dagger \Dfbd H,\\
&\vec{h}^Tt^{1_R}D_\mu\vec{h}=-\frac{1}{2}(\widetilde H^\dagger D_\mu H-D_\mu H^\dagger\widetilde H),\quad \vec{h}^Tt^{2_R}D_\mu\vec{h} =\frac{i}{2}(H^\dagger D_\mu\widetilde H+D_\mu\widetilde H^\dagger H).
\end{split}
\ee
where the $\Dfbd$ is defined as:
\be
H^\dagger\Dfbd H\equiv H^\dagger(D_\mu H)-(D_\mu H^\dagger)H;\quad H^\dagger\sigma^{a_L}\Dfbd H\equiv H^\dagger\sigma^{a_L}(D_\mu H)-(D_\mu H^\dagger)\sigma^{a_L}H.
\ee
The quartet top partner fields, $\Psi_\4$ can be decomposed as two $SU(2)_L$ doublets with hypercharge $Y=1/6$, $7/6$ as follows:
\be
\label{eq:doublets}
\Psi_\4 = P\begin{pmatrix}Q_X\\Q\end{pmatrix}, \qquad Q=\begin{pmatrix}T\\ B\end{pmatrix}_{1/6}, \quad Q_X=\begin{pmatrix}X_{5/3}\\ X_{2/3}\end{pmatrix}_{7/6},
\ee 
with the same $P$ matrix as defined in \Eq{eq:matrixp}. The SM fermions are assumed to be embedded in the $\5_{X}$ representation of $SO(5)\times U(1)_X$ with hypercharge given by $Y= T^{3_R} + X$. We only consider the top sector in our paper.
For the SM $SU(2)$ doublet $q_L =(t_L,b_L)^T$,  we have the embedding:
\be\label{qL5_definition}
q_L^{\textbf{5}}=\frac{1}{\sqrt{2}}\begin{pmatrix}ib_L&b_L&it_L&-t_L&0\end{pmatrix}^T_{2/3}=P_\5\begin{pmatrix}0,0,t_L,b_L,0\end{pmatrix}^T,\quad P_\5=\begin{pmatrix}P&0_{4\times1}\\ 0_{1\times4}& 1\end{pmatrix}.
\ee
The $q_L^\5$ formally transforms under the $\mG \in SO(5)$ and $\mG_X\in U(1)_X$ as $q_L^{\5} \rightarrow \mG\otimes\mG_X q_L^{\5} $.  For the right-handed top quark, we will consider two possibilities: $t_R$ as an elementary filed or as a massless bound state of the strong sector. In the first case, we also embed it in the representation of  $\textbf{5}_{2/3}$:
\be\label{tR5_definition}
t_R^{\textbf{5}}=\begin{pmatrix}0&0&0&0&t_R^{\rm (P)}\end{pmatrix}^T_{2/3}.
\ee
For the fully composite right-handed top quark, we assume that it is  a singlet of $SO(4)$, denoted as $t_R^{\rm (F)}$
and its interactions preserve the non-linearized $SO(5)$. We denote those two treatments as partially and fully composite $t_R$ scenario, respectively. 

All  the effective Lagrangian in MCHMs can be rewritten in terms of the doublet notation easily using \Eq{d_and_e_decompose}, \Eq{h2H}, \Eq{eq:doublets}, \Eq{qL5_definition} and \Eq{tR5_definition}. The full results are tedious, thus we will not list them here; however, their LO expansions in  $H^\dagger H/f^2$ order will be listed and discussed in Appendix~\ref{app:models}.

\section{The models}
\label{app:models}

In this section, we briefly describe the models considered in our paper (see Refs.~\cite{Greco:2014aza,DeSimone:2012fs,Contino:2011np}).  We focus on the minimal coset $SO(5)\times U(1)_X/SO(4)\times U(1)_X$ of the strong sector, where the Higgs bosons are the pseudo-Nambu-Goldstone bosons associated with this global symmetry breaking.  

\subsection{The models involving $\rho_L (3,1)$ and quartet top partners $\Psi_\4 (2,2)$: LP(F)$_4$}\label{app:rhoL}

We start from the models involving the $\rho_L$ and the quartet top partners $\Psi_\4$. The Lagrangian of the strong sector reads:
\be
\label{eq:LagrhoL}
\begin{split}
\mathcal{L}^{\text{L}_\4}=&-\frac{1}{4}\rho_{\mu\nu}^{a_L}\rho^{a_L\mu\nu} +\frac{m_{\rho_L}^2}{2g_{\rho_L}^2}(g_{\rho_L}\rho_\mu^{a_L}-e_\mu^{a_L})^2+\bar\Psi_{\textbf{4}}\gamma^\mu i\left(\nabla_\mu-ig_1\frac{2}{3}B_\mu\right)\Psi_{\textbf{4}}-M_{\textbf{4}}\bar\Psi_{\textbf{4}}\Psi_{\textbf{4}} \\
&+c_1\bar\Psi_{\textbf{4}}\gamma^\mu t^{a_L}\Psi_{\textbf{4}}(g_{\rho_L}\rho^{a_L}_\mu-e^{a_L}_\mu),
\end{split}
\ee
where the field strength of the spin-1 resonance is defined as
\be\label{rhoL_fs}
\rho_{\mu\nu}^{a_{L}}=\partial_\mu\rho^{a_{L}}_\nu-\partial_\nu\rho^{a_{L}}_\mu
+g_{\rho_L}\epsilon^{a_Lb_Lc_L}\rho_\mu^{b_L}\rho_\nu^{c_L}.
\ee
The Yukawa interactions between strong and elementary sector are:
\be\label{eq:PF4}\begin{split}
\mathcal{L}^{\text{P}_\4}=&y_L f \bar{q}_L^{\textbf{5}I}U_{Ij}\Psi_{\textbf{4}}^j+y_R f \bar{t}_R^{\textbf{5}I}U_{Ij}\Psi_{\textbf{4}}^j +\text{h.c.},\\
\mathcal{L}^{\text{F}_\4}=&(c_2\bar\Psi_{\textbf{4}}^i\gamma^\mu id^i_\mu t_R^{\rm (F)} +\text{h.c.})
+(y_{L} f\bar{q}_L^{\textbf{5}I}U_{Ij}\Psi_{\textbf{4}}^j+ y_{2L} f \bar{q}_L^{\textbf{5}I}U_{I5}t_R^{\rm (F)}+\text{h.c.}).
\end{split}\ee
The fully Lagrangian is then written as~\cite{Greco:2014aza,DeSimone:2012fs,Contino:2011np}
\be\label{LP4_lll}
\mathcal{L}^{\text{LP}_\4}=\mathcal{L}^{\text{L}_\4}+\mathcal{L}^{\text{P}_\4};\quad\mathcal{L}^{\text{LF}_\4}=\mathcal{L}^{\text{L}_\4}+\mathcal{L}^{\text{F}_\4},
\ee
where we omitted the SM Lagrangians for the quark fields $q_L$ and $t_R$. 
Note that the CCWZ covariant  objects $e_\mu^a$ include the SM gauge fields:
\beq
\begin{split}
e_\mu^{a_L}&=g_2 W_\mu^{a_L}-\frac{i}{f^2}H^\dagger\frac{\sigma^{a_L}}{2}\Dfbd H + \cdots, \\
e_\mu^{3_R}&=g_1 B_\mu-\frac{i}{2f^2} H^\dagger\Dfbd H + \cdots
\end{split}
\eeq
and we have written the formulae in terms of SM Higgs doublet $H$ (see Appendix~\ref{app:CCWZ} for the definition and derivation). Note that the SM gauge interactions don't preserve the non-linearly realized $SO(5)$ symmetry and provide the explicit breaking, thus will contribute to the Higgs potential at one-loop level. The term with coefficient $c_1$ involves the direct coupling between the $\rho_L$ and the quartet top partners at the order of $g_{\rho_L}$. As discussed in Ref.~\cite{Greco:2014aza}, this interaction will have an important impact on the phenomenology of $\rho_L$ especially when $m_{\rho_L} > 2 M_\4$ and decaying into two top partners are allowed.  In most of the case, we will choose $c_1 = 1$ as our benchmark point.

Note that the mass term for the $\rho_L$ in Eq.~(\ref{eq:LagrhoL}) will induce a linear mixing between them and the SM $W_\mu$ gauge bosons before EWSB. Diagonalizing the mass matrix will lead to the  partial compositeness of $\mO( g_2/g_{\rho_L})$ for the $W$ bosons. As a result, the SM $SU(2)_L$ gauge coupling will be redefined as follows:
\beq
\label{eq:gL}
 \frac{1}{g^2} = \frac{1}{g_2^2} + \frac{1}{g_{\rho_L}^2},
\eeq
and the $W$-mass at the LO is given by (see Appendix~\ref{app:mass} for detail):
\beq
M_W^2 = \frac14 g^2 v^2,\quad v = f \sin\frac{\left<h\right>}{f} = 246~\GeV.
\eeq
Due to the linear mixing, the mass of the $\rho_L$ will also be modified as follows:
\beq
M_{\rho_L}^2 = m_{\rho_L}^2 \left(1 + \frac{g_2^2}{g_{\rho_L}^2}\right).
\eeq
Note  that this direct mixing mass term will also lead to contribution to  $\hat S$-parameter  in the low energy observable. Actually, integrating out the $\rho_L$ at the LO, we will obtain the $\mO_W$ operator  (see Ref.~\cite{Greco:2014aza}), which leads to the contribution to the $\hat{S}$ parameter~\cite{Barbieri:2004qk}:
\beq
\hat{S} = \frac{M_W^2}{g_{\rho_L}^2 f^2}.
\eeq
The $\rho_L$ resonance will be coupled to SM fermions universally with strength of $\mO(g^2/g_{\rho_L})$ due to the linear mixing. The non-universality comes from the linear mixing between the SM fermions and corresponding composite partners. Since the mixing is the source of the SM fermion masses after EWSB, it is roughly the order of the fermion Yukawa couplings. Thus we expect that only the third generation mixings (especially the top quark) have the important impact on  phenomenology of the $\rho_L$, which is the reason we only focus on the top sector. 

For the partially composite right-handed top quark scenario, we have two parameters $y_L$, $y_R$ controlling the mixing between $q_L$, $t_R^{\rm (P)}$ and the top partner $\Psi_\4$. Similar to the SM gauge bosons, there will be direct mixing between $q_L$ and the composite $SU(2)_L$ doublet $Q $ before EWSB proportional to $y_L$:
\beq
y_L f \bar{q}_L Q_R +{\rm h.c.},
\eeq
where the doublet $Q=(T,B)^T$ is defined in \Eq{eq:doublets}. This motives us to define a left-handed mixing angle $\theta_L$ as follows:
\beq
\label{eq:lmix}
\tan\theta_L = \frac{y_L f }{M_\4},
\eeq
which measures the partial compositeness of the SM fermions $q_L$. Due to the linear mixing, the mass formulae for the fermionic resonances before EWSB are given by:
\beq
\label{eq:masstp}
M_Q = \sqrt{M_\4^2 + y_L^2 f^2}, \quad M_{Q_X} = M_\4.
\eeq
Note that $y_L$ breaks the $SO(4)$ explicitly and will contribute to the $\hat T$ parameter at the loop level, thus can't be too large. In contrast, $t_R^{\rm (P)}$ is an $SO(4)$ singlet so that $y_R$ term preserves the custodial symmetry  can in principle can be large~\cite{Ghosh:2015wiz}. For the fully composite $t_R^{\rm (F)}$,  besides the mixing between $q_L$ and $\Psi_\4$ (denoted also as $y_L$), we can write a direct coupling $y_{2L}$ between $q_L$ and $t_R^{\rm (F)}$. This term provides  the main source of top quark mass.  Since $t_R^{\rm (F)}$ belongs to the strong sector, there are also direct interactions between it and the composite resonances, which are written as the $c_2$ term in the $\mL^{\text{F}_\4}$.  As  discussed in \Ref{DeSimone:2012fs}, this strong interaction term provides the dominant contribution to decay of the top partners, especially when the mixing parameters are small.

Note that it will be very useful to rewrite the Lagrangian in terms of SM $SU(2)_L\times U(1)_Y$ notation, where the SM gauge symmetries are manifest. By using the formulae of the Goldstone matrix $U$ and the $d_\mu, e_\mu$ in the Appendix~\ref{app:CCWZ}, we can write the Lagrangian $\mathcal{L}^{\text{L}_\4}$ using the doublet notation as follows:
\be
\label{eq:LPdoublet}
\begin{split}
\mathcal{L}^{\text{L}_\4}=&-\frac{1}{4}\rho_{\mu\nu}^{a_L}\rho^{a_L\mu\nu}+\frac{a_{\rho_L}^2}{2}f^2\left(g_{\rho_L}\rho_\mu^{a_L}-g_2W_\mu^{a_L}+\frac{i}{f^2} H^\dagger\frac{\sigma^{a_L}}{2} \Dfbd H\right)^2\\
&+\bar Q(\gamma^\mu iD_\mu-M_\4)Q+ \bar Q_X(\gamma^\mu iD_\mu-M_\4)Q_X\\
& - \frac{i}{4f^2}\left(\bar Q_X\gamma^\mu \sigma^{a_L} Q_X+\bar Q\gamma^\mu\sigma^{a_L}Q\right)H^\dagger\sigma^{a_L}\Dfbd H\\
&- \frac{i}{4f^2}\left(\bar Q_X\gamma^\mu Q_X-\bar Q\gamma^\mu Q\right) H^\dagger \Dfbd H
+\left(\frac{i}{4f^2}\bar Q\gamma^\mu Q_X H^\dagger \Dfbd\widetilde H+\text{h.c.}\right)\\
&+c_1\left(\bar Q\gamma^\mu\frac{\sigma^{a_L}}{2}Q + \bar Q_X\gamma^\mu\frac{\sigma^{a_L}}{2}Q_X\right) \left(g_{\rho_L}\rho_\mu^{a_L}-g_2W_\mu^{a_L}+\frac{i}{f^2} H^\dagger\frac{\sigma^{a_L}}{2} \Dfbd H\right)+ \cdots,
\end{split}
\ee
where the $\cdots$ denotes the higher order terms in $H^\dagger H/f^2$ and we have defined the $\mO(1)$ parameter $a_{\rho_L}$ as in \Ref{Contino:2011np}:
\beq
\label{eq:arhoL}
a_{\rho_L} = \frac{m_{\rho_L}}{g_{\rho_L} f}.
\eeq
From the dimension-six operators involving the top partners and the Higgs fields, we can see that generally the gauge couplings  of the top partners are modified at the $\mO(\xi)$ after EWSB. 
Note that there is an accidental parity symmetry $P_{LR}$ in the kinetic Lagrangian for the quartet top partner defined as~\cite{Agashe:2006at}:
\beq
P_{LR}^{(4)} = \text{diag} (-1,-1,-1,1),
\eeq
and the couplings between eigenstates of this parity ($X_{5/3}, B$) and the SM $Z$ gauge bosons will not obtain  any modification after EWSB. This can be easily  seen by using the formulae for the currents in the vacuum:
\beq
\begin{split}
i H^\dagger \Dfbd H &\to  -\frac{ \left<h\right>^2}{2} \left(g_2 W_\mu^3 - g_1 B_\mu\right), \\
i H^\dagger \sigma^{a_L} \Dfbd H &\to  \frac{\left<h\right>^2}{2} \left(g_2 W_\mu^{a_L} - g_1 \delta^{a3}B_\mu\right),\\
\end{split}
\eeq
remembering that $T^{3_L}(X_{5/3}) = T^{3_R}(X_{5/3})  = 1/2$ and    $T^{3_L}(B) = T^{3_R}(B)= -1/2$. This is important because $Z B_L \bar B_L$ are not modified by the Higgs VEV means that after the mixing between $b_L$ and $B_L$, the $Zb_L\bar b_L$ remains the same as the SM canonical couplings~\footnote{There are universal modification to the SM $Z\bar{f}f$ due to the $\rho_L - W$ mixing terms or $\hat{S}$ parameter by integrating out the $\rho_L$, but they are suppressed by $(g/g_{\rho_L})^2 \xi$.}.

Similarly, we can write the elementary-composite mixing Lagrangian $\mL^{P_4}$ in the doublet notation:
\be
\label{eq:P4doublet}
\mathcal{L}^{\text{P}_\4}=y_Lf \left( \bar{q}_L Q_R +\frac{1}{2f^2} \bar{q}_L\widetilde H  \,  (H^\dagger Q_{XR} - \tilde{H}^\dagger Q_R ) \right) +y_R \left(\bar Q_L\widetilde H t_R^{\rm (P)}- \bar Q_L^X H t_R^{\rm (P)} \right) + \text{h.c.}
\ee
where we only keep the leading terms in the expansion of $H^\dagger H/f^2$.
We can see clearly  that after EWSB only the mass matrix in the top sector obtains corrections of $\mO(y_L f \xi, y_R v)$, while for the charge $-1/3$ and charge-$5/3$ resonances,  their mass formulae are not modified~\footnote{Since we don't include the right-handed bottom quark mixings with bottom partners, the bottom quark remains massless. }. After EWSB, the top mass is given by:
\beq
\label{eq:mtp4}
M_t =\frac{y_R v s_{\theta_L}}{\sqrt{2}} + \cdots,
\eeq
where $s_{\theta_L}$ denotes $ \sin\theta_L$ defined in Eq.~(\ref{eq:lmix}).
The EWPT at the LEP prefers $y_L \lesssim y_R$, thus $y_R$ mixing term is dominant. In the unitary gauge, this term becomes:
\beq\label{TX23_mixing}
\frac{y_R}{\sqrt{2}} (\left<h\right> + h) \left(\bar{T}_L - \bar{X}_{2/3 L}\right) t_R^{\rm (P)}.
\eeq
So in the large $y_R$ limit,  there will be a top partner (the heavier one) in the mass eigenstate, which will primarily decay into $th$ and the other one will primarily decay into $tZ$.  See Appendix~\ref{app:mm} for detail, where  we summarize the mass matrices and mass formulae. As we will discuss below, in our consideration, we will focus on the region $y_R \gtrsim 1$, this effect will not be manifest. 
For the fully composite $t_R^{\rm (F)}$ case, we have:
\be\label{eq:F4doublet}\begin{split}
\mathcal{L}^{\text{F}_\4}=
&-c_2\frac{\sqrt{2}}{f}(\bar Q_{XR}\gamma^\mu t_RiD_\mu H-\bar Q_R \gamma^\mu t_RiD_\mu\widetilde H+\text{h.c.}).\\
&+y_Lf \left( \bar{q}_L Q_R +\frac{1}{2f^2} \bar{q}_L\widetilde H  \,  (H^\dagger Q_{XR} - \tilde{H}^\dagger Q_R )  \right) -y_{2L} \bar q_L\widetilde H t_R^{\rm (F)} + \text{h.c.},\\
\end{split}
\ee
The top mass to the leading order is given by:
\beq
\label{eq:mtf4}
M_t =  \frac{y_{2L} c_{\theta_L} v}{\sqrt{2}}  +\cdots,
\eeq
where $c_{\theta_L}$ denotes $ \cos\theta_L$ defined in Eq.~(\ref{eq:lmix}). So that the top Yukawa coupling is mainly determined by $y_{2L}$, which is different with partially composite $t_R^{\rm (P)}$ case. 
 
\subsection{The models involving $\rho_R (1,3)$ and quartet top partners $\Psi_\4 (2,2)$: RP(F)$_\4$}
\label{app:rhoR}
For the $\rho_R$ models, the effective Lagrangians read:
\be
\label{eq:LagrhoR}
\begin{split}
\mathcal{L}^{\text{R}_\4} = &-\frac{1}{4}\rho_{\mu\nu}^{a_R}\rho^{a_R\mu\nu} +\frac{m_{\rho_R}^2}{2g_{\rho_R}^2}(g_{\rho_R}\rho_\mu^{a_R}-e_\mu^{a_R})^2 +\bar\Psi_{\textbf{4}}\gamma^\mu i\left(\nabla_\mu-ig_1\frac{2}{3}B_\mu\right)\Psi_{\textbf{4}}-M_{\textbf{4}}\bar\Psi_{\textbf{4}}\Psi_{\textbf{4}} \\
&+c_1\bar\Psi_{\textbf{4}}\gamma^\mu t^{a_R}\Psi_{\textbf{4}}(g_{\rho_R}\rho^{a_R}_\mu-e^{a_R}_\mu);
\end{split}
\ee
where the definition of $\rho_{\mu\nu}^{a_R}$ is the same as in \Eq{rhoL_fs} with $(L\to R)$. The effective Lagrangians in models RP(F)$_\4$ are given by:
\be
\mathcal{L}^{\text{RP}_\4} =\mathcal{L}^{\text{R}_\4}+\mathcal{L}^{\text{P}_\4};\quad\mathcal{L}^{\text{RF}_\4}=\mathcal{L}^{\text{R}_\4}+\mathcal{L}^{\text{F}_\4},
\ee
where the Lagrangians $\mL^{{\rm P(F)}_\4}$ are the same as in \Eq{eq:LagrhoL}. In terms of doublet notation, we have:
\be\begin{split}
\label{eq:rhoR}
\mathcal{L}^{\rm R_\4}=&-\frac{1}{4}\rho_{\mu\nu}^{a_R}\rho^{a_R\mu\nu}+\frac{a_{\rho_R}^2}{2}f^2\left(g_{\rho_R}\rho_\mu^{3_R}-g_1B_\mu+\frac{1}{2f^2}H^\dagger i\Dfbd H\right)^2\\
&+\frac{a_{\rho_R}^2}{2}f^2\left(\sqrt{2}g_{\rho_R}\rho_\mu^{-_R}-\frac{1}{2f^2}H^\dagger i\Dfbd\widetilde H\right)\left(\sqrt{2}g_{\rho_R}\rho_\mu^{+_R}-\frac{1}{2f^2}\widetilde H^\dagger i\Dfbd H\right)\\
&+c_1\left(\frac{1}{2}\bar Q_X\gamma^\mu Q_X-\frac{1}{2}\bar Q\gamma^\mu Q\right)\left(g_{\rho_R}\rho_\mu^{3_R}-g_1B_\mu+\frac{1}{2f^2}H^\dagger i\Dfbd H\right)\\
&+c_1\frac{1}{2}\bar Q\gamma^\mu Q_X\left(\sqrt{2}g_{\rho_R}\rho_\mu^{-_R}-\frac{1}{2f^2}H^\dagger i\Dfbd \widetilde H\right) + \text{h.c.} + \cdots,
\end{split}\ee
where we only show the terms involving the $\rho_R$ and defined:
\beq
\label{eq:arhoR}
a_{\rho_R} = \frac{m_{\rho_R}}{g_{\rho_R} f}.
\eeq
Note that similar with $\rho_L$, there is a direct mixing between $\rho^{3R}_\mu $ and the  hypercharge field $B_\mu$. So the $U(1)_Y$ gauge coupling is redefined as follows:
\beq
\label{eq:gpR}
\frac{1}{g^{\prime 2}} = \frac{1}{g_1^2} + \frac{1}{g_{\rho_R}^2},
\eeq
and the $Z$-mass to the LO is given by:
\beq
\label{eq:gR}
M_Z^2 = \frac{g^2 + g^{\prime 2}}{4} v^2, \quad g = g_2,
\eeq
Note that this direct mixing mass term will also lead to contribution to  $\hat S$-parameter in the  low energy observable: integrating out the $\rho_R$ will result in the $\mO_B$ operator and  
\beq
\hat{S} = \frac{M_W^2}{g_{\rho_R}^2 f^2}.
\eeq

As can been seen from Eq.~(\ref{eq:rhoR}), for the neutral resonance $\rho^{3_R}$, it has the universal coupling of $\mO(g^{\prime 2}/g_{\rho_R})$ to the SM fermions, while for the charged $\rho_R$, its coupling arise from $\mO(\xi)$. This makes $\rho_R^0$ more produced at the LHC than the charged one and thus the most stringent constraint on the $\rho_R$ models comes from the neutral spin-1 resonance searches. Because of  the smallness of $U(1)_Y$ gauge coupling $g^\prime$ compared with $SU(2)_L$ gauge coupling $g$, its constraints are weaker than $\rho_L$.  For the direct interactions with the fermionic resonances (the $c_1$ term),  they are similar to the $\rho_L$ interactions except that the charged currents are between $Q$ and $Q_X$.

\subsection{The models involving $\rho_X(1,1):$ XP(F)$_\4$ and   XP(F)$_\1$}
\label{app:rhoX}

For the models involving the $\rho_X$ and the quartet $\Psi_\4$, the Lagrangian containing the $\rho_X$ are given by:
\be
\label{eq:LagrhoX}\begin{split}
\mathcal{L}^{\rm X_\4 }=&-\frac{1}{4}\rho_{X\mu\nu}\rho_X^{\mu\nu}+\frac{m_{\rho_X}^2}{2 g_{\rho_X}^2}(g_{\rho_X}\rho_{X\mu}-g_1B_\mu)^2+\bar\Psi_{\textbf{4}}\gamma^\mu i\left(\nabla_\mu-ig_1\frac{2}{3}B_\mu\right)\Psi_{\textbf{4}}-M_{\textbf{4}}\bar\Psi_{\textbf{4}}\Psi_{\textbf{4}}\\
&+c_1\bar\Psi_\4\gamma^\mu \Psi_\4 (g_{\rho_X}\rho_{X\mu}-g_1B_\mu),
\end{split}
\ee
where $\rho_{X\mu\nu}=\partial_\mu\rho_{X\nu}-\partial_\nu\rho_{X\mu}$, and 
\be\label{eq:rhox4}\begin{split}
\mathcal{L}^{\rm XP_\4}=&\mathcal{L}^{\rm X_\4 }+\mathcal{L}^{\rm P_\4 },\\
\mathcal{L}^{\rm XF_\4}=&\mathcal{L}^{\rm X_\4 }+\mathcal{L}^{\rm F_\4 }+c_1'\bar t_R^{\rm (F)}\gamma^\mu t_R^{\rm (F)} (g_{\rho_X}\rho_{X\mu}-g_1B_\mu).
\end{split}\ee
where the Lagrangians $\mL^{{\rm P(F)}_\4}$ are the same as in \Eq{eq:LagrhoL}.
Similar to $\rho^{3_R}_\mu$, $\rho_{X\mu}$ is mixing with the hypercharge gauge field $B_\mu$, thus will have a universal coupling of $\mO(g^{\prime 2}/g_{\rho_X})$ to the SM elementary fermions. The $U(1)_Y$ gauge coupling $g^\prime$ is redefined as:
\beq
\label{eq:gpX}
\frac{1}{g^{\prime 2}} = \frac{1}{g_1^2} + \frac{1}{g_{\rho_X}^2}.
\eeq
Similar to the case of $\rho_{L,R}$, we will also define the $\mO(1)$ parameter $a_{\rho_X}$ as follows:
\beq
\label{eq:arhoX}
a_{\rho_X} = \frac{m_{\rho_X}}{g_{\rho_X}f}.
\eeq
$\rho_X$ will not contribute to $\hat S$-parameter because of its singlet nature, but will contribute to the $Y$-parameter (defined in Ref.~\cite{Barbieri:2004qk}) as follows:
\beq\label{rho_X-Y}
Y = \frac{2 g^{\prime 2} M_W^2}{g^2_{\rho_X} m_{\rho_X}^2}.
\eeq
The extra suppression factor $(g'/g_{\rho_X})^2$ will make the constraint on the mass of the $\rho_X$ from EWPT much weaker than $\rho_{L,R}$. For the case of fully composite right-handed top quark, a direct interaction term between $\rho_X$ and $t_R^{\rm (F)}$ can be written down. The coefficient is denoted as $c_1'$ in \Eq{eq:rhox4}. This term is special in the sense that it can affect the decay of $\rho_X$ and also can lead to a new production mechanism of $\rho_X$: $t\bar{t}$ fusion.  The decay of $\rho_X$ into a pair of top quark will result in four top final states, which can be probed using  the SSDL final state~\cite{Liu:2015hxi}. 

Finally, we consider the models involving $\rho_X$ and the singlet $\Psi_\1$. The Lagrangian involving the heavy resoances read:
\be
\label{eq:LagrhoX}\begin{split}
\mathcal{L}^{\rm X_\1 }=&-\frac{1}{4}\rho_{X\mu\nu}\rho_X^{\mu\nu}+\frac{m_{\rho_X}^2}{2 g_{\rho_X}^2}(g_{\rho_X}\rho_{X\mu}-g_1B_\mu)^2+\bar\Psi_{\textbf{1}}i\slashed{D}\Psi_{\textbf{1}}-M_{\1}\bar\Psi_{\textbf{1}}\Psi_{\textbf{1}}\\
&+c_1\bar\Psi_\1\gamma^\mu \Psi_\1 (g_{\rho_X}\rho_{X\mu}-g_1B_\mu),
\end{split}\ee
The mixing term is given by:
\be\begin{split}
\mathcal{L}^{\rm P_\1}=&y_L f ( \bar{q}_L^\5)^I U_{I5} \Psi_{\1 R} +y_Rf  (\bar{t}_R^{\5})^I U_{I5} \Psi_{\1 L}+\text{h.c.},\\
\mathcal{L}^{\rm F_\1}=&y_L f ( \bar{q}_L^\5)^I U_{I5} \Psi_{\1 R} +y_{2L} f ( \bar{q}_L^\5)^I U_{I5} t_R^{\rm (F)} + \text{h.c.},
\end{split}\ee
and the effective Lagrangians in models XP(F)$_\1$ are:
\be\label{eq:rhox1}\begin{split}
\mathcal{L}^{\rm XP_\1 }=&\mathcal{L}^{\rm X_\1 }+\mathcal{L}^{\rm P_\1 },\\
\mathcal{L}^{\rm XF_\1 }=&\mathcal{L}^{\rm X_\1 }+\mathcal{L}^{\rm F_\1 }+\left(c_1'\bar t_R^{\rm (F)}\gamma^\mu t_R^{\rm (F)}+c_1''(\bar t_R^{\rm (F)}\gamma^\mu \Psi_{\1R}+\text{h.c.})\right)(g_{\rho_X}\rho_{X\mu}-g_1B_\mu).
\end{split}\ee
Note that here besides the $c'_1$ term, we also have the non-diagonalized interaction, i.e. the $c''_1$ term. The mixing term between the elementary SM quarks and the composite fields can be rewritten  in terms of doublet notation. The results read:
\be\begin{split}
\label{eq:MXdoublet}
\mathcal{L}^{\rm P_\1}& = -y_L \bar{q}_L\widetilde H\widetilde T_R +y_Rf \bar{t}_R^{\rm (P)} \widetilde T_L+\text{h.c.},\\
\mathcal{L}^{\rm F_\1}&= - y_L\bar{q}_L\widetilde H\widetilde{T}_R + y_{2L} \bar{q}_L\widetilde H t_R^{\rm (F)}+\text{h.c.}.
\end{split}\ee
For the model XP$_\1$, the linear mixing term between $t_R^{\rm (P)}$ and  the singlet $\widetilde T$ will lead to the partial compositeness of the right-handed top quark with mixing angle $\theta_R$:
\beq
\label{eq:rmix}
\tan\theta_R = \frac{y_R f}{M_\1}.
\eeq
The top partner mass and the top mass will become:
\beq
\label{eq:mtp1}
M_t = \frac{y_L v s_{\theta_R}}{\sqrt{2}}+\cdots, \quad M_{\widetilde T} = \sqrt{M_\1^2 + y_R^2 f^2}.
\eeq
For the fully composite $t_R^{\rm (F)}$, the top mass is simply:
\beq
\label{eq:mtf1}
M_t = \frac{y_{2L} v }{\sqrt{2}} + \cdots, 
\eeq
In both $\text{XP}_\1$ and $\text{XF}_\1$ models, the $y_L$ mixing term controls the top partner $\widetilde{T}$ decay, as this is the leading term with trilinear interactions violating the top partner fermion number. By using the Goldstone equivalence theorem, we can easily see the following branching ratios for the decay of the singlet $\widetilde{T}$:
\beq\label{TS_decay}
\Br(\widetilde T \rightarrow bW) \simeq 2 \Br(\widetilde T \rightarrow t h) \simeq 2 \Br(\widetilde T \rightarrow tZ) \simeq 50\%,
\eeq
where the factor $2$  in the branching ratios comes from the $\sqrt{2}$ suppression of the real scalar fields compared with complex scalar fields.

\section{The mass matrices and the mass eigenstates}\label{app:mm}
\label{app:mass}

Before EWSB, the mixing between the composite resonances and SM particles can be easily and exactly solved, as stated in Appendix~\ref{app:models} of this paper. However, after EWSB, i.e. $\langle \vec{h}\rangle=(0,0,0,\langle h\rangle)^T$, all particles with the same electric charge and spin will be generally mixed, and it is impossible to analytically resolve the mixing matrices exactly. In this section, we list all mass matrices after EWSB, and use perturbation method to derive the mass eigenvalues up to $\xi = v^2/f^2$ level.

\subsection{The spin-1 resonances}

Due to the SM gauge quantum number, $\rho_L^{a_L}$ mixes with $W^{a_L}$, while $\rho_R^{3_R}$ and $\rho_X^0$ mix with $B$ before EWSB, and the mixing angles are determined by $\tan\theta_\rho=g_{\rm SM}/g_\rho$. The VEV of Higgs will provide $\mO(\xi)$ modifications to such pictures. Below, we will give the mass eigenvalues up to $\xi$ level for the vector bosons.

\subsubsection{The $\rho_L(\3,\1)$ resonance}

After EWSB, the mass terms of vector bosons are
\be
\mathcal{L}^{\rm L_\4}\supset\begin{pmatrix}W_\mu^-&\rho^-_{L\mu}\end{pmatrix}M_{L\pm}^2\begin{pmatrix} W^{+\mu}\\ \rho_L^{+\mu}\end{pmatrix}+\frac{1}{2}\begin{pmatrix} B_\mu& W_\mu^3& \rho^3_{L\mu}\end{pmatrix}M_{L0}^2\begin{pmatrix}B^\mu\\ W^{3\mu}\\ \rho_L^{3\mu}\end{pmatrix},
\ee
where
\be
M_{L\pm}^2=\left(
\begin{array}{cc}
 \frac{1}{4} f^2 g_2^2 \left(a_{\rho_L}^2 \left(-\xi +2 \sqrt{1-\xi }+2\right)+\xi
   \right) & -\frac{1}{2} a_{\rho_L}^2 f^2 g_2 g_{\rho_L} \left(\sqrt{1-\xi
   }+1\right) \\
 -\frac{1}{2} a_{\rho_L}^2 f^2 g_2 g_{\rho_L} \left(\sqrt{1-\xi }+1\right) &
   a_{\rho_L}^2 f^2 g_{\rho_L}^2 \\
\end{array}
\right),
\ee
and
\be\begin{split}
&M_{L0}^2=\\&{\footnotesize \left(
\begin{array}{ccc}
 \frac{1}{4} f^2 g_1^2 \left(\xi -a_{\rho_L}^2 \left(\xi +2 \sqrt{1-\xi }-2\right)\right)
   & \frac{1}{4} (a_{\rho_L}^2-1) f^2 g_1 g_2 \xi  & \frac{1}{2} a_{\rho_L}^2 f^2
   g_1 g_{\rho_L} \left(\sqrt{1-\xi }-1\right) \\
 \frac{1}{4} (a_{\rho_L}^2-1) f^2 g_1 g_2 \xi  & \frac{1}{4} f^2 g_2^2
   \left(a_{\rho_L}^2 \left(-\xi +2 \sqrt{1-\xi }+2\right)+\xi \right) & -\frac{1}{2} a_{\rho_L}^2
   f^2 g_2 g_{\rho_L} \left(\sqrt{1-\xi }+1\right) \\
 \frac{1}{2} a_{\rho_L}^2 f^2 g_1 g_{\rho_L} \left(\sqrt{1-\xi }-1\right) &
   -\frac{1}{2} a_{\rho_L}^2 f^2 g_2 g_{\rho_L} \left(\sqrt{1-\xi }+1\right) &
   a_{\rho_L}^2 f^2 g_{\rho_L}^2 \\
\end{array}
\right)}.
\end{split}\ee
By using $\xi$ as the expanding parameter, we can diagonalize above matrices perturbatively. Up to $\xi$ order, the mass eigenvalues of the SM gauge bosons are
\be
M_W^2=\frac{g_2^2 g_{\rho_L}^2}{4(g_2^2+g_{\rho_L}^2)}f^2\xi=\frac{g^2}{4}f^2\xi,\quad M_Z^2=\frac{1}{4}\left(g_1^2+\frac{g_2^2 g_{\rho_L}^2}{g_2^2+g_{\rho_L}^2}\right)f^2\xi=\frac{g^2+g'^2}{4}f^2\xi,
\ee
and the photon is massless, due to the residual electromagnetic gauge invariance. Note that the $\hat T$-parameter is 0, as expected. For the spin-1 resonances, the mass eigenvalues are
\be\begin{split}
M_{\rho_L^\pm}^2=M_{\rho_L^0}^2=&a_{\rho_L}^2 f^2 (g_2^2 + g_{\rho_L}^2) +\frac{[(1-2a_{\rho_L}^2)g_2^4-2a_{\rho_L}^2 g_{\rho_L}^2 g_2^2]f^2\xi}{4(g_2^2 + g_{\rho_L}^2)}\\
=&\frac{g_{\rho_L}^2}{g_{\rho_L}^2-g^2}m_{\rho_L}^2-\frac{g^2\xi}{4}\left(\frac{2m_{\rho_L}^2-g^2f^2}{g_{\rho_L}^2-g^2}\right).
\end{split}\ee

\subsubsection{The $\rho_R(\1,\3)$ resonance}

We  can obtain the mass terms from the Lagrangian as follows:
\be
\mathcal{L}^{\rm R_\4}\supset\begin{pmatrix}W_\mu^-&\rho^-_{R\mu}\end{pmatrix}M_{R\pm}^2\begin{pmatrix} W^{+\mu}\\ \rho_R^{+\mu}\end{pmatrix}+\frac{1}{2}\begin{pmatrix} B_\mu& W_\mu^3& \rho^3_{R\mu}\end{pmatrix}M_{R0}^2\begin{pmatrix}B^\mu\\ W^{3\mu}\\ \rho_R^{3\mu}\end{pmatrix},
\ee
where
\be
M_{R\pm}^2=\left(
\begin{array}{cc}
 \frac{1}{4} f^2 g_2^2 \left(\xi -a_{\rho_R}^2 \left(\xi +2 \sqrt{1-\xi }-2\right)\right)
   & \frac{1}{2} a_{\rho_R}^2 f^2 g_2 g_{\rho_R} \left(\sqrt{1-\xi }-1\right) \\
 \frac{1}{2} a_{\rho_R}^2 f^2 g_2 g_{\rho_R} \left(\sqrt{1-\xi }-1\right) &
   a_{\rho_R}^2 f^2 g_{\rho_R}^2 \\
\end{array}
\right),
\ee
and
\be\begin{split}
&M_{R0}^2=\\&{\footnotesize \left(
\begin{array}{ccc}
 \frac{1}{4} f^2 g_1^2 \left(a_{\rho_R}^2 \left(-\xi +2 \sqrt{1-\xi }+2\right)+\xi
   \right) & \frac{1}{4} (a_{\rho_R}^2-1) f^2 g_1 g_2 \xi  & -\frac{1}{2} a_{\rho_R}^2
   f^2 g_1 g_{\rho_R} \left(\sqrt{1-\xi }+1\right) \\
 \frac{1}{4} (a_{\rho_R}^2-1) f^2 g_1 g_2 \xi  & \frac{1}{4} f^2 g_2^2
   \left(\xi -a_{\rho_R}^2 \left(\xi +2 \sqrt{1-\xi }-2\right)\right) & \frac{1}{2} a_{\rho_R}^2
   f^2 g_2 g_{\rho_R} \left(\sqrt{1-\xi }-1\right) \\
 -\frac{1}{2} a_{\rho_R}^2 f^2 g_1 g_{\rho_R} \left(\sqrt{1-\xi }+1\right) &
   \frac{1}{2} a_{\rho_R}^2 f^2 g_2 g_{\rho_R} \left(\sqrt{1-\xi }-1\right) &
   a_{\rho_R}^2 f^2 g_{\rho_R}^2 \\
\end{array}
\right)}.
\end{split}\ee
The masses eigenvalues can be derived as the series of $\xi$, and we list the terms up to $\xi$ order here. For SM gauge bosons, the results are
\be\label{SM_MR}
M_W^2=\frac{g_2^2}{4}f^2\xi=\frac{g^2}{4}f^2\xi,\quad M_Z^2=\frac{1}{4}\left(g_2^2+\frac{g_1^2 g_{\rho_R}^2}{g_1^2+g_{\rho_R}^2}\right)f^2\xi=\frac{g^2+g'^2}{4}f^2\xi,
\ee
and the photon is massless. For the composite vector resonances, the results are $M_{\rho_R^\pm}^2=m_{\rho_R}^2$, and
\be\begin{split}
M_{\rho_R^0}^2=&a_{\rho_R}^2 f^2 (g_1^2 + g_{\rho_R}^2) +\frac{[(1-2a_{\rho_R}^2)g_1^4-2a_{\rho_R}^2 g_{\rho_R}^2 g_1^2]f^2\xi}{4 (g_1^2 + g_{\rho_R}^2)}\\
=&\frac{g_{\rho_R}^2}{g_{\rho_R}^2-g'^2}m_{\rho_R}^2-\frac{g'^2\xi}{4}\left(\frac{2m_{\rho_R}^2-g'^2f^2}{g_{\rho_R}^2-g'^2}\right).
\end{split}\ee

\subsubsection{The $\rho_X(\1,\1)$ resonance}

For the $\rho_X$, the mass matrices read: 
\be\label{MX4_vector}
\mathcal{L}^{\rm X_\4},\mathcal{L}^{\rm X_\1}\supset W_\mu^-M_{X\pm}^2W^{+\mu}+\frac{1}{2}\begin{pmatrix} B_\mu& W_\mu^3& \rho_{X\mu}\end{pmatrix}M_{X0}^2\begin{pmatrix}B^\mu\\ W^{3\mu}\\ \rho_X^\mu\end{pmatrix},
\ee
where
\be
M_{X\pm}^2=\frac{g_2^2f^2\xi}{4},\quad M_{X0}^2=\left(
\begin{array}{ccc}
 \frac{1}{4} f^2 g_1^2 (4 a_{\rho_X}^2+\xi ) & -\frac{1}{4} f^2 g_1 g_2 \xi 
   & -a_{\rho_X}^2 f^2 g_1 g_{\rho_X} \\
 -\frac{1}{4} f^2 g_1 g_2 \xi  & \frac{1}{4} f^2 g_2^2 \xi  & 0 \\
 -a_{\rho_X}^2 f^2 g_1 g_{\rho_X} & 0 & a_{\rho_X}^2 f^2 g_{\rho_X}^2 \\
\end{array}
\right).
\ee
The $W^\pm$'s are already mass eigenstates because there are no charged vector bosons mixing with them. Up to $\xi$ order, the SM gauge bosons have the same mass eigenvalues as \Eq{SM_MR}, while the $\rho_X^0$ has mass
\be\label{MX4_vector_mass}
M_{\rho_X^0}^2=a_{\rho_X}^2 f^2 (g_1^2 + g_{\rho_X}^2) +\frac{g_1^4f^2\xi}{4 (g_1^2 + g_{\rho_X}^2)}
=\frac{g_{\rho_X}^2}{g_{\rho_X}^2-g'^2}m_{\rho_X}^2+\frac{g'^2\xi}{4}\frac{g'^2f^2}{g_{\rho_X}^2-g'^2}.
\ee

\subsection{The fermionic resonances}

In this section, we consider the $SO(4)$ quartet and singlet spin-$1/2$ resonances, and for each case we discuss both the partially and fully composite $t_R$ scenarios. The $X_{5/3}$ does not mix with any particles in SM, because of its exotic charge. In the quartet case, the mixing between $b_L$ and $B_L$ is not affected by the EWSB and has been exactly solved in Appendix~\ref{app:models}; while in the singlet case, $b_L$ quark has no mixing in the unitary gauge (in our massless $b$ approximation). Below we just discuss the mass matrices of charge-$2/3$ fermions.

\subsubsection{The $\Psi_\4(\2,\2)$ resonance}

In the quartet case, the charge-$2/3$ mass term of top sector is
\be
\mathcal{L}^{\rm P(F)_\4}\supset-\begin{pmatrix}\bar{t}&\bar{T}&\bar X_{2/3}\end{pmatrix}_LM_{2/3}^{\rm P(F)_\4}\begin{pmatrix}t^{\rm (P,F)}\\ T\\ X_{2/3}\end{pmatrix}_R+\text{h.c.},
\ee
where the mass matrices are
\be\begin{split}
M_{2/3}^{\rm P_\4}=&\begin{pmatrix}
0&-\frac{y_Lf}{2}(1+\sqrt{1-\xi})&-\frac{y_Lf}{2}(1-\sqrt{1-\xi})\\ -\frac{y_Rf\sqrt{\xi}}{\sqrt{2}}&M_\4&0\\
\frac{y_Rf\sqrt{\xi}}{\sqrt{2}}&0&M_\4
\end{pmatrix},\\
M_{2/3}^{\rm F_\4}=&\begin{pmatrix}
\frac{y_{2L}f\sqrt{\xi}}{\sqrt{2}}&-\frac{y_Lf}{2}(1+\sqrt{1-\xi})&-\frac{y_Lf}{2}(1-\sqrt{1-\xi})\\ 0&M_\4&0\\
0&0&M_\4
\end{pmatrix}.
\end{split}\ee
Those $M_{2/3}^{\rm P(F)_\4}$'s are not symmetric. Thus, instead of diagonalization, we should do the singular value decomposition, i.e. finding unitary matrices $U_t$ and $V_t$ such that $U_t^\dagger M_{2/3}^{\rm P(F)_\4}V_t$ is diagonal. Up to $\xi$ level, for partially composite $t_R^{\rm (P)}$ scenario we have
\be\label{mass_P}\begin{split}
&M_t=\frac{y_Ly_Rf^2\sqrt{\xi}}{\sqrt{2}\sqrt{M_\4^2+y_L^2f^2}},\\
&M_{T}=\sqrt{f^2 y_L^2+M_\4^2}+\frac{M_\4^2 y_R^2f^2\xi}{4 \left(f^2
   y_L^2+M_\4^2\right)^{3/2}},\quad M_{X_{2/3}}=M_\4+\frac{y_R^2f^2\xi}{4 M_\4},
\end{split}\ee
while for fully composite $t_R^{\rm (F)}$ scenario we have
\be\label{mass_F}\begin{split}
&M_t=\frac{M_\4 y_{2L}f\sqrt{\xi}}{\sqrt{2} \sqrt{f^2
   y_L^2+M_\4^2}},\\ 
&M_{T}=\sqrt{f^2 y_L^2+M_\4^2}-\frac{\left(M_\4^2-\left(y_{2L}^2-y_L^2\right) f^2\right)y_L^2f^2\xi}{4 \left(f^2
   y_L^2+M_\4^2\right)^{3/2}},\quad M_{X_{2/3}}=M_\4.
\end{split}\ee
In this scenario, the lightest charge-$2/3$ top partner $X_{2/3}$ has degenerate mass with $X_{5/3}$ up to $\xi$ order.

\subsubsection{The $\Psi_\1(\1,\1)$ resonance}

The fermion mass term is
\be
\mathcal{L}^{\rm P(F)_\1}\supset-\begin{pmatrix}\bar{t}&\overline{\widetilde T}\end{pmatrix}_L
M_{2/3}^{\rm P(F)_\1}\begin{pmatrix}t^{\rm (P,F)}\\ \widetilde T\end{pmatrix}_R+\text{h.c.},
\ee
where
\be
M_{2/3}^{\rm P_\1}=\begin{pmatrix}0&\frac{y_Lf\sqrt{\xi}}{\sqrt{2}}\\
-y_Rf\sqrt{1-\xi} &M_\1\end{pmatrix},\quad M_{2/3}^{\rm F_\1}=\begin{pmatrix}\frac{y_{2L}f\sqrt{\xi}}{\sqrt{2}}&\frac{y_Lf\sqrt{\xi}}{\sqrt{2}}\\
0 &M_\1\end{pmatrix}.
\ee
Singular value decomposition is used to find the mass eigenvalue, and up to $\xi$ order for P$_\1$,
\be
M_t=\frac{y_Ly_Rf^2\sqrt{\xi}}{\sqrt{2}\sqrt{M_\1^2+y_R^2f^2}},\quad
M_{\widetilde T}=\sqrt{f^2 y_R^2+M_\1^2}+\frac{M_\1^2y_L^2 f^2\xi}{4 \left(f^2y_R^2+M_\1^2\right)^{3/2}};
\ee
and for F$_\1$,
\be
M_t=\frac{y_{2L}f\sqrt{\xi}}{\sqrt{2}},\quad
M_{\widetilde T}=M_\1+\frac{y_L^2f^2\xi}{4 M_\1} .
\ee

\section{The NNLO cross sections for QCD pair production  of the top partners}\label{app:NNLO}

In this appendix, we list the cross section for the QCD pair production of the top parters. They are calculated using {\tt Top++2.0} package, at NNLO level with next-to-next-to-leading logarithmic soft-gluon resummation~\cite{Czakon:2011xx,Czakon:2013goa,Czakon:2012pz,Czakon:2012zr,Baernreuther:2012ws,Cacciari:2011hy}. The results are shown in Table~\ref{tab:ffbarNNLO}.

\begin{table}[h]
\centering
\begin{tabular}{|c|c|c|c|c|c|c|c|c|}
 \hline
Mass [TeV] & 1.0 & 1.2 & 1.4 & 1.6 & 1.8 & 2.0 & 2.2 & 2.4 \\ \hline
XS $@$ 13 TeV  [fb] & 42.9 & 11.5 & 3.48 & 1.13 & $0.386$ & $0.135$ & $0.0482$ & $0.0172$ \\ \hline
\end{tabular} 
\begin{tabular}{|c|c|c|c|c|c|c|c|}
 \hline
Mass [TeV] & 1.5 & 2.0 & 2.5 & 3.0 & 3.5 & 4.0 & 4.5 \\ \hline
XS $@$ 27 TeV  [fb]  & 61.9 & 9.43 & 1.91 & $0.455$ & $0.120$ & $0.0332$ & $0.00947$ \\ \hline
\end{tabular} 
\begin{tabular}{|c|c|c|c|c|c|c|}
 \hline
Mass [TeV] & 2 & 4 & 6 & 8 & 10 & 12 \\ \hline
XS $@$ 100 TeV  [fb]  & 858 & 19.8 & 1.68 & $0.244$ & $0.0467$ & $0.0107$ \\ \hline
\end{tabular} 
\caption{The NNLO cross sections for QCD pair production of the top partners at various collision energies of $pp$ collider.}
\label{tab:ffbarNNLO}
\end{table}

\section{The  extrapolating method}
\label{app:method}
In this appendix, we sketch the method we used to extrapolate the existing searches to the future high luminosity or high energy LHC. We refer the reader to \Ref{Thamm:2015zwa} for the detailed description of the method. The basic assumption of the method is that the same number of background events in the signal region of two searches with different luminosity and collider energy  will result in the same upper limit on the number of  signal events. To be specific, from an existing resonance search at collider energy $\sqrt{s_0}$ with integrated luminosity $L_0$, we can obtain the  95\% CL upper limit  on the  $\sigma\times\text{Br}$ for a given channel for the mass $m_\rho^0$, which is denoted as $[\sigma\times\text{Br}]^{95\%}(s_0,L_0;m_\rho^0)$. Note that the range of $m_{\rho}^0$ maybe different from different measurements. For each possible $m_\rho^0$ at collider energy $\sqrt{s_0}$ and luminosity $L_0$, we obtain the corresponding $m_\rho$ at collider energy $\sqrt{s}$ and luminosity $L$ with the same number of background in the small mass window around the resonance masses by solving the following equation:
\beq
\label{eq:beq}
B(s,L; m_\rho) = B(s_0,L_0;m_\rho^0).
\eeq
Then  the  95\% CL upper limit  on the $\sigma\times \text{Br}$ for the resonance mass $m_{\rho}$  at collider energy $\sqrt{s}$ with luminosity $L$ can be obtained as follows: 
\be
[\sigma\times\text{Br}]^{95\%}(s,L;m_\rho)=\frac{L_0}{L}[\sigma\times\text{Br}]^{95\%}(s_0,L_0;m^0_\rho).
\ee
For an explicit model, the $\sigma\times\text{Br}$ can be calculated and are functions of some model parameters $X$.  We can obtain the exclusion region in the parameter space $X$ as follows:
\beq
[\sigma\times \text{Br}](s, m_\rho, X) > [\sigma\times\text{Br}]^{95\%}(s,L;m_\rho).
\eeq
Note that \Eq{eq:beq} can be further expressed as an identity involving the parton luminosities associated with the background~\cite{Thamm:2015zwa}:
\be
\sum_{\{i,j\}}c_{ij}\frac{d\mathcal{L}_{ij}}{d\hat s}(m_\rho;\sqrt{s})=\frac{L_0}{L}\sum_{\{i,j\}}c_{ij}\frac{d\mathcal{L}_{ij}}{d\hat s}(m_\rho^0;\sqrt{s_0}), \quad c_{ij} \simeq \hat{s} \hat{\sigma}_{ij};
\ee
where $d\mathcal{L}_{ij}/d\hat s$ is the parton luminosity defined as~\cite{Thamm:2015zwa,Quigg:2009gg}:
\be
\frac{d\mL_{ij}}{d\hat s}(\sqrt{\hat{s}}, \sqrt{s})= \frac{1}{s}\int_{\hat s/s}^1\frac{dx}{x} f_i(x,\sqrt{\hat{s}})f_j(\frac{\hat s}{xs},\sqrt{\hat{s}}).
\ee
We have chosen the factorization scale to be the partonic center-of-mass energy $\sqrt{\hat{s}}$.
Note that if the signal and the main background come from the same parton initial states, the method is the same as in Ref.~\cite{Hinchliffe:2015qma}.

For the QCD pair production of top partners, we have chosen an invariance mass square window around $(2M_F)^2$, where $M_F$ is the mass of the top partner under consideration. This adjustment makes use of the fact that the heavy fermion pair is mainly produced at threshold. For single production (e.g. $tW$ or $tZ$ fusion) of fermion resonance, although there is no invariance mass peak in such channels, we still use extrapolation method in the invariance mass square at $(M_F+M_t)^2$ to set an estimate limit.

\bibliographystyle{apsrev}
\bibliography{references}

\end{document}